%% file: journalist-ideology.tex
\documentclass[12pt,hyperfootnotes=false, hidelinks=true]{article}
\usepackage{mathptmx}
\usepackage[latin9]{inputenc}
\usepackage{verbatim}
\usepackage{cprotect}
\usepackage{float}
\usepackage{amsmath}
\usepackage{amsthm}
\usepackage{amssymb}
\usepackage{graphicx}
\usepackage{geometry}
\usepackage{setspace}
\usepackage[authoryear]{natbib}
\usepackage[]
 {hyperref}

\makeatletter
\theoremstyle{plain}
\newtheorem{prop}{\protect\propositionname}
\theoremstyle{plain}
\newtheorem{assumption}{\protect\assumptionname}
\theoremstyle{plain}
\newtheorem{lem}{Lemma}
\theoremstyle{remark}
\newtheorem{rem}{\protect\remarkname}
\theoremstyle{plain}
\newtheorem{cor}{\protect\corollaryname}

\usepackage{amsmath,amsthm,bm}
\usepackage{setspace}
\usepackage{titlesec}
\titlespacing*{\section}{0pt}{0.2\baselineskip}{0.2\baselineskip}
\titlespacing*{\subsection}{0pt}{0.2\baselineskip}{0.2\baselineskip}
\titlespacing*{\subsubsection}{0pt}{0.2\baselineskip}{0.2\baselineskip}
\titleformat{\section}{\normalfont\large\bfseries}{\thesection}{1em}{}
\titleformat{\subsection}{\normalfont\normalsize\bfseries}{\thesubsection}{1em}{}
\titleformat{\subsubsection}{\normalfont\itshape}{\thesubsubsection}{1em}{}

\usepackage{booktabs}
\usepackage{array}

\usepackage{datetime}
\usepackage{pdflscape}
\usepackage[english]{babel}
\usepackage{times}

\usepackage[small]{caption}
\usepackage[bottom,hang,flushmargin]{footmisc}
\usepackage[caption=false,position=top]{subfig}
\usepackage{enumitem}
\setlist[itemize]{noitemsep, topsep=0pt}

\usepackage{graphicx}
\usepackage{pgfplots}
\pgfplotsset{compat = newest}
\usetikzlibrary{positioning,arrows.meta}
\usepgfplotslibrary{fillbetween}

\DeclareMathOperator{\Var}{Var}

\DeclareMathOperator{\E}{E}

\PassOptionsToPackage{pdftex,hidelinks}{hyperref}
\usepackage{hyperref}
\usepackage{natbib}

\usepackage{etoolbox}
\AtBeginDocument{}

\newlength{\mybibhang}
\apptocmd{\thebibliography}{%
  \setlength{\itemindent}{-\mybibhang}%
  \setlength{\leftmargin}{\mybibhang}%
}{}{}

\makeatother

\providecommand{\assumptionname}{Assumption}
\providecommand{\corollaryname}{Corollary}
\providecommand{\propositionname}{Proposition}
\providecommand{\remarkname}{Remark}

\input{input/scalars.tex}

\providecommand{\E}{\mathbb{E}}
\providecommand{\Var}{\mathrm{Var}}
\providecommand{\Lam}{\Lambda}

\begin{document}
\title{\noindent\textbf{Journalist Ideology and the Production of News:
Evidence from Movers}}
\author{\noindent Levi Boxell, \textit{Independent}\textbf{\thanks{E-mail:\ leviboxell@gmail.com, jacob.conway@chicagobooth.edu, nicolas.longuetmarx@berkeley.edu. We thank Elliott Ash, John
Dalton, Pascaline Dupas, Matthew Gentzkow, Andrew Hall, Tommy Leung,
Jesse Shapiro, Isaac Sorkin, and seminar participants at Stanford University, the University of Chicago, the Monash-Warwick-Zurich-CEPR Text-as-Data Workshop, and the IHS Summer Research
Fellowship for their feedback and suggestions. We thank Peng Zhang, Valentina Zvereva, and Donghang Zou for outstanding research assistance. We also thank Brandon
Williams for his excellent support with using ProQuest TDM Studio.
This material is based upon work supported by the National Science
Foundation Graduate Research Fellowship Program under Grant No. DGE-1656518.
Any opinions, findings, and conclusions or recommendations expressed
in this material are those of the authors and do not necessarily reflect
the views of the National Science Foundation. We also acknowledge
funding from the Institute for Humane Studies.}}\\
Jacob Conway, \textit{University of Chicago}\textbf{}\\
Nicolas Longuet-Marx, \textit{University of California, Berkeley}\textbf{}\\
\textbf{}\\
}
\date{October 2026}
\begin{spacing}{1.15}
\maketitle
\vspace{-2\baselineskip}
\begin{abstract}
\noindent{}%

What role do journalists play in determining the political slant of
the news they produce? We develop and estimate a model where journalists and newspaper
outlets contract over both slant and wages. The model implies a set
of conditions under which we can consistently estimate the role of
journalist preferences in driving the observed variation in slant
across outlets by leveraging journalist transitions between outlets.
To measure slant, we train a transformer-based, machine learning model
using articles tweeted by politicians and apply it to a full-text
database of \SummaryAllOutletsArticlesMillionsFloor{}+ million newspaper articles published in the US between
2013 and 2018. Under our model and identifying assumptions,
our estimates (a) reject the hypothesis that journalists have zero
ideological preferences over the content they produce and (b) imply
that \StructuralSlantVarianceJournalistWithSelectionTotalPercentRounded\% of the variation in slant can
be explained by journalists. When journalists move across outlets, their average slant shifts by \EventStudytzeroTfive\% of the gap between their destination and origin outlet averages.
In counterfactuals that reweight journalist party composition, moving from the observed left-leaning distribution to an even distribution of Democrats, Republicans, and non-partisans shifts average article slant
in a more conservative direction, but only by \SelEqualThirdsFixedSD{} standard deviations in slant. %
\bigskip{}
\end{abstract}
\end{spacing}
\pagebreak{}

\begin{spacing}{1.4}

\section{Introduction}\label{sec:introduction}

In the United States, \JournalistDemShareWeaverHardcodedfloor\ percent of journalists identify as a Democrat
whereas only \JournalistRepShareWeaverHardcodedfloor\ percent identify as a Republican (\citealt{Weaver2019AmericanJournalistDigitalAge}).\footnote{In 1971, \JournalistRepShareWeaverSeventyoneHardcoded\ percent of surveyed journalists in the US identified
with the Republican party and \JournalistDemShareWeaverSeventyoneHardcoded\ percent identified with the Democratic
party. By 2013, only \JournalistRepShareWeaverHardcoded\ percent of surveyed journalists identified
with the Republican party whereas \JournalistDemShareWeaverHardcoded\ percent continued to identify
with the Democratic party (\citealt{Weaver2019AmericanJournalistDigitalAge}). When focusing on Washington
correspondents, the partisan skew is more striking---with \JournalistVotingtoClintonSharePovich\ percent
of surveyed correspondents indicating they voted for Clinton in the
1992 election. The distribution of self-reported partisan identification
among journalists is mirrored by the overall distribution of political
contributions from members of the news and media industry (\citealt{Bonica2013IdeologyInterests}). See also Figure \ref{fig:weaver}.

Academics and media critics have also expressed concerns about an
overarching liberal media bias in the US---with these concerns being
expressed in a diverse set of outlets including \emph{The Washington
Post}, \emph{The Wall Street Journal}, and \emph{The Quarterly Journal
of Economics} (e.g., \citealt{Groseclose2005MeasureMediaBias,Wemple2017DearMainstreamMedia,Sauter2020WSJLean}).} In a 2019 Gallup survey, 42 percent of U.S. adults described the national news media as too liberal, compared with 13 percent who described it as too conservative. At the same time, trust in the mass media has fallen sharply: the share of Americans reporting a great deal or fair amount of trust declined from 51 percent in 2000 to 28 percent in 2025, and from 47 percent to 8 percent among Republicans (\citeauthor{Gallup2000MediaTrust} \citeyear{Gallup2000MediaTrust}; \citeyear{Gallup2019MediaIdeology}; \citeyear{Gallup2025MediaTrust}).

One proposed explanation is that the political preferences of journalists themselves affect the content produced by news organizations.\footnote{Matthew Yglesias, co-founder of Vox, stated ``\emph{I think content is getting
dragged left because that's the labor market (and to an extent the audience) rather than as
deliberate strategy}'' (\citealt{Smith2021NoahpinionYglesias}). See also
\citet{Groseclose2011LeftTurn}. High-profile exits of certain journalists are also linked to
within-outlet political representation. For example, in 2017, \emph{The New York Times} hired
Bari Weiss to diversify the paper's opinion page. Yet, in 2020, Weiss
resigned---citing ``\emph{constant bullying by colleagues who disagree with my views}''
(\citealt{Weiss2020ResignationLetter}).} Prior academic work provides theoretical mechanisms through which journalists' preferences can generate persistent media bias (e.g., \citealt{Baron2006PersistentMediaBias}). Whether these mechanisms operate empirically, and to what extent individual journalists shape the political content they produce, remains an open question.

Answering this question has proven difficult because journalists sort across outlets. More liberal journalists may work for more liberal news organizations, while more conservative journalists may sort into more conservative outlets. At the same time, outlets differ in their readership, editorial policies, and incentives to produce politically slanted content.

In this paper, we develop and estimate a model where journalists and newspaper
outlets contract over both slant and wages. This framework allows us to empirically examine the extent to which individual journalists shape the ideological content of the articles they write and the market provision of slant as a whole.

We start by building a simple model of the journalist labor market
and the production of news. Our model highlights how observed slant
is a weighted combination of three components: journalist preferences,
outlet average preferences, and a search-specific outlet shock. Our
model provides testable predictions for whether journalists have preferences
over slant, and it provides guidance for estimating the magnitude
of these preferences given potential selection in job transitions
induced by search-specific shocks to outlet preferences.

In order to take the model-informed estimators to the data, we construct
a novel measure of media bias leveraging recent advances in machine
learning. Using a new dataset of articles tweeted by politicians,
we fine-tune RoBERTa (a transformer-based model of human language)
to predict the political party of the politician who shared a given
article. Our fine-tuned model performs well on both in-sample cross-validation
tests and out-of-sample comparisons to human coders, and it represents
a significant improvement over bag-of-words methods used previously
in the literature.

We then apply our measure of slant to predict the partisan slant of
\SummaryAllOutletsArticlesMillionsFloor{}+ million articles published between 2013 and 2018 from
almost \SummaryAllOutletsOutletsCeilHundreds\ newspapers in the ProQuest US Newsstream database (accessed
via TDM Studio). Importantly, the ProQuest database also includes
the byline for each article---allowing us to construct an author-level
panel dataset that tracks the ideological slant of the articles written
by a given journalist as they move across outlets. 

Our empirical strategy for the observational data builds on the rich
literature in labor economics that estimates worker and firm wage
fixed effects (e.g., \citealt[AKM]{Abowd1999HighWage}) or worker
preferences (e.g., \citealt{Sorkin2018RankingFirms}) using employee
transitions between firms.\footnote{Our results
are consistent with a persistent journalist component in measured slant.
Estimated journalist fixed effects also differ by recorded party affiliation
in the voter-linked sample. We interpret these patterns through a model of
ideological preferences, but the empirical designs do not isolate those
preferences as the causal mechanism. Other persistent attributes, such as
source networks, beat expertise, writing style, or reporting skills, may
contribute to these patterns and may themselves be related to ideology.}

We start by estimating an event-study specification using a sample
of consistent movers.\footnote{Among journalists, there are a wide variety of work arrangements including
both freelancers and salaried employees. In our event study analysis,
we constrain our analysis to likely salaried employees by focusing
on journalists who wrote $T$ or more articles at the origin outlet
before switching and writing $T$ or more articles at the destination
outlet.} Under our model and assumptions, if we are able to reject the hypothesis
that journalists close the entire origin-destination gap between outlets,
we can also reject the hypothesis that journalists have zero preferences
over the slant they produce. In other words, if journalists simply
write content that mirrors the preferred slant of an outlet and have
zero ideological preferences, the gap in slant between the origin
and destination outlets would close entirely when journalists switch
outlets. On the other hand, if journalists have strong ideological
preferences, we would expect there to be little-to-no change in observed
slant when a journalist switches outlets. 

Under these assumptions, our event-study estimates reject the hypothesis that journalists have
zero ideological preferences over the slant they produce. Using our
sample of consistent movers, journalists close roughly \EventStudytzeroTfive\ percent
of the gap in average slant between the origin outlet and the destination
outlet. The 95 percent confidence intervals on our event study estimates
exclude both 0 percent and 100 percent.

We also use the AKM fixed effect framework to decompose the variation
in article-level slant using journalist and outlet fixed effects.
In our preferred specification, restricted to the set of movers and using a leave-out correction, we
estimate that journalist fixed effects explain \KSSJournalistExplainedAKM\ percent of the observed
variation in slant, outlet fixed effects explain \KSSOutletExplainedAKM\ percent, and sorting
between journalists and outlets explains another \KSSCovarianceExplainedAKM\ percent.\footnote{The remaining variation is explained by time fixed effects, other
covariance terms, and the residual.} 

We also examine the relationship between journalist fixed effects
and other observable characteristics of journalists. As we would expect,
the distribution of journalist fixed effects for Republican journalists
is further to the right (i.e., more conservative) than the distribution
for other journalists. Women and non-white journalists tend to have
more liberal fixed effects than men and white journalists.

AKM decompositions restrict complementarities between journalists and outlets. As a robustness check, we re-estimate the decomposition using the framework of \citet{Bonhomme2019DistributionalFramework}, which discretizes journalists and outlets into types and allows for nonparametric interactions between the two. The estimated interaction component is negligible: about \BLMInteractionVarianceSharePct{} percent of the variation in slant.

While interesting, we view these decomposition results as primarily descriptive
given the search-specific shocks to outlet preferences highlighted
in our model which may violate the conditional mean independence
assumption of these models. 

We derive a new estimator to provide causal estimates of the
\emph{between }outlet variation in slant driven by journalists.
By focusing on only the \emph{between} outlet variation
in slant, we are able to provide causal estimates under a weaker set
of assumptions than the conditional mean independence assumption of
the AKM model.\footnote{Moreover, we may be particularly interested in the role of journalists
in driving the \emph{between} outlet variation in slant, rather than
both the \emph{between }and \emph{within }outlet variation. For example,
if consumers are choosing between outlets to subscribe to, between
outlet variation in slant may be more relevant for understanding how
variation in slant affects consumer welfare.} Rather than assuming no selection on the search-specific shocks (as
the AKM model assumes), we allow movers to be selected on these shocks,
but require that, within each origin--destination pair, the average
shock deviation from the outlet mean is the same at the origin and
destination.

Our preferred estimates from this specification suggest journalists
explain \JournalistExplainedOutletDecompResidualizedPercent\ percent of the observed variation in average slant between
outlets. This is broadly similar in magnitude to the \ConsumerExplainedGSMediaBias\ percent of outlet-level variation
in slant attributed to consumers in \citet{Gentzkow2010WhatDrivesMediaBias}.

Finally, we model selection directly, adapting our framework into an empirical model of journalists and outlets contracting over slant. In the estimated model, journalists explain \StructuralSlantVarianceJournalistWithSelectionTotalPercentRounded{} percent of the variation in article slant. We use the estimated model to conduct counterfactuals that change the party composition of the journalist pool to be representative of the population. Setting the pool to equal thirds of Democrats, Republicans, and non-partisans moves mean slant by only \SelEqualThirdsFixedSD{} standard deviations.

Overall, these results suggest that while journalists have ideological preferences that influence the production of slant, they play a relatively moderate role in driving the market provision of slant.%

This paper relates to several strands of literature.

First, our paper relates to a literature on media bias. Within this
literature, our paper makes both a methodological and a substantive
contribution. 

Methodologically, previous work has used bag-of-words or citation
based measures of media bias (e.g., \citealt{Gentzkow2010WhatDrivesMediaBias,Groseclose2005MeasureMediaBias}).
We leverage recent advances in machine learning, BERT-based transformer
models, to measure media bias and show how these more advanced methodologies
can show substantial improvement in predicting the human perceived
slant of an article. This builds on separate work using recent advances
in machine learning to measure non-verbal media bias based on the
facial expression of politicians in images used on news websites (\citealt{Boxell2022SlantedImages}). 

We also make substantive contributions to the literature on media
bias. Much of the prior empirical work on media slant has focused
on demand-driven motives for media bias (e.g., \citealt{Gentzkow2010WhatDrivesMediaBias,Martin2017BiasCableNews,WidmerEtAl2026MediaSlantContagious}),
conglomeration and owner effects (e.g., \citealt{Martin2019LocalNewsNationalPolitics}),
or professional norms (e.g, \citealt{Baum2008NewMediaPolarization,Shapiro2016SpecialInterests}). 
Recent work also uses article-level slant measures to study polarization
in news consumption on social media (\citealt{BraghieriEtAl2025ArticleLevelSlant}). 

The three closest related papers empirically are \citet{CageEtAl2025PoliticalBiasMedia}, \citet{Srinivasan2021JournalistsDriveMediaSlant},
and \citet{Xu2024NewsBiasFinancialJournalists}.\footnote{\citet{Baron2006PersistentMediaBias} showed theoretically how journalist
discretion and preferences could drive persistent media bias.} \citet{CageEtAl2025PoliticalBiasMedia} use an AKM-style decomposition
to show that much of the variation in slant on French television and
radio shows can be explained by host- or journalist-specific fixed
effects. In contemporaneous work, \citet{Srinivasan2021JournalistsDriveMediaSlant}
finds smaller average responses to outlet slant in cross-outlet
comparisons of journalists' writing at eight US online news outlets.
Our broader sample and event-study focus on sustained outlet transitions
may contribute to the different estimates. \citet{Xu2024NewsBiasFinancialJournalists} also leverages
journalist turnover to examine how financial journalists' social
networks influence the favorability of coverage towards firms in the
US.\cprotect\footnote{\citet{BarberaSood2018FollowYourIdeology} find a strong correlation
between journalist campaign contributions and the ideological slant
of the content they produce. See also \citet{Hassell2020NoLiberalMediaBias}
and \citet{Graves2016FactCheckingFieldExperiment}.%
} 

Our model, however, highlights the limits of the AKM-style variance
decomposition estimates given the potential for selection into journalist
transitions. We solve this selection issue by providing novel estimators
and associated conditions with which we can recover the role
of journalist preferences in driving observed variation in slant across
outlets. Overall, our empirical results corroborate \citet{CageEtAl2025PoliticalBiasMedia}
and \citet{Xu2024NewsBiasFinancialJournalists} in finding important
effects of journalists, but in the broader US print newspaper market
as opposed to French broadcasting or US financial news.

Second, our paper relates to a literature on employee preferences
and compensating differentials. Some work examines the importance
of compensating differentials in general (e.g., \citealt{Sorkin2018RankingFirms}),
whereas other work focuses on more specific dimensions such as a firm's
corporate social responsibility behavior (e.g., \citealt{Hedblom2019CSR})
or the co-partisanship of co-workers (e.g., \citealt{Gift2015PoliticsHiring,Colonnelli2024PoliticsAtWork}).
Interpreted through our model, our results provide evidence on employee preferences over the ideological
content of their production in the news industry.

\section{Model of News Production and Empirical Implications}\label{sec:model_production}

We build a simple model of news production with heterogeneous journalists
and outlets.

\subsection{Journalists}

Journalist $i$ has utility from a job at outlet $j$
\[
u_{ijt}(x_{ij},\omega_{ij})=\omega_{ij}-\beta_{i}(x_{ij}-\delta_{i})^{2}+\epsilon_{ijt}
\]
where $x_{ij}$ is the average slant of articles written, $\delta_{i}$
is the bliss point for journalist $i$, $\omega_{ij}$ is an unobserved
utility transfer from the outlet to the journalist (e.g., wages and
benefits). The outside option gives utility
$\omega_{i0t}+\epsilon_{i0t}$. The shocks $\epsilon_{ijt}$ and
$\epsilon_{i0t}$ are alternative-specific type-1 EV shocks that are
independent across individuals, alternatives, and time. Note that for a given match,
utility can be transferred by giving journalists assignments closer
to their bliss point $\delta_{i}$ or increasing the unobserved utility
transfer (e.g., wages).

\subsection{Outlets}

An outlet $j$ has preferences over their portfolio of journalists
$A_{j}$
\[
u_{j}(A_{j})=R_{j}(A_{j})-\sum_{i\in A_{j}}\omega_{ij}
\]
where 
\[
R_{j}(A_{j})=\sum_{i\in A_{j}}\left[\nu_{j}-\beta_{j}(x_{ij}-\delta_{j}-\xi_{ij})^{2}\right].
\]
Here, $\nu_j$ is fixed revenue per journalist, $\delta_j$ is the outlet's baseline
slant ideal point, $\beta_j$ measures its strength of slant preferences, and
$\xi_{ij}$ is a shock independent across \emph{searches}.

Outlets have preferences over revenue $R_{j}(A_{j})$ minus the wage
transfers to journalists $\sum_{i\in A_{j}}\omega_{ij}$.

\subsection{Matching and Contracting}

There is an initial distribution of journalists across outlets. Each
period, firms draw a vector of search-specific preference shocks $\xi_{j}$
of length $s_{j}$. Journalists are then matched to firm-searches,
where journalist $i$ is matched to firm $j$'s search $s$ with some
probability $p_{ijs}(\cdot)$.\footnote{If journalist $i$ is matched
to outlet $j$'s search $s$, we write $\xi_{ij}=\xi_{js}$.}
We do not take a stand on the inputs
to $p_{ijs}(\cdot)$---it could include previous values of slant
$x$, previous wages $\omega$, previous firms $j$, and/or previous
or current search-specific shocks $\xi$. 

The prospective outlet $j$ observes the journalist's current contract
$(x_c,\omega_c)$, type $(\delta_i,\beta_i)$, and outside option
$\omega_0$, and offers a contract $(x,\omega)$.\footnote{We suppress
journalist and time subscripts on contract terms and outside options
when there is no ambiguity.} The journalist then draws the alternative-specific
shocks described above, which are independent of variables realized
before the choice, and chooses whether to remain at current outlet $c$,
move to prospective outlet $j$, or take outside option $0$.
We assume journalist's output is perfectly observable to the
firm, contract enforcement is costless, and both journalists and outlets
are myopic.

Given a match with journalist $i$ who has current contract $(x_c,\omega_c)$,
the prospective outlet $j$ chooses a contract $(x,\omega)$ that solves
\[
\max_{x,\omega}\underbrace{\Delta U_{j}(x,\omega)}_{\text{Change in Profit}}\times\underbrace{M_{i}(x,x_c,\omega,\omega_c,\omega_0)}_{\text{Prob. Journalist Accepts Offer}},
\]
where 
\[
M_{i}(x,x_c,\omega,\omega_c,\omega_0)=\frac{\exp\left(\omega-\beta_{i}(x-\delta_{i})^{2}\right)}{\exp(\omega_0)+\exp\left(\omega-\beta_{i}(x-\delta_{i})^{2}\right)+\exp\left(\omega_c-\beta_{i}(x_c-\delta_{i})^{2}\right)}.
\]
and

\[
\Delta U_{j}(x,\omega)=\nu_j-\beta_j(x-\delta_j-\xi_{ij})^2-\omega.
\]

\begin{prop}\label{prop:optimal_contract}
If $\beta_i,\beta_j\geq0$ and $\beta_i+\beta_j>0$, the outlet's unique
optimal contract has slant $x_{ij}^{*}=\frac{\beta_{i}\delta_{i}+\beta_{j}(\delta_{j}+\xi_{ij})}{\beta_{i}+\beta_{j}}$
and wage $\omega_{ij}^{*}=h_{ij}(x_{ij}^{*},x_c,\omega_c,\omega_0)$,
where $h_{ij}$ is determined by the journalist and outlet primitives.

\textbf{Proof}: See Appendix \ref{sec:appendix}.
\end{prop}
Holding the journalist and outlet primitives fixed, the optimal wage
can depend on the journalist's current contract $(x_c,\omega_c)$ and
outside option $\omega_0$, whereas the optimal slant contract cannot.

The distribution of $\xi_{ij}$ across \emph{accepted} matches can differ
from its distribution across \emph{searches}. For example, if journalists have strong slant preferences
and if journalists are all more conservative than the most conservative
outlet, then journalists will be more likely to accept a slant contract
with a higher $\xi_{ij}$, potentially raising the mean shock among accepted matches.
In this sense, \emph{realized} variation in $\xi$ can also be a function
of journalist preferences.

Zero journalist slant preferences eliminate this acceptance-stage
selection, but do not by themselves eliminate selection into the offered
pool.
\begin{lem}\label{lem:acceptance_nonselection}
For any outlet $j$ with $\beta_j>0$, if $\beta_i=0$, then, conditional on the journalist's
current wage and outside option, accepting outlet $j$'s offer does not
further select on $\xi_{ij}$: the conditional mean of $\xi_{ij}$ among
journalists who accept outlet $j$'s offer equals its conditional mean
among journalists who receive outlet $j$'s offer.

\textbf{Proof}: See Appendix \ref{sec:appendix}.
\end{lem}
\begin{prop}\label{prop:no_preferences}
For any outlet $j$ with $\beta_j>0$, if $\beta_i=0$ for all $i$ and
\[
\mathbb{E}\left[\delta_j+\xi_{ij}\mid \omega_c,\omega_0,
i\text{ receives }j\text{'s offer}\right]=\mathbb{E}[\delta_j+\xi_{ij}],
\]
then
\[
\mathbb{E}[x_{ij}^*\mid i\text{ accepts }j\text{'s offer}]=\mathbb{E}[\delta_j+\xi_{ij}],
\]
where the unconditional expectation is across searches.

\textbf{Proof}: See Appendix \ref{sec:appendix}.
\end{prop}
The lemma shows that zero journalist slant preferences prevent acceptance
from adding further selection on outlet preferences beyond what exists at the offer stage. The proposition's conditional mean restriction then assumes no selection on mean outlet preference at the offer stage, leading to a continued lack of selection at the acceptance stage given zero journalist slant preferences.

\section{Data\label{sec:data}}

Our data is derived from three primary sources: articles shared by
politicians on Twitter, full-text newspaper articles from ProQuest,
and voter registration data from L2. We use the politician-shared
articles to build a labeled training dataset, which allows us to train
a machine learning model for predicting an article\textquoteright s
political slant. We use the full-text data on newspaper articles to
track journalists as they move across outlets and to measure how the
slant of their content varies as they move. And we use the voter registration
data to determine the ground-truth partisan affiliation of journalists
in our data set. 

\subsection{Articles Tweeted by Politicians}

We construct a labeled dataset of news articles shared by either Republican
or Democratic politicians. The labeled nature of this dataset (either
Republican or Democratic) is important for training our measure of
media slant later. 

We start with a list of active Twitter accounts by politicians from
ProPublica's Politwoops dataset released in July 2019. We then scraped
the tweets from each username listed---giving us the tweet history
from more than \NPoliticianUsernames\ usernames associated with politicians. 

From the dataset of politician tweets, we locate all URLs tweeted
by politicians---giving more than \NTweetedUniqueURLsMillionsFloor\ million unique tweeted URLs.\footnote{We exclude urls including ``youtu.be,'' ``pscp.tv,'' ``facebook.com,''
``instagram.com,'' ``patreon.com,'' and ``google.com.''} We then randomly sample \NSampledURLs\ URLs and scrape the text associated
with each URL. Not all URLs are successfully scraped. We obtain \NArticlesNonEmptyRows\
URLs with some text. We further exclude URLs with either (a) \MinArticleChars\
or less characters or (b) no author detected by our scraper. We use LLMs to identify and remove non-news content, and to assign IPTC media topics to news articles.\footnote{Additional detail on our data processing can be found in Appendix Section \ref{sec:data_processing}. Information on our topic assignments specifically can be found in Appendix Section \ref{sec:topics}.} This
gives a final dataset of \Nfinaltweets\ non-unique URLs with text. \NfinaltweetsRep\
of these URLs were shared by Republican politicians, and \NfinaltweetsDem\ of
these URLs were shared by Democratic politicians. Appendix Figure
\ref{fig:twitter_tabulate_plot} summarizes the outlet-level distribution
of these article tweets by the party of the politician who shared them.

\subsection{ProQuest Full-Text News}

We also use a full-text database of news articles from more than \SummaryAllOutletsOutlets\
US newspapers, including major US daily newspapers such as \emph{The
New York Times} and \emph{The Wall Street Journal}. The full-text
nature of the data enables us to improve upon bag-of-word or keyword-based
approaches to measuring media slant (e.g., \citealt{Gentzkow2010WhatDrivesMediaBias})
by applying recent machine learning advances that leverage textual
context (e.g., BERT or RoBERTa), thereby allowing us to construct
more accurate measures of slant at the article level. Moreover, the
large number of outlets allows us to track journalists as they move
across outlets and provides insights across both national and local
newspaper markets. 

We obtain the full text for all articles from the ProQuest US Newsstream
(accessed via TDM Studio) published between January 2013 and December
2018.%
{} This provides more than \SummaryAllOutletsArticlesMillionsFloor\ million articles. For most of the analysis,
we restrict to articles labeled as ``News'' by ProQuest.

For articles with multiple authors on the same byline, we reshape
the data to be at the author-article level. So, an article that is
written by two authors would appear twice in our dataset---once for
each author.%
\footnote{We assume any individual with the same name is the same person, and
we clean author names by replacing common female and male nicknames
using the list of nicknames from \href{https://github.com/uguryi/abeR/tree/master/inst/extdata}{https://github.com/uguryi/abeR/tree/master/inst/extdata},
which is derived from \citet{Abramitzky2021AutomatedLinking}. } (From here-on, when we refer to an ``article,'' we mean an ``article-author''
pair.) We restrict attention to authors with more than \MinAuthorArticlesHardcoded\ articles
in our data and perform various cleaning steps in an attempt to remove
wire services and syndicated articles, and to handle outlet mergers.\footnote{Additional details on duplicate removal and other data-processing steps appear in Appendix Section \ref{sec:data_processing}. We combine evening and day editions, city-specific editions,
online editions, and outlet mergers into a single outlet where known.}

After the restrictions, we are left with a dataset of \SummaryAllOutletsArticles\ articles
from \SummaryAllOutletsAuthors\ authors published in \SummaryAllOutletsOutlets\ outlets. Table \ref{tab:summary}
provides summary statistics for our sample. Roughly a third of the
authors in our dataset write for the same outlet throughout our analysis
period, while the remaining two-thirds of authors switch outlets at
some point. %
However, there appears to be heterogeneity in the share
of movers versus non-movers across outlets. For example, many authors
at \emph{The Wall Street Journal }never write for another outlet,
whereas the vast majority of authors at \emph{The Chicago Tribune
}switch outlets at some point. These differences may be a combination
of differences in the use of freelancers versus staff writers, job
stability, syndication agreements, and the use of wire services.

\subsection{Voter Registration Data}

We also use voter registration data from L2 to match journalists to
their party registration status. To match journalists, we first apply
a similar name cleaning procedure to the L2 data. We then assign a
journalist a specific party if matched on first and last name to a
set of voters with a unique \emph{party, }after restricting to voters\emph{
}in the same state as the outlet headquarters. This allows a journalist
to be matched to multiple voters in a given state, as long as those
voters all have the same party affiliation. The party assignment is
done at the author-outlet level---thus allowing authors to change
party affiliation as they change outlets.%

Of the \NMatchedJournalists\ journalists we matched to voter registration
records (covering \JournalistLTwoMatchedShare\ percent of the cleaned author
sample), \JournalistDemShare\ percent registered as Democrats, \JournalistNonPartisanShare\ percent as non-partisan,
and \JournalistRepShare\ percent as Republican. Appendix Table
\ref{tab:linkage_summary} compares all journalists to the subset
linked to L2 records, while Appendix Table \ref{tab:party_summary}
summarizes journalist characteristics by party affiliation.

\section{Measuring Political Slant and Stylized Facts}

We use the politician-shared articles as a labeled dataset to fine-tune
an existing transformer-based machine learning model (RoBERTa) to
predict the political party of the politician sharing each article
based on the article\textquoteright s text. Our fine-tuned RoBERTa
model performs well in the leave-out validation set and is a significant
improvement over previous methods when evaluated on a separate human-labeled
dataset of political news articles. 

\subsection{RoBERTa Model}

To measure political slant, we use the RoBERTa model framework (\citealt{liu2019roberta}%
). The RoBERTa model builds on the previously successful BERT transformer
architecture (\citealt{devlin2019bert}%
). (See \citealt{vaswani2017attention}%
{} for more on transformers.) 

BERT stands for Bidirectional Encoder Representations from Transformers
and is an unsupervised pre-trained model. To start, BERT uses word
embeddings for each token from a \BERTVocabSizeHardcoded\ token vocabulary taken from
\citet{Wu2016GNMT}%
. These word embeddings are then split into sequences at the sentence
level (each sequence having two sentences). The BERT model is then
trained under a dual objective function. First, the BERT model will
randomly ``mask'' or hide certain tokens (i.e., words) and the training
objective is to successfully predict the masked tokens based on the
surrounding context.%
{} Second, the BERT model is tasked with predicting the next sentence
in a sequence. A key contribution of BERT is the ability to use the
left \emph{and }right context simultaneously in making these masked
word predictions, as opposed to the unidirectional or left-to-right
predictions used in many other language models (e.g., OpenAI's GPT).
The BERT model was pre-trained on data from the BooksCorpus (\citealt{zhu2015aligning}%
) and English Wikipedia, giving a combined total of \BERTPretrainingWordsBillionsHardcoded\ billion words. 

After pre-training the BERT model on the masked word task, the model
can be subsequently fine-tuned on various downstream tasks, such as
next sentence prediction, classification, or question answering.
When released in 2018, BERT achieved state-of-the-art performance
on many of the standard NLP tasks including both sentence-level and
token-level tasks. The base BERT model has \BERTBaseLayersHardcoded\ layers, a hidden size
of \BERTBaseHiddenSizeHardcoded, and \BERTBaseAttentionHeadsHardcoded\ self-attention heads---giving a model with \BERTBaseParamsMillionsHardcoded\ million
parameters. The large version of the BERT model has \BERTLargeParamsMillionsHardcoded\ million parameters. 

RoBERTa, or robustly optimized BERT approach, is one of several refinements
to the BERT model. It improves on BERT's performance by pre-training
on more data and larger batches, training on longer sequences, dynamically
changing the masking pattern, and removing the next sentence prediction
objective. These modifications resulted in significant improvements
over BERT across multiple NLP tasks.

\subsection{Fine-Tuning RoBERTa and Measurement of Slant}

The RoBERTa model is a pre-trained model for masked word prediction.
We use the base RoBERTa model available in the HuggingFace library
(\href{https://huggingface.co/roberta-base}{https://huggingface.co/roberta-base}). 

Our objective is to accurately predict the \emph{perceived} political
slant of an article. That is, how liberal or conservative would a
reader think a given article is? To achieve this objective, we use
our dataset of articles tweeted by politicians to fine-tune the RoBERTa
model.

We label each article according to the party of the politician who tweeted it.
We then randomly select \TrainSplitPercent\
percent of the articles to be the training dataset and the remaining
articles to be the validation dataset.\footnote{Since an article can be tweeted by more than one politician and we
allow duplicate articles in our dataset in general, we exclude all
articles in the validation dataset that match the text of an article
in the training dataset. This helps prevent cross-contamination of
our training and validation datasets.} To avoid class imbalance in training,
we reweight observations so that articles tweeted by Democratic and Republican
politicians receive equal total weight within each tweet month; see Appendix
Section \ref{sec:tweet-linked-article-data} for detail.

Since the RoBERTa architecture assumes a sequence length of 512 tokens
(where one token is roughly one word), we restrict the initial sections
of each article when they are longer than the required sequence length
and we ``pad'' articles that are shorter than required. Our fine-tuning
task is a sequence classification task with two labels (Republican
or Democratic). We use the default training parameters from the HuggingFace
library for RoBERTa (e.g., cross-entropy loss), except that we use:
a batch size of \RoBERTaEffectiveBatchSize, \RoBERTaAttentionDropoutPercent\ percent attention dropout, \RoBERTaHiddenDropoutPercent\ percent hidden
dropout, a learning rate of $\RoBERTaLearningRate$, a linear learning
rate scheduler with \RoBERTaWarmupSteps\ step warmup, and \RoBERTaTrainingEpochs\ training epochs. These
hyper-parameters were chosen by a manual search over values. 

We choose the model weights that achieve the lowest cross-entropy
loss in the validation dataset at the end of each epoch, which was
achieved on the fourth epoch. The training loss was \RoBERTaTrainingLoss, the validation
loss was \RoBERTaValidationLoss, and the balanced accuracy in the validation dataset was \TweetClassificationValidationBalancedAccuracy\ percent, with a ROC-AUC of \TweetClassificationValidationROCAUC. Appendix Figure
\ref{fig:monthly_oos_accuracy} reports validation balanced accuracy
by tweet month, and Appendix Table \ref{tab:slant_measure_comparison} summarizes
the highest-contribution tokens in the fine-tuned model.\footnote{Appendix
Table \ref{tab:slant_measure_comparison} also compares the
highest-contribution tokens, model diagnostics, and key results across our
baseline and two alternative models that use an LLM to anonymize geography and
other potential location proxies in the politician-tweeted articles used as
training data. See Appendix Section \ref{sec:slant-measure-comparison} for
details.}

Using the fine-tuned RoBERTa model, we then predict the partisanship
of every article in our ProQuest dataset.\footnote{One potential concern is that the model
could learn non-slant text features predictive of outlet identity,
rather than slant itself. To assess this concern, we compare out-of-sample predictive performance for models trained on article-stratified vs.\ outlet-stratified splits of a common sample; for the latter, a given outlet appears only in the training data or only in the held-out data for that split, but not both. Appendix Figure \ref{fig:outlet_stratification} reports the results. Across five split assignments, outlet-stratified mean balanced accuracy is
\OutletStratificationHeldoutBalancedAccuracyPct\ percent vs. \OutletStratificationMixedBalancedAccuracyPct\ percent for
article-stratified splits ($p=\OutletStratificationHeldoutMinusMixedBalancedAccuracyGapWelchPValue$
for the difference); mean ROC-AUC is \OutletStratificationHeldoutROCAUC\ and
\OutletStratificationMixedROCAUC, respectively
($p=\OutletStratificationHeldoutMinusMixedROCAUCGapWelchPValue$).
Average performance is somewhat (but not always statistically significantly) lower for outlet-stratified metrics. This difference could reflect learning about outlet-specific
features related or unrelated to human-perceived slant, or other
differences between the splits, such as the composition of the held-out
samples.}

\subsection{External Validation of Slant Measure}
\label{sec:slant-external-validation}

In addition to the leave-out validation dataset of articles tweeted
by politicians, we validate our measure of slant by comparing to human-coded
measures of slant at the article level from \citet{Budak2016FairBalanced}. 

\citet{Budak2016FairBalanced} dataset contains a sample of over \BudakSampleArticlesRoundedDownTenThousand\
political news articles from \BudakSampleOutlets\ popular news outlets in 2013. Each
article was then rated by MTurk coders on a 5-point likert scale for
its favorability towards the Republican Party and, separately, for
its favorability towards the Democratic Party. Each article was assigned
to two coders, one who observed the outlet name and one who did not.
Following \citet{Budak2016FairBalanced}, our validation measure uses
the rating from the coder who did not observe the outlet name, with
human-perceived partisanship defined as the relative difference in
perceived favorability towards the Republican Party versus the Democratic
Party.

For a random subset of these articles, we scraped the underlying text
from the source URL and predicted the slant using three different
methods (a) our RoBERTa slant measure described above, (b) a FastText
model (\citealt{Joulin2017BagOfTricks}%
) that we also train on the politician-tweeted articles, and (c) a
phrase-based classifier following \citet{Gentzkow2019MeasuringGroupDifferences},
re-estimated on the same article training sample as RoBERTa. We use its
Republican posterior probability under equal party priors. Panel A of Figure \ref{fig:human_compare_rank} reports the Spearman
rank correlation for each of the three text-based, machine-coded measures
of slant and the human-coded slant measure from \citet{Budak2016FairBalanced}.
While all three machine-coded methods have a significant positive
correlation with the human-coded measure, our RoBERTa-based slant
measure has the highest rank correlation.

Panel B of Figure \ref{fig:human_compare_rank} reports results from
a partitioning exercise where we examine the ability of each method
to accurately partition the articles in half based on their relative
partisan leanings. Our RoBERTa slant measure accurately categorizes
over \BudakRobertaPartitionAccuracy\ percent of articles that are perceived to be clearly partisan
by human coders---again, performing better than these other methods.

Overall, we believe these results suggest that both (a) our RoBERTa-based
slant measure is performing well in capturing text-based variation
that humans would perceive as political slant, and (b) our RoBERTa-based
slant measure is a significant improvement over previous methods used
in the literature.

Appendix \ref{sec:appendix-1} reports additional results comparing
our model's performance against human-coded articles, as well as comparisons
to previous measures of slant or partisanship in the literature.

\subsection{Stylized Facts }

\subsubsection{Sorting of Journalists}

In Figure \ref{fig:outlet_slant_partisanship}, we compare the average
slant at an outlet to the share of partisan journalists matched in
the L2 data who are Republicans. Overall, there is a strong, positive,
and significant correlation between the average slant of outlets and
the share of Republican journalists. 

\emph{The New York Times }and \emph{The Wall Street Journal }are interesting
reference points. While both outlets are located in New York City,
and thus can pull from the same potential set of employees, the share
of Republicans at \emph{The Wall Street Journal }is roughly twice
that of \emph{The New York Times}. And, correspondingly, the average
slant at \emph{The Wall Street Journal }is to the right of the average
slant at\emph{ The New York Times. }

\subsubsection{\emph{The New York Times} and \emph{The Wall Street Journal}}

We examine the distribution of author-level slant at \emph{The New
York Times} and \emph{The Wall Street Journal }across all authors
and separately for the set of authors who have written for both outlets.

Panel A of Figure \ref{fig:nyt_vs_wsj} plots the distribution of
author-level slant at \emph{The New York Times} and \emph{The Wall
Street Journal. }Consistent with prior suggestion, we see that the
average slant of all authors at \emph{The New York Times }is shifted
to the left relative to the average slant of authors at \emph{The
Wall Street Journal}. 

Panel B of Figure \ref{fig:nyt_vs_wsj} is the same as Panel A, except
it restricts attention to the set of authors who have written for
both outlets. Among this set of authors, we see that the distributions
of slant at \emph{The New York Times }and \emph{The Wall Street Journal
}are more similar. Through the lens of our model, the set of authors
who have written at both outlets either have significantly different
ideal points $\delta_{i}$ than the general set of authors at \emph{The
Wall Street Journal }and/or have large search-specific shocks $\xi$. 

Lastly, Appendix Figure \ref{fig:opeds} examines the average slant
of select politicians when writing op-eds at \emph{The New York Times
}and \emph{The Wall Street Journal. }As we would expect, among these
politicians, the average slant of Republican politicians is to the
right of the distribution at both outlets relative to the average
slant of Democratic politicians.

\subsubsection{Trends in Slant}

We also examine trends between 2013 and 2018 in slant at key outlets,
as well as the newspaper market overall. Figure \ref{fig:slant_trends}
shows a secular shift to the left over time in slant for key outlets
as well as the market as a whole. While this trend is gradual for
the market as a whole, the 2016 election marks a turning point for
many national news outlets. Several national US dailies exhibit a
rapid shift leftwards after the 2016 election including for \emph{The
Wall Street Journal---}an outlet typically viewed as more favorable
towards Republicans. Of the major national US dailies examined, it
is interesting to note that while \emph{USA Today} had similar levels
of slant to \emph{The Washington Post }in 2013, it reaches parity
with \emph{The Wall Street Journal }in the aftermath of the 2016 election---highlighting
how outlets differentially responded to the election. \emph{USA Today}
has a higher share of Republican journalists than the other outlets
considered (see Figure \ref{fig:outlet_slant_partisanship}). %

\section{Evidence from Journalist Movers}

We now turn to examining the role of journalists in determining the
slant of the content they produce.

We start by providing preliminary evidence on journalist content preferences
using standard mover event study frameworks and AKM variance decompositions.
We provide guidance on the extent to which we can consider these estimates
causal, as opposed to descriptive, in light of our model.

To make progress empirically, we assume that journalists and outlets
are only heterogeneous in their bliss points $\delta$ and have common
strength of preferences. That is, $\beta_{i}=\beta$ for all $i$
and $\beta_{j}=1$ for all $j$, where $\beta_{j}=1$ is a normalization.
Therefore,

\[
x_{ij}^{*}=\frac{\beta}{\beta+1}\delta_{i}+\frac{1}{\beta+1}(\delta_{j}+\xi_{ij}).
\]
Note that, in this case, the average observed slant at an outlet is
also additively separable
\[
\bar{x}_{j}=\frac{\beta}{\beta+1}\bar{\delta}_{A(j)}+\frac{1}{\beta+1}\delta_{j}+\frac{1}{\beta+1}\bar{\xi}_{j}
\]
where each barred term is calculated over the full sample by averaging
author--article rows within outlet $j$. Recall that $A(j)$ is the
portfolio of journalists working at firm $j$.

\subsection{Mover Event Studies\label{subsec:Mover-Event-Studies}}

Consider a mover from $j$ to $j'$. Then
\[
x_{ij'}-x_{ij}=\frac{1}{\beta+1}\left[\left(\delta_{j'}+\xi_{ij'}\right)-\left(\delta_{j}+\xi_{ij}\right)\right].
\]
Ideally, we would be able to estimate 
\[
\hat{x}_{it}=\alpha_{q}+d_{m(i,t)}+\sum_{l=-T,l\neq-1}^{T-1}\rho_{l}\left[\left(\delta_{d(q)}+\xi_{i,d(q)}\right)-\left(\delta_{o(q)}+\xi_{i,o(q)}\right)\right]{\bf 1}_{\{t=l\}}+\epsilon_{it},
\]
where $\hat{x}_{it}$ is the estimated slant of the article at event
time $t$ for move $q$ involving journalist $i$, $t=0$ is the
first article at the destination outlet, $\alpha_q$ are move-pair
fixed effects, and $d_{m(i,t)}$ are year-by-month fixed effects.
$o(q)$ and $d(q)$ denote the origin and destination outlets. We omit
$t=-1$, so each $\rho_l$ is measured relative to the last article
at the origin outlet.

The difference between $\rho_{l}$ before and after the move would
be an estimate of $\frac{1}{\beta+1}$. However, the outlet preference
terms $\left(\delta_{j(i,t)}+\xi_{i,j(i,t)}\right)$ are unobserved.
We estimate the following event-study specification.\footnote{Before estimation
in the mover event-study and outlet-level-decomposition exercises (Section
\ref{sec:outlet-decompositions}), we residualize
article slant by publication-month means estimated once on the full eligible
article universe. Throughout those exercises, $x$ and $\bar{x}$ denote article
outcomes and outlet means constructed from the resulting measure.}%
\[
x_{it}=\alpha_{q}+\sum_{l=-T,l\neq-1}^{T-1}\rho_{l}\left[\bar{x}_{d(q)}^{(-i)}-\bar{x}_{o(q)}^{(-i)}\right]{\bf 1}_{\{t=l\}}+\epsilon_{it},
\]
replacing the outlet preferences with observed average slant at the
destination and origin outlets. We calculate both outlet means over the
full sample using the author--article panel, but remove every article on
which the moving journalist appears.\footnote{For a coauthored article,
we remove every author-expanded row associated with that article.}
The same destination-minus-origin gap scales every
pre- and post-move event-time indicator, so the coefficients are expressed
throughout as fractions of the observed outlet gap.

Let $m_j=\mathbb{E}[\delta_j+\xi_{ij}]$ denote outlet $j$'s mean preference
across searches. In general, $\mathbb{E}[\bar{x}_{d(q)}^{(-i)}-\bar{x}_{o(q)}^{(-i)}]\not=m_{d(q)}-m_{o(q)}$
and so our estimates $\hat{\rho}$ do not have precise mappings to
the underlying model primitives. However, if journalists have zero slant
preferences, selection into and out of outlets is balanced, and other technical
conditions hold,\footnote{Proposition \ref{prop:no_preferences} establishes
balance for newly accepted offers; see Appendix \ref{app-sec-event-study-bridge}
for the additional conditions.} then the population benchmark is $\rho_l=1$ for
$l\geq0$. Under our model and identifying assumptions, by testing for whether $\hat{\rho}_{t\geq0}=1$, we are testing
for whether journalists have preferences over the ideological content which
they produce, i.e., $\beta>0$. Thus, while this exercise
does not allow us to pin down the relative magnitude of $\beta,$
it is still informative about the existence of journalist preferences.
An important caveat is that unobserved shocks $\xi$ still function
as measurement error in the independent variable, which could attenuate
our estimates of $\rho$.

In estimating the event study specification, we restrict to ``consistent''
moves which we define as when a journalist has written $T$ or more
articles consecutively at the origin outlet before writing $T$ or
more articles consecutively at the destination outlet.\footnote{Prior to defining a ``consistent'' move, we drop outlet spells with
fewer than \EventStudyMinSpellArticlesWord\ articles to reduce the potential influence
of syndications and news wires. Appendix Tables \ref{tab:consistent_mover_summary}
and \ref{tab:move_summary} report journalist- and move-level summary
statistics for the resulting mover samples.} Our baseline uses $T=5$;
we retain $T=10$ as a horizon robustness check. Lastly, to improve precision of our estimates, we weight observations
by the square of the relative move size between outlets $(\bar{x}_{d(q)}^{(-i)}-\bar{x}_{o(q)}^{(-i)})^{2}$.
Figure \ref{fig:movedist} plots the distribution of move sizes $\bar{x}_{d(q)}^{(-i)}-\bar{x}_{o(q)}^{(-i)}$
for our estimating sample of consistent movers when $T=5$ and $T=10$. As we
would like to see, the distribution of moves is approximately symmetric
around zero and has many ``large'' moves.

Our event study estimates allow us to reject the hypothesis that $\rho_{0}=1$
and, under our model and identifying assumptions, reject the hypothesis that $\beta=0$. Figure \ref{fig:eventstudy}
reports the estimates of $\rho$ from our event study specification
when $T=5$ and when $T=10$.
All preperiod coefficients in these two specifications
are individually insignificant at the 5 percent level. Upon move, both specifications
show a quick transition from approximately zero, reaching
\EventStudyPostMoveLevelTfive\ at $t=0$ in the baseline. The magnitude of this shift suggests journalists close
roughly \EventStudytzeroTfive\ percent of the observed slant gap between outlets when they move between outlets. Appendix Figure \ref{fig:eventstudy_single_author} shows a qualitatively similar pattern after restricting to articles with a single author.\footnote{Appendix Figure \ref{fig:eventstudy_topic} follows the main event-study specification after replacing article slant with the mean slant of the article's topic, calculated across all eligible articles using the same publication-month adjustment. The resulting coefficients are near zero. See Section \ref{sec:topics} for details on topic classification.}
We see no evidence of learning behaviors---with
the estimates being relatively flat across time in the post-move window
of analysis. If we were to interpret our event study coefficient estimates
as estimates of $\frac{1}{\beta+1}$, they would imply $\beta\approx\EventStudyBetaApproxTfive$.

In Figure \ref{fig:asymmetry}, Panel A repeats our estimation separately
for the set of movers who move to a more conservative outlet (e.g.,
$(\bar{x}_{d(q)}^{(-i)}-\bar{x}_{o(q)}^{(-i)})>0$) and the set of movers who move
to a more liberal outlet (e.g., $(\bar{x}_{d(q)}^{(-i)}-\bar{x}_{o(q)}^{(-i)})<0$).
Panel B examines the relationship between the change in average outlet
slant $\bar{x}_{d(q)}^{(-i)}-\bar{x}_{o(q)}^{(-i)}$ with the change in average
journalist slant using pooled pre- and post-move mean changes, with the
fitted line constrained through the origin. Both panels suggest a high degree
of symmetry across the direction and size of journalist moves across
outlets. Appendix Figure \ref{fig:eventstudy_party_four_groups} reports
analogous event-study estimates by party-affiliation group.

\subsection{AKM Decompositions}
\label{subsec-akm}
In this section, we continue to examine the role of journalists across
a broader set of movers using a variance decomposition of journalist
and outlet fixed effects following \citet[AKM]{Abowd1999HighWage}. 

To highlight the appropriate interpretation of our AKM estimates,
we can write and re-arrange the optimal slant contract as

\[
x_{ij}^{*}=\underbrace{\frac{\beta}{\beta+1}\delta_{i}}_{\gamma_{i}}+\underbrace{\frac{1}{\beta+1}\delta_{j}}_{\alpha_{j(i)}}+\underbrace{\frac{1}{\beta+1}\xi_{ij}}_{\epsilon_{ij}}
\]
and so, 
\[
x_{ij}^{*}=\gamma_{i}+\alpha_{j(i)}+\epsilon_{ij}.
\]
As explained in \citet{Abowd1999HighWage}, the outlet fixed effects
$\alpha_{j}$ are identified by the set of journalists moving between
outlets. In order for our estimates of the journalist and outlet fixed
effects ($\gamma_{i}$ and $\alpha_{j(i,t)}$) to be unbiased, we
need AKM's conditional mean independence assumption to hold 
\[
\mathbb{E}[\epsilon_{ij}|i\in A(j)]=0.
\]

If the conditional mean independence assumption holds, then we can
use the leave-out estimator proposed by \citet[KSS]{Kline2020LeaveOutVariance}
to consistently estimate the variance of the journalist fixed effects
$V(\gamma_{i})$, which is the combination of the variation in journalist
ideal points and their relative strength of preferences 
\[
V(\gamma_{i})=\left(\frac{\beta}{\beta+1}\right)^{2}V(\delta_{i}).
\]

However, given the importance of selection in our model, we do not
expect the conditional mean independence assumption to hold. As such,
we interpret our AKM variance decompositions as a descriptive
exercise familiar to many readers---not as estimates of underlying
model primitives.

In practice, we estimate
\[
\hat{x}_{ijt}=\gamma_{i}+\alpha_{j(i,t)}+\tau_{t}+e_{ijt}
\]
to allow for systematic temporal variation in slant due to the timeliness
of different topics, where each $\hat{x}_{ijt}$ is the slant of an
article published by journalist $i$ at outlet $j$ in month $t$,
$\gamma_{i}$ are journalist fixed effects, $\alpha_{j(i,t)}$ are
outlet fixed effects, $\tau_{t}$ are time fixed effects, and $e_{ijt}$
is the error term.\footnote{In Appendix \ref{app-sec-blm}, we use the BLM framework of \citet{Bonhomme2019DistributionalFramework} to allow for interactions between journalist types and outlet classes. The interaction residuals account for \BLMInteractionVarianceSharePct\% of observed calendar-adjusted article variance.}

Relative to our event study sample, we are less restrictive in the
set of articles we include. We still drop outlet spells with fewer than
\MonthlyPanelMinSpellArticlesWord\ articles (to reduce the influence of syndicated
articles and news wires) and journalists with more than \EventStudyMaxMovesCap\ moves
(to reduce the influence of freelancers), but make no other restrictions after the
restrictions defined in Section 3. This will likely still include some freelancers,
salaried journalists, and other writers. We use these observations
directly in the AKM and KSS specifications, with publication-month
fixed effects. In total, there are \ConnectedSampleJournalists\ journalists
(with \ConnectedSampleMovers\ of those being movers), \ConnectedSampleOutlets\ outlets, and \ConnectedSampleObservations\ observations
in the connected sample.\footnote{In the AKM analysis, movers are journalists observed at more than one retained outlet.}%
{} 

Table \ref{tab:akm_kss_table} reports the variance decomposition
estimates using both the plug-in estimator (Panel A) as well as the
leave-out variance estimator proposed by \citet{Kline2020LeaveOutVariance},
which addresses the limited mobility bias that is often present in
these employee-employer transition datasets (Panel B). Both panels use
the leave-out connected set of outlets. The different
columns report estimates on various samples of the data. 

Panel B is our preferred set of estimates.
Our baseline specification (Column 1 of Panel B) estimates \KSSJournalistExplainedAKMColOne\ percent
of the variation in article slant is explained by journalist
fixed effects, \KSSOutletExplainedAKMColOne\ percent is explained by outlet fixed effects, and
\KSSCovarianceExplainedAKMColOne\ percent is explained by the covariance between journalist and outlet
fixed effects. These results are reasonably robust when the estimation
sample includes both movers and stayers (e.g., Columns 1, 4, and 5
of Panel B), with the exception of the specification where we replace
outlet fixed effects with parent company fixed effects (Column
10 of Panel B) or restrict to ``top'' outlets (Column 6 of Panel B).
We report complementary evidence on ownership changes in the appendix.\footnote{Appendix
Figures \ref{fig:ownership_event_study}, \ref{fig:ownership_event_study_variation},
and \ref{fig:ownership_event_study_alignment_variation} report ownership-change
event-study analyses and the underlying collapsed DID variation. Appendix
Section \ref{app-sec-ownership-event-study} describes the specifications.
These exercises examine whether target-outlet slant changes around ownership
changes in ways related to the acquiring owner's portfolio.}

When our estimation sample is restricted to a set of movers only (Columns
2, 3, 7, 8, and 9 of Panel B), the estimated shares attributed to
journalist and outlet fixed effects vary across the alternative mover
samples.%
{} However, even in the most restrictive set of movers (Columns 8 and
9 of Panel B), journalist fixed effects still explain \KSSJournalistExplainedAKMColEight\ percent
and \KSSJournalistExplainedAKMColNine\ percent of the variation in slant,
respectively.

When restricting the sample to spells around large moves (absolute outlet
slant gaps greater than 0.1), Column (3) still estimates
a role for journalist fixed effects. Column (8), which uses the large-move
subset of the $T=5$ event study sample defined by the event-study gap cutoff, reports a larger role for journalist
fixed effects than Column (3), and a smaller role for outlet fixed effects.

It is also interesting to note that both sorting and the variance
of outlet fixed effects is larger among low-tier outlets. When restricting
to the top 20 outlets by circulation, our estimated correlation between
$\gamma_{i}$ and $\alpha_{j}$ is \KSSCorrelationTopAKM\ (Column 6, Panel B) compared
to \KSSCorrelationAKMColOne\ in our full sample (Column 1, Panel B). These results could
be consistent with a model where top-tier outlets differentiate themselves
vertically, whereas low-tier outlets differentiate themselves horizontally. 

Figure \ref{fig:akm_hetero} plots the distribution of estimated journalist
fixed effects for various subsets of journalists. Panel A reports
the distribution for Republican, Democratic, and non-partisan journalists,
where we see that the distribution of fixed effects for Republican
journalists is significantly to the right of other journalists. Panels B and C suggest that women have
more left-leaning fixed effects than men and non-white journalists
have more left-leaning fixed effects than white journalists.

\subsection{Slant Under Counterfactual Journalist Party Compositions}\label{sec:akm_counterfactuals}

These heterogeneous AKM estimates by party affiliation also allow us to conduct a simple partial-equilibrium
counterfactual exercise, analyzing how average slant would change under different distributions of journalists' party affiliations. 
We can decompose our average estimated slant as:
\[
\bar{x} = \left( \sum s_{p} \bar{\gamma}_{p} \right) + \bar{\alpha} + \bar{\tau}
\]
where $s_{p}$ and $\bar{\gamma}_{p}$ denote the share and average journalist fixed effect among those with party affiliation $p$, while $\bar{\alpha}$ and $\bar{\tau}$ denote overall averages of outlet and time fixed effects.\footnote{By construction, the residual ($e_{ijt}$) averages to zero across articles.} This allows us to analyze changes in mean slant as journalist party-affiliation shares vary, holding fixed the average journalist fixed effect by party as well as the overall average outlet and time fixed effects.

Appendix Figure \ref{fig:counterfactual_mean_slant} reports the resulting counterfactual
means together with benchmark observed means for selected outlets.
Because the reweighting exercise is defined over journalists with Democratic,
Republican, or non-partisan affiliations, the natural baseline is
the observed mean slant in the non-missing-party subsample. Relative
to that baseline, moving from the observed party distribution to an
even one-third Democratic, one-third Republican, and one-third non-partisan
distribution shifts mean slant in a more conservative direction by
\AKMArticleWeightedEqualThirdsGapRaw\ points, or only \AKMArticleWeightedEqualThirdsGapSd\
standard deviations of article-level slant. Even the most extreme
comparison, moving from an all-Democratic to an all-Republican journalist
distribution, changes mean slant by only \AKMArticleWeightedRepMinusDemGapRaw\
points, or \AKMArticleWeightedRepMinusDemGapSd\ standard deviations.
These shifts are therefore modest relative to the variance in slant across articles and to the dispersion in benchmark outlet means shown in the figure. In this reweighting exercise, changes in the composition of journalist party affiliations generate modest changes in average slant, holding within-party mean journalist effects, outlet effects, and time effects fixed.\footnote{Note that this partial equilibrium exercise holds the supply of outlets and the number of articles per outlet fixed. Changing journalist ideologies could lead to the expansion or entry of ideologically aligned outlets, changes in journalist-outlet bargaining strength, or other effects which might plausibly generate changes larger than our estimates.}

\section{Outlet-level Decompositions}\label{sec:outlet-decompositions}

Both the event study and AKM results are limited by their ability
to address the selection driven by the search-specific shock process
in our model. In this section, we relax the conditional mean independence
assumption by focusing on estimating the role of journalists in driving
the observed variation in \emph{outlet-level} slant.

\subsection{Model-Driven Specification}

Recall that the average observed slant at an outlet is 
\[
\bar{x}_{j}=\frac{\beta}{\beta+1}\bar{\delta}_{A(j)}+\frac{1}{\beta+1}\delta_{j}+\frac{1}{\beta+1}\bar{\xi}_{j},
\]
 Let 
\[
\psi_{j}=\frac{\beta}{\beta+1}\bar{\delta}_{A(j)}
\]
denote the ``journalist effect'' at outlet $j$. 

Our goal is to estimate $\operatorname{Var}_{p}(\psi_{j})$.\footnote{Throughout this subsection,
baseline outlet-level variances and covariances use fixed weights $p_j$ proportional
to each outlet's full-sample number of articles and the
same fixed connected outlet support.}

Our identification strategy uses a movers identification strategy
to difference out most of the non-$\psi_{j}$ components, and we provide
conditions under which we can ignore the remaining selection. Let $q$
index observed moves. For move $q$, let $i(q)$ denote the journalist
and let $j=o(q)$ and $j'=d(q)$ denote the origin and destination outlets.
Write $\xi_{qj}$ and $\xi_{qj'}$ for the search shocks on the corresponding
origin and destination contracts. Then
\[
x_{i(q)j'}-x_{i(q)j}=\frac{1}{\beta+1}\left[\left(\delta_{j'}+\xi_{qj'}\right)-\left(\delta_{j}+\xi_{qj}\right)\right],
\]
 and we can write the difference in average slant between outlets
$j'$ and $j$ as 
\[
\bar{x}_{j'}-\bar{x}_{j}=\frac{\beta}{\beta+1}(\bar{\delta}_{A(j')}-\bar{\delta}_{A(j)})+\frac{1}{\beta+1}(\delta_{j'}-\delta_{j})+\frac{1}{\beta+1}(\bar{\xi}_{j'}-\bar{\xi}_{j}).
\]
Let $y_{q}=(\bar{x}_{j'}-\bar{x}_{j})-(x_{i(q)j'}-x_{i(q)j})$. Combining
the two equations above, we get 
\[
y_{q}=\frac{\beta}{\beta+1}(\bar{\delta}_{A(j')}-\bar{\delta}_{A(j)})+\frac{1}{\beta+1}(\bar{\xi}_{j'}-\bar{\xi}_{j})-\frac{1}{\beta+1}(\xi_{qj'}-\xi_{qj}),
\]
which we can re-arrange and relabel as
\[
y_{q}=\underbrace{\frac{\beta}{\beta+1}\bar{\delta}_{A(j')}}_{\psi_{j'}^{d}}-\underbrace{\frac{\beta}{\beta+1}\bar{\delta}_{A(j)}}_{\psi_{j}^{o}}-\underbrace{\frac{1}{\beta+1}\left[(\xi_{qj'}-\bar{\xi}_{j'})-(\xi_{qj}-\bar{\xi}_{j})\right]}_{e_{q}}
\]
to get 
\begin{equation}
y_{q}=\psi_{j'}^{d}-\psi_{j}^{o}-e_{q}.\label{eq:twowayfe}
\end{equation}
Note that $\psi_{j}^{d}=\psi_{j}^{o}$ for a given outlet $j$. 

Intuitively, $x_{i(q)j'}-x_{i(q)j}$ differences out journalist $i(q)$'s fixed
preferences $\delta_{i(q)}$ that remain the same across outlets, thereby
leaving the difference in outlet components $(\delta_{j'}+\xi_{qj'})-(\delta_{j}+\xi_{qj})$.
Then, by subtracting $x_{i(q)j'}-x_{i(q)j}$ from $\bar{x}_{j'}-\bar{x}_{j}$,
we difference out the outlet components that remain the same for all
journalists (i.e., $\delta_{j}$). Thus, leaving the difference in
journalist components $\psi_{j}$ between origin and destination outlets,
as well as the difference in the structural errors for movers and
non-movers at each outlet, $\left[(\xi_{qj'}-\bar{\xi}_{j'})-(\xi_{qj}-\bar{\xi}_{j})\right].$

We would like to estimate equation (\ref{eq:twowayfe}) using origin
and destination fixed effects among the sample of movers and to construct
$\operatorname{Var}_{p}(\hat{\psi}_{j}^{d})$, an estimate of
$\operatorname{Var}_{p}(\psi_{j})$.

Two empirical issues remain.

First, it is well known that plug-in estimates of quadratic forms,
such as variances, are biased (e.g., \citealt{Kline2020LeaveOutVariance}).
Motivated by this concern, we base our estimator on the covariance between
origin and destination fixed effects rather than either estimated variance.

Second, our estimates of $\psi_{j}^{d}$ and $\psi_{j}^{o}$ are potentially
biased themselves if they are correlated with $e_{q}$---which is
likely under any form of journalist preferences over slant or selection
in job transitions. We postulate a set of conditions for which we
can derive consistent estimates for $\operatorname{Var}_{p}(\psi_{j})$ below.

Let $T_{qj}^{o}$ be an indicator for $o(q)=j$, and let $T_{qj}^{d}$
be an indicator for $d(q)=j$. The supported move pairs are
\[
\mathcal{P}=\{(j,k):\Pr(o(q)=j,d(q)=k)>0\}.
\]

\medskip{}

\begin{assumption}\label{assump:pair_conditional_selection}
(Pair-Conditional Difference Balance)\footnote{This condition permits
nonzero, pair-specific selection in both shock deviations as
long as their conditional means balance. When the weights are fixed
independently of the shocks, the standard AKM conditions of conditional
mean independence and exogenous mobility imply this condition.}
For every $(j,k)\in\mathcal{P}$, conditional on the fixed full-sample
article-production weights $w$,
\[
\mathbb{E}\left[(\xi_{qk}-\bar{\xi}_{k})
-(\xi_{qj}-\bar{\xi}_{j})\mid o(q)=j,d(q)=k,w\right]=0.
\]
\end{assumption}
\medskip{}
Let $N_m$ denote the number of observed moves, let ${\bf Y}$ stack
$y_q$, and let $T_j^o$ and $T_j^d$ stack the corresponding indicators
across moves.
Consider the linear model 
\[
\mathbb{E}[{\bf Y}|{\bf T}]={\bf T}\theta
\]
where ${\bf T}=[-T_{1}^{o},...,-T_{J}^{o},T_{1}^{d},...,T_{J-1}^{d}]$,
a $N_{m}\times(2J-1)$ matrix and $\theta$ is a $(2J-1)$ parameter
vector which we can partition as $\theta=[\theta^{o},\theta^{d}]'$.
Then 
\[
\hat{\theta}=({\bf T}'{\bf T})^{-1}{\bf T'Y}
\]
is the OLS estimator. When the same moves estimate both role effects,
shared sampling errors can enter their covariance and bias the plug-in
estimate upward. We therefore cross-fit the covariance (see also
\citealt{Kline2020LeaveOutVariance}).
For each repeated split $s$, we partition journalists into two groups,
$A$ and $B$, and estimate equation (\ref{eq:twowayfe}) separately in
each group. Let $\hat{\theta}^{r,h,s}$ denote the estimate for role
$r\in\{o,d\}$ from group $h\in\{A,B\}$. Our estimator is
\[
\widehat C_{\mathrm{CF}}=\frac{1}{S}\sum_{s=1}^{S}\frac{1}{2}\left[
\operatorname{Cov}_{p}(\hat{\theta}^{d,A,s},\hat{\theta}^{o,B,s})
+\operatorname{Cov}_{p}(\hat{\theta}^{d,B,s},\hat{\theta}^{o,A,s})
\right].
\]
Each covariance therefore combines role-effect fits estimated on
disjoint sets of movers.
\begin{prop}\label{prop:pair_conditional_covariance}
If pair-conditional difference balance holds, $\widehat C_{\mathrm{CF}}$
is a consistent estimator for $\operatorname{Var}_{p}(\psi_{j})$.\footnote{See
Appendix \ref{sec:appendix} for normalization details and additional
regularity conditions.}

\textbf{Proof}: See Appendix \ref{sec:appendix}.\medskip{}
\end{prop}
\begin{rem}
$\operatorname{Var}_{p}(\hat{\theta}^{d})$ or
$\operatorname{Var}_{p}(\hat{\theta}^{o})$ would also be consistent
estimators for $\operatorname{Var}_{p}(\psi_{j})$, as would the full-sample
plug-in covariance $\operatorname{Cov}_{p}(\hat{\theta}^{d},
\hat{\theta}^{o})$. We prefer the cross-fit covariance in finite samples
because the disjoint halves avoid reusing the same mover observations
in both role-effect fits while preserving the same probability limit.

The proposition conditions on the realized full-sample weights. It also
applies when estimated weights converge to fixed population production
shares and the stated moment and regularity conditions
hold jointly.

\medskip{}
The pair-conditional difference balance assumption is admittedly strong.
However, it is weaker than the standard conditional mean independence
and exogenous mobility assumptions commonly used in AKM and related
models. Intuitively, while the standard AKM assumptions imply no selection
on the search-specific shocks, our assumption allows movers on a given
route to be selected on both their origin and destination shocks, but
requires those two forms of selection to balance on average.

Agreement between the origin and destination estimates is a necessary
implication of the model and can provide a partial diagnostic.\footnote{Disagreement
may reflect a failure of pair-conditional difference balance or sampling
error, while agreement alone does not establish the assumption.} %
\par
Under certain conditions, $\operatorname{Var}_{p}(\psi_j)$ is a lower bound for
the variance in observed outlet-level slant driven by journalist effects
\emph{and }selection.
\end{rem}
\begin{assumption}
(Non-negative Covariance) The covariance between the average journalist
ideal point and the average structural error at outlet $j$ is non-negative.
That is, $\operatorname{Cov}_{p}(\bar{\delta}_{A(j)},\bar{\xi}_{j})\geq0$.
\end{assumption}
\begin{cor}
If the pair-conditional difference balance and non-negative covariance
assumptions hold,
then $\operatorname{Var}_{p}(\psi_{j}+\frac{1}{\beta+1}\bar{\xi}_{j})
\geq\operatorname{Var}_{p}(\psi_{j})$ and
$\widehat C_{\mathrm{CF}}$
consistently estimates the lower-bound value $\operatorname{Var}_{p}(\psi_j)$.

\textbf{Proof}: Note that 
\[
\operatorname{Var}_{p}(\psi_{j}+\frac{1}{\beta+1}\bar{\xi}_{j})
=\operatorname{Var}_{p}(\psi_{j})
+\operatorname{Var}_{p}(\frac{1}{\beta+1}\bar{\xi}_{j})
+2\operatorname{Cov}_{p}(\frac{\beta}{\beta+1}\bar{\delta}_{A(j)},
\frac{1}{\beta+1}\bar{\xi}_{j})
\geq\operatorname{Var}_{p}(\psi_{j}),
\]
with the last inequality holding due to the non-negative covariance
assumption. Proposition 3 then implies $\widehat C_{\mathrm{CF}}$
consistently estimates the lower-bound value
$\operatorname{Var}_{p}(\psi_j)$.\medskip{}
\medskip{}
\end{cor}
\subsection{Results}

To visualize each of these differences in $y_{q}$, we partition outlets into
above- and below-median groups based on average slant using our full dataset and
compute $\bar{x}_{d(q)}$ and $\bar{x}_{o(q)}$. We then compute
$x_{i(q),d(q)}$ and $x_{i(q),o(q)}$ for the set of moves from above-median
outlets to below-median outlets, and vice-versa, from our baseline $T=5$
event study dataset as used in Section \ref{subsec:Mover-Event-Studies}.\footnote{This descriptive construction is distinct from the event-study regression
estimand, which uses leave-focal-author-out outlet gaps and squared-gap
weights.}
(Note that this implicitly restricts to larger moves.)

Figure \ref{fig:descriptive_event_study} plots the results. The ``Outlet
Gap'' in each panel of Figure \ref{fig:descriptive_event_study}
reports $(\bar{x}_{j'}-\bar{x}_{j})$, the difference in average slant
between destination and origin outlet. The ``Change on Move'' in each panel
reports $(x_{ij'}-x_{ij})$, the average change in slant for a journalist that
moves from the origin to the destination outlet.

If outlets in the set of above-median outlets have homogeneous preferences
(and likewise for below-median outlets), then the difference between
the ``Outlet Gap'' and the ``Change on Move'' for each panel would
give us estimates of $\psi^{d}$ and $-\psi^{o}$ for one set of outlets
normalized relative to the other set. Using the below-median outlets
as the reference group, our example implies $\hat{\psi}^{d}=(\OutletGapRight-\ChangeOnMoveRight)=\OutletPsiDAboveMedian$
and $\hat{\psi}^{o}=-(\OutletGapLeft+\ChangeOnMoveLeftAbs)=\OutletPsiOAboveMedian$ for the above-median set
of outlets. Together, the two role effects account for about
\DescriptiveEventStudyJournalistEffectsSharePct\ percent\footnote{$(\OutletPsiDAboveMedian+\OutletPsiOAboveMedian)/
[\OutletGapRight-(\OutletGapLeft)]\approx \DescriptiveEventStudyJournalistEffectsShare$.} of the difference in average
slant between the two outlet groups.

The results in Figure \ref{fig:descriptive_event_study} are from
a restrictive set of authors and estimate first-moment differences
across sets of outlets. Table \ref{tab:outlet_decomp} provides our
cross-fit variance decomposition, $\widehat C_{\mathrm{CF}}/
\operatorname{Var}_{p}(\bar{x}_{j})$, using outlets identified in every
split half.

Our baseline estimate in Column 1 of Table \ref{tab:outlet_decomp}
suggests \JournalistExplainedOutletDecompResidualizedPercent\ percent of the outlet-level
variation in slant can be explained by journalist effects, with a
95 percent confidence interval of
\OutletDecompCrossfitCILowResidualizedPercent--\OutletDecompCrossfitCIHighResidualizedPercent\ percent
from a bootstrap that reweights journalist clusters.\footnote{Without cross-fitting, the plug-in estimate
is \OutletDecompNaturalPluginResidualizedPercent\ percent on its natural
\OutletDecompNaturalPluginOutlets-outlet connected support and
\OutletDecompMatchedPluginResidualizedPercent\ percent on the same
\OutletDecompCrossfitOutlets-outlet support as the cross-fit estimate.
The equal-outlet-weight cross-fit estimate is
\OutletDecompEqualOutletWeightsResidualizedPercent\ percent, with a 95 percent
confidence interval of \OutletDecompEqualOutletWeightsCILowResidualizedPercent--\OutletDecompEqualOutletWeightsCIHighResidualizedPercent
\ percent.}
A wild cluster bootstrap test of the model-implied equality between
the origin and destination effects yields
$p=\OutletDecompEqualityPValueResidualized$ (see Appendix
\ref{sec:outlet_decomp_inference}).

We can compare our baseline to the estimated \ConsumerExplainedGSMediaBias\ percent of observed
variation in outlet-level average slant that is explained by consumer
demand in \citet{Gentzkow2010WhatDrivesMediaBias}. Although the two
estimates are not perfectly comparable, taken at face value they suggest
that journalist effects are nearly as important as consumer demand in
explaining the variation in observed slant at the outlet level. %

\section{Modeling Selection}\label{sec:modeling-selection}

In this section, we model selection directly, adapting the framework of Section~\ref{sec:model_production} into an
estimable model by restricting preferences, grouping outlets into classes, and using nested stay/move choices.

Journalist $i$ has a slant ideal point $\delta_i\in(0,1)$ and aversion to mismatch $\beta$; outlets are
grouped into $K=5$ classes,\footnote{We group outlets with similar distributions of journalists' average slant during their spells at the outlet. This captures both the level of slant and the variation within an outlet. The classes are ordered from most liberal to most conservative by their average slant.} class $k$ having a median preference $\delta_k\in(0,1)$ before selection, and outlet aversion is
normalized to $\beta_j=1$. The optimal contract offered at article choice $t$ by class $k$ is
\begin{equation}
x^*_{ikt} \;=\; \frac{\beta\,\delta_i + \delta_k+\xi_{ikt}}{\beta+1},
\label{eq:contract}
\end{equation}
where $\delta_k+\xi_{ikt}=\Lam(b_k+\zeta_{ikt})$, $b_k=\log\!\big(\delta_k/(1-\delta_k)\big)$, and $\Lam(u)=e^{u}/(1+e^{u})$. This is the optimal contract of Proposition~\ref{prop:optimal_contract} with $\beta_i=\beta$ and $\beta_j=1$. The log-odds shock $\zeta_{ikt}$ keeps offers and contracts between 0 and 1; $\xi_{ikt}$ is the corresponding slant-scale deviation from $\delta_k$.

We also write:
\begin{equation}
\delta_i = \Lam\!\big(a_{p(i)} + \tau_{\delta,p(i)}\,v_i\big),\ \ v_i\sim N(0,1);
\qquad
\zeta_{ikt}\overset{\mathrm{iid}}{\sim} N(0,\tau_\zeta^2),
\label{eq:latents}
\end{equation}
where $p(i)\in\{\text{Dem, Rep, NP, NA}\}$\footnote{$NP$=Non partisan, $NA$ corresponds to journalists that we could not match to the voter file, either because of name discrepancy or because they are not registered to vote. We treat them as their own party.} is journalist $i$'s party, $a_p$ and
$\tau_{\delta,p}$ capture party-specific heterogeneity in slant. The type $\delta_i$ is fixed over time. Before each article is written, one shock $\zeta_{ikt}$ is drawn for every outlet class, and the journalist observes these draws before deciding whether to stay or to move to another outlet. If they move to $k$, their contract carries the accepted $\xi_{ikt}$ throughout their time at that outlet. Their own match is not redrawn while they stay.

For a spell entered at $t$ in class $k$, observed article slant contains a month-level unobservable and measurement error:
$x_{ia}=x^*_{ikt}+\mu_{m(i,a)}+e_{ia}$, $e_{ia}\sim N(0,\sigma_e^2)$.
Write $\hat{x}_{ia}=x_{ia}-\mu_{m(i,a)}$ for calendar-adjusted article slant.

For each article, journalist $i$ chooses among $\{\text{stay}\}\cup\{\text{move to class }0,\dots,4\}$,
with systematic utilities:
\begin{align}
u^S_{it} &= s + \kappa_n\,\log(1+n_{it})
 - \beta\,(\bar{x}_{it}-\delta_i)^2
&&\text{(stay)}
\nonumber\\
u^M_{ikt} &= \alpha_k + \gamma\,\mathbf{1}\{k=c_{it}\} - \kappa_d\,\log(1+d_{ikt})
\nonumber\\
&\quad - \tfrac{\beta}{(\beta+1)^2}\,\big[\delta_i-(\delta_k+\xi_{ikt})\big]^2
&&\text{(move to $k$)}
\label{eq:utils}
\end{align}
with $c_{it}$ the current outlet class and $\gamma$ a same-class premium. The stay utility rises with prior article count in the current spell
($n_{it}$).\footnote{Tenure is
unknown for spells already in progress at the start of the panel; for these articles the tenure
term is replaced by a separately estimated stay level $s_c$. These spells account for most
articles, since every journalist present at the panel's start begins in one.} $d_{ikt}$ is the
distance in km from the current outlet's city to the nearest outlet of class $k$ (own outlet
excluded), so $\kappa_d$ captures the cost of moving to a distant outlet.

The coefficient $\beta/(\beta+1)^2$ is the cost for the journalist of writing slant at an outlet of class $k$ evaluated at the anticipated contract: moving to $k$ means writing $x^*$ from \eqref{eq:contract} and
$\beta\,(x^*-\delta_i)^2=\tfrac{\beta}{(\beta+1)^2}\big[\delta_i-(\delta_k+\xi_{ikt})\big]^2$. In estimation, we proxy the current contract by $\bar{x}_{it}$, the observed calendar-adjusted mean over the full current spell. This proxy stays fixed within the spell.

We use a nested logit with one stay option and a common nest of five destinations,
with nesting parameter $\lambda\in(0,1]$. Writing
$I_{it}=\lambda\log\sum_{h=0}^4\exp(u^M_{iht}/\lambda)$, the choice probabilities are
\[
P(\text{stay})=\frac{\exp(u^S_{it})}{\exp(u^S_{it})+\exp(I_{it})},
\qquad
P(k)=\frac{\exp(I_{it})}{\exp(u^S_{it})+\exp(I_{it})}
\frac{\exp(u^M_{ikt}/\lambda)}{\sum_{h=0}^4\exp(u^M_{iht}/\lambda)}.
\]

Ideology plays three roles: (1) in writing through $\beta/(\beta+1)$; (2) in whether to move
through $\beta/(\beta+1)^2$; and (3) in where to move through
$\beta/\big((\beta+1)^2\lambda\big)$.

\subsection*{Estimation}

Estimation proceeds in two steps. The first step recovers the amount of measurement noise
$\sigma_e$ through stayers; the remaining 24 parameters $\theta=\big(\beta,\;\allowbreak\{a_p,\tau_{\delta,p}\}_{p},\;\allowbreak b_1,\dots,b_4,\;\allowbreak
\lambda,\;\allowbreak\gamma,\;\allowbreak\alpha_1,\dots,\alpha_4,\;\allowbreak s,\;\allowbreak\kappa_n,\;\allowbreak s_c,\;\allowbreak\kappa_d,\;\allowbreak\tau_\zeta\big)$
are estimated by maximum simulated likelihood with $\sigma_e$ held fixed. We normalize the class-$0$ location $b_0$ to recover the observed mean slant at class-$0$ outlets,
\begin{equation}
\bar{x}_0 \;=\; \tfrac{\beta}{\beta+1}\,\big(\textstyle\sum_p s_{0p}\,\E[\delta_i\mid p]\big)
 + \tfrac{1}{\beta+1}\,\E\big[\Lam(b_0+\zeta)\big],
\label{eq:anchor}
\end{equation}
where $s_{0p}$ are the class-$0$ party shares. The year-month fixed effects $\mu_m$ are
absorbed by residualizing article slant on the calendar month before estimation.

\textbf{First Step} Articles written by stayers within a spell belong to the same contract, so all the dispersion around the spell mean is noise; the within-spell variance of article slant therefore estimates the noise variance,
\begin{equation}
\widehat{\sigma}_e^2 \;=\;
\frac{\sum_\ell \sum_{a\in \ell}\big(\hat{x}_{ia}-\bar{x}_{\ell}\big)^2}{\sum_\ell (n_\ell-1)},
\end{equation}
where $\ell$ runs over all journalists' spells, $n_\ell$ is the number of articles in
spell $\ell$, and $\bar{x}_\ell$ is its calendar-adjusted mean.

\textbf{Second Step} Conditional on the accepted draw, all articles in an entered spell are drawn around the same contract. Let $m_{it}$ indicate a move at article $t$ to destination $k_{it}$, $\mathcal A_{it}$ its entered spell's articles, and $\boldsymbol{\zeta}_{it}$ the five offer shocks. The likelihood can be written as:
\begin{equation}
\begin{aligned}
L_i(\theta) = \int \prod_t \bigg[\int
&P(\text{stay}\mid\delta,\boldsymbol{\zeta})^{\,1-m_{it}}\\
&{}\times\Big[P(k_{it}\mid\delta,\boldsymbol{\zeta})\,\textstyle\prod_{a\in\mathcal A_{it}} f(\hat{x}_{ia}\mid k_{it},\delta,\boldsymbol{\zeta})\Big]^{m_{it}}
\,dG(\boldsymbol{\zeta})\bigg]\,dF_{p(i)}(\delta),
\end{aligned}
\label{eq:likelihood}
\end{equation}
where $f$ is the normal density of calendar-adjusted article slant around the accepted contract, $F_{p}$ is the ideal-point distribution implied by~\eqref{eq:latents}, and $G$ the distribution of the offer draws.\footnote{The mobility likelihood conditions on the observed initial outlet and treats spell means as fixed. Spells with unobserved entry contribute to mobility but have no writing-density term.}

The component of a journalist's slant that persists across their own spells recovers the journalists' component $(\frac{\beta}{(\beta+1)})^2\mathrm{Var}(\delta)$; since $\delta_i$ enters writing with weight $\beta/(\beta+1)$ but choices with weight $\beta/(\beta+1)^2$, comparing how strongly that persistent component predicts writing versus where journalists move separates $\beta$ from the dispersion of $\delta$. Finally, the variation coming from movers entering the same outlet class with the same previous contract but who write at different levels there identifies $\tau_\zeta$. We report analytical standard errors, clustered at the journalist level.

\subsection*{Estimates}

We estimate that journalists carry a weight of $\beta/(\beta+1)= \SelW$ in the slant contract. The gap in mean ideal points between journalists registered with
the Democrats and with the Republicans is \SelIdealGap{} in slant units, which is relatively small compared to the \SelLadderSpan{}
distance between the most liberal and the most conservative outlet classes; non-partisan
journalists and those we could not match to the voter file are closer to Democrats than to
Republicans. The small nesting parameter $\lambda$ makes the destination choice, conditional on moving, highly sensitive to journalists' ideology, while the decision to move at all is dominated by the mobility shock, with ideology playing a more moderate role.

At the estimated parameters, we can obtain the average accepted log-odds draw among movers,
$\E[\zeta\mid\text{accepted offer}]=+0.04$; it varies from $-0.17$ at the most conservative
class to $+0.30$ at the most liberal class. The mean absolute value of
$\E[\zeta\mid\text{accepted offer}]$ across movers is $0.24$: each journalist accepts draws closer to
their own ideal, liberal journalists lower ones and conservative journalists higher ones.

Appendix Table~\ref{tab:structural_article_variance} decomposes calendar-adjusted
article slant in the model's stationary distribution at observed party composition
and fixed outlet-class shares. The journalist, outlet-class baseline, and accepted
offer deviation components have variances equal to
\StructuralSlantVarianceJournalistTotalPercent\%,
\StructuralSlantVarianceOutletClassTotalPercent\%, and
\StructuralSlantVarianceOfferShockTotalPercent\% of total model variance;
independent article noise accounts for \StructuralSlantVarianceArticleNoiseTotalPercent\%.
All three signed covariance terms are reported separately.\footnote{The AKM shares use observed raw article variance, while the BLM shares use observed calendar-adjusted article variance in its first-move sample.}

To combine the direct journalist component with associated offer selection, we calculate
\[
\Var\!\left(
\frac{\beta}{1+\beta}\delta_i+
\frac{1}{1+\beta}\E[\xi_{ikt}\mid\delta_i]
\right).
\]
When calculated over the model's stationary distribution of published articles, this journalist component accounts for \StructuralSlantVarianceJournalistWithSelectionTotalPercentRounded\% of the total model-implied variance of article-level slant.

\subsection*{Counterfactuals}

We conduct a few counterfactual scenarios, changing the party composition of the journalist pool and assessing the extent to which that impacts the distribution of slant. We implement extreme counterfactual scenarios where we replace the population of journalists to be 100\% Democratic, 100\% Republican, or 100\% non-partisan, as well as more realistic ones with equal thirds of Democrats, Republicans, and non-partisans. We let journalists of each type move according to the estimated process until the allocation
across outlets reaches an equilibrium, holding outlet positions and all estimated parameters
fixed. We report two types of counterfactuals: (A) \emph{outlet sizes
fixed} and (B)
\emph{sizes free} (outlets hire whoever comes, so the allocation of jobs across classes
responds to the pool). Figure~\ref{fig:cf_contributions} reports the change in mean slant under each scenario and its
decomposition into the direct journalist effect, selected offers, and (under (B)) outlet-class
composition. Note that under (B), even without changing the pool of journalists, we would get a small change in mean slant, since the observed distribution of journalists across outlets is not an equilibrium, so we compare the counterfactual results changing the pool to the equilibrium we would get without changing the pool.

Three patterns stand out. First, with outlet sizes fixed, changing the
composition of journalists only marginally impacts equilibrium slant: even the full $100\%$ Dem
to $100\%$ Rep swing is \SelSwingFixed{} in slant (\SelSwingFixedSD{} article-level standard
deviations).\footnote{Changing to a balanced 1/3 each of Democrat-, Republican-, and non-partisan-written articles shifts average slant by just \SelEqualThirdsFixedSD{} article-level standard deviations.} Second, when letting outlet size vary freely, the same swing is \SelSwingFree{}
(\SelSwingFreeSD{} standard deviations), with the outlet-class composition channel accounting
for \SelRepCompShareFree\% of the $100\%$ Rep effect as the share of journalists at the two
most conservative classes rises from \SelConsShareDemB\% to \SelConsShareRepB\% of
employment. Third, outlet polarization, measured by comparing classes, changes very little ---
the gap between the most conservative and the most liberal class stays between \SelGapMin{} and
\SelGapMax{} across every scenario and closure --- so journalists' partisanship contributes
little to outlet polarization.

\section{Conclusion}

\noindent This paper makes three primary contributions.

First, we provide an important advance in measuring media slant---with
our fine-tuned RoBERTa model achieving a significant improvement in
measuring human-perceived political slant relative to prior methods---and
document new stylized facts regarding media slant.

Second, we provide novel evidence on the role journalists play in
determining the political slant of the news they produce. Under our model and identifying assumptions, our estimates
reject the hypothesis that journalists have zero ideological preferences
over the content they produce and imply that \StructuralSlantVarianceJournalistWithSelectionTotalPercentRounded{} percent of
variation in slant can be explained by journalists. Our estimates also allow us to calculate counterfactual mean slant under alternative journalist party compositions. We estimate that shifting from the current distribution of journalist party affiliations to a more even or more representative distribution would have relatively modest effects on average slant, relative to the overall observed variation in slant and holding the supply of outlets fixed.

Third, to the extent our results speak to worker preferences over
the slant they produce, they provide evidence on the importance of
non-pecuniary worker preferences in driving sorting across firms. 

There are several caveats to our analysis. First, our analysis requires
strong assumptions about the selection of journalists across outlets.
Violations of these assumptions will bias our estimates. Second, our
notion of the role of journalists focuses on journalists as individuals.
Group dynamics or the spillover effect of \emph{other }journalists, would be
attributed to the outlet. If these dynamics are important, our counterfactual exercises
would be understating the role of journalists. Our estimates do not fully separate
ideological preferences from other persistent journalist attributes. Third, we treat outlet preferences as a single composite, which bundles the influences of media owners, advertisers, editors, and readers, among others. Future research should continue to examine the effects of these potential limitations on our results.

\end{spacing}

\pagebreak{}

\bibliographystyle{chicago}
{\inputencoding{utf8x}\bibliography{journalist-ideology}}

\pagebreak{}

\input{figures.tex}

\clearpage\input{descriptive_tables_filled}\begin{landscape}\input{akm_kss_table_filled}

\end{landscape}

\pagebreak\input{outlet_decomp_table_filled}\pagebreak{}

\begin{table}[H]
\caption{Simulated Maximum Likelihood Estimates\label{tab:key_estimates}}

\begin{centering}
\input{modeling_selection_key_table}
\par\end{centering}
\smallskip{}

{\small Notes: The table reports coefficients estimated by simulated maximum likelihood.
Standard errors are clustered by journalist and are conditional on the first-step
$\widehat{\sigma}_e$. Ideal points are in slant units; the full parameter vector is reported
in Table~\ref{tab:full_estimates}.}{\small\par}
\end{table}

\appendix

\section{Theoretical Appendix\label{sec:appendix}}

\subsection{Proof of Proposition \ref{prop:optimal_contract}}

Write $M_i=M_i(x,x_c,\omega,\omega_c,\omega_0)$ and suppress its
arguments below. The first-order conditions for the outlet's problem
with respect to the wage and slant are
\[
\begin{aligned}
M_i&=\Delta U_j(x,\omega)\frac{\partial M_i}{\partial\omega},\\
2\beta_j(x-\delta_j-\xi_{ij})M_i
&=\Delta U_j(x,\omega)\frac{\partial M_i}{\partial x}.
\end{aligned}
\]
For the logit acceptance probability,
$\partial M_i/\partial\omega=M_i(1-M_i)$ and
$\partial M_i/\partial x=-2\beta_i(x-\delta_i)M_i(1-M_i)$.
Because $M_i\in(0,1)$, we can therefore divide the slant first-order
condition by the wage first-order condition to obtain
\[
\begin{aligned}
2\beta_j(x-\delta_j-\xi_{ij})
&=\frac{\partial M_i/\partial x}{\partial M_i/\partial\omega}
=-2\beta_i(x-\delta_i),\\
x_{ij}^{*}&=\frac{\beta_i\delta_i+\beta_j(\delta_j+\xi_{ij})}
{\beta_i+\beta_j}.
\end{aligned}
\]

For any fixed $x$, the derivative of the outlet's objective with respect
to $\omega$ has the sign of
\[
\nu_j-\beta_j(x-\delta_j-\xi_{ij})^2-\omega-1
-\frac{\exp\left(\omega-\beta_i(x-\delta_i)^2\right)}
{\exp(\omega_0)+\exp\left(\omega_c-\beta_i(x_c-\delta_i)^2\right)}.
\]
This continuous expression is strictly decreasing from positive infinity
as $\omega\to-\infty$ to negative infinity as $\omega\to\infty$.
It therefore has a unique root $\omega^*(x)$. Because the objective's
derivative is positive below this root and negative above it,
$\omega^*(x)$ is the unique globally optimal wage conditional on $x$.

After maximizing over $\omega$, the envelope theorem and the wage
first-order condition imply that the derivative of the outlet's maximized
objective with respect to $x$ is
\[
-2M_i\left(x,\omega^*(x)\right)
\left[\beta_j(x-\delta_j-\xi_{ij})+\beta_i(x-\delta_i)\right].
\]
Because $M_i>0$ and the expression in brackets has positive slope
$\beta_i+\beta_j$, the outlet's maximized objective increases below
$x_{ij}^*$ and decreases above it. Therefore the outlet's problem has
the unique global solution $(x_{ij}^*,\omega_{ij}^*)$, where
$\omega_{ij}^*=\omega^*(x_{ij}^*)$. This defines
$h_{ij}(x_{ij}^{*},x_c,\omega_c,\omega_0)$.
\textbf{QED.}

\subsection{Proof of Lemma \ref{lem:acceptance_nonselection}}

When $\beta_i=0$, Proposition \ref{prop:optimal_contract} gives
\[
x_{ij}^*=\delta_j+\xi_{ij}.
\]
At this slant, the outlet's slant loss is zero, and the wage first-order
condition is
\[
\nu_j-\omega_{ij}^*
=1+\frac{\exp(\omega_{ij}^*)}{\exp(\omega_0)+\exp(\omega_c)}.
\]
As shown in the proof of Proposition \ref{prop:optimal_contract}, this
equation has a unique solution, which is the globally optimal wage at
$x_{ij}^*$. Thus, conditional on $(\omega_c,\omega_0,j)$, the optimal
wage does not depend on $\xi_{ij}$. Because the alternative-specific
choice shocks are drawn independently after the offer,
\[
\begin{aligned}
&\Pr(i\text{ accepts }j\text{'s offer}\mid
i\text{ receives }j\text{'s offer},\omega_c,\omega_0,\xi_{ij})\\
&\qquad=\frac{\exp(\omega_{ij}^*)}
{\exp(\omega_0)+\exp(\omega_c)+\exp(\omega_{ij}^*)},
\end{aligned}
\]
which does not depend on $\xi_{ij}$. Acceptance therefore does not
further select on $\xi_{ij}$ conditional on the journalist's current
wage, outside option, and prospective outlet. \textbf{QED.}

\subsection{Proof of Proposition \ref{prop:no_preferences}}

When $\beta_i=0$, Proposition \ref{prop:optimal_contract} gives
$x_{ij}^*=\delta_j+\xi_{ij}$. Lemma \ref{lem:acceptance_nonselection}
and the stated conditional mean restriction therefore imply
\[
\begin{aligned}
&\mathbb{E}[x_{ij}^*\mid \omega_c,\omega_0,
i\text{ accepts }j\text{'s offer}]\\
&\qquad=\mathbb{E}[\delta_j+\xi_{ij}\mid \omega_c,\omega_0,
i\text{ receives }j\text{'s offer}]=\mathbb{E}[\delta_j+\xi_{ij}].
\end{aligned}
\]
Taking expectations over current wages and outside options among accepted
matches gives
\[
\mathbb{E}[x_{ij}^*\mid i\text{ accepts }j\text{'s offer}]=\mathbb{E}[\delta_j+\xi_{ij}].
\]
\textbf{QED.}

\subsection{Proof of Proposition \ref{prop:pair_conditional_covariance}}

Let $\mathcal J$ be the fixed set of outlets that appear as both origins
and destinations in the regression, and let $J=|\mathcal J|$. The
corresponding bipartite support graph has one origin and one destination
node for each outlet, with an edge for each supported move pair among
these outlets. Assume this graph is connected.

For each $j\in\mathcal J$, let $p_j$ be its full-sample number of articles
normalized to sum to one across $\mathcal J$.

Without loss of generality, let $r\in\mathcal J$ be the reference outlet
and omit its destination indicator. For each move $q$, let ${\bf t}_q'$
collect the origin and destination indicators under this normalization;
${\bf T}$ stacks these move-level rows. Conditional on the fixed weights
$w$, define
\[
\theta^o=(\psi_j-\psi_r)_{j\in\mathcal J},\qquad
\theta^d=(\psi_j-\psi_r)_{j\in\mathcal J\setminus\{r\}},
\]
and let $\theta=[(\theta^o)',(\theta^d)']'$. Equation
(\ref{eq:twowayfe}) gives ${\bf Y}={\bf T}\theta-{\bf e}$.
The outlet means entering ${\bf Y}$ are estimated on the full article
sample. Assume their estimation error is $o_p(1)$ uniformly over the
fixed set $\mathcal J$; by Slutsky's theorem, it suffices to establish
the result at their population limits.
Let $N_m$ denote the number of observed moves. Assume that the common
support $\mathcal J$ is eventually fixed and that, in the full sample and
separately within every half of each fixed split (using the corresponding
half-sample move count), every supported move pair has positive limiting
frequency, $e_q$ has a finite second moment, and the sample averages of
${\bf t}_q{\bf t}_q'$ and ${\bf t}_qe_q$ converge in probability to their
expectations.

For every supported pair $(j,k)$, pair-conditional difference balance
implies
\[
\begin{aligned}
\mathbb{E}[e_q\mid o(q)=j,d(q)=k,w]
&=\frac{1}{\beta+1}\left\{
\mathbb{E}[\xi_{qk}-\bar\xi_k\mid o(q)=j,d(q)=k,w]\right.\\
&\qquad\left.-\mathbb{E}[\xi_{qj}-\bar\xi_j\mid o(q)=j,d(q)=k,w]
\right\}=0.
\end{aligned}
\]
Because ${\bf t}_q$ is determined by the origin--destination pair,
$\mathbb{E}[{\bf t}_qe_q\mid w]=0$. The assumed law of large numbers therefore
gives
\[
\frac{{\bf T}'{\bf e}}{N_m}\overset{p}{\longrightarrow}0.
\]
Connected support and positive limiting pair frequencies imply that
$G=\mathbb{E}[{\bf t}_q{\bf t}_q'\mid w]\succ0$. The assumed law of large
numbers gives $N_m^{-1}{\bf T}'{\bf T}\overset{p}{\longrightarrow}G$.
Therefore,
\[
\hat\theta-\theta
=-\left(\frac{{\bf T}'{\bf T}}{N_m}\right)^{-1}
\left(\frac{{\bf T}'{\bf e}}{N_m}\right)
\overset{p}{\longrightarrow}0.
\]

Under these half-specific conditions, the same argument applies separately
to every half of each fixed split, and the two complete role-effect vectors converge to
$(\psi_j-\psi_r)_{j\in\mathcal J}$. Because $J$ and $p$ are fixed, the
continuous mapping theorem therefore gives, for every split $s$,
\[
\begin{aligned}
\operatorname{Cov}_{p}(\hat\theta^{d,A,s},\hat\theta^{o,B,s})
&\overset{p}{\longrightarrow}\operatorname{Var}_{p}(\psi_j),\\
\operatorname{Cov}_{p}(\hat\theta^{d,B,s},\hat\theta^{o,A,s})
&\overset{p}{\longrightarrow}\operatorname{Var}_{p}(\psi_j).
\end{aligned}
\]
Their symmetric average, and then the average over the fixed number of
splits, has the same probability limit. The reference outlet cancels
from every covariance, so its choice does not affect the result.
\textbf{QED.}

\subsection{Event-Study Interpretation under Zero Preferences}
\label{app-sec-event-study-bridge}

In this subsection, we discuss the relationship between Proposition
\ref{prop:no_preferences} and our event-study specification. The proposition analyzes
newly accepted offers. It does not by itself establish the stock, mover-selection,
or measurement conditions needed to interpret the event-study coefficients. Let
$\hat d_m$ be estimated once on the full eligible article universe and subtracted
from all event-study outcomes and outlet means. Throughout this subsection,
$x_{it}$ and $\bar{x}_{j}$ denote the resulting residualized quantities. Under
$\beta=0$, write the residualized article outcome for move $q$ as
\[
x_{it}=\alpha_q
+(m_{d(q)}-m_{o(q)})\mathbf{1}_{\{t\geq0\}}+u_{it},
\]
where $u_{it}$ collects deviations of the mover's structural shocks from their
search means and the article-level
error $\epsilon_{it}$. Suppose that, on the selected-mover support, the implemented
full-period, author--article-weighted residualized leave-focal-author-out gap satisfies
\[
\bar{x}_{d(q)}^{(-i)}-\bar{x}_{o(q)}^{(-i)}
\overset{p}{\longrightarrow}m_{d(q)}-m_{o(q)};
\]
$u_{it}$ is mean zero conditional on the full selected regression design at
each event time; and the implemented squared-gap-weighted fixed-effects regression
satisfies the usual law of large numbers and full-rank conditions. A sufficient
route to this gap condition is balance in initial outlet stocks and among
subsequent entrants, with retention, separation, article-production weights,
and the leave-focal-author-out construction preserving that balance, while the
pooled publication-month residualizer consistently removes the common calendar
component and mean-zero leave-out errors satisfy a law of large numbers as outlet
article samples grow. These are maintained restrictions beyond Proposition
\ref{prop:no_preferences}.

Under these conditions and $\beta=0$, standard weighted least-squares
consistency implies that the event-study coefficients converge in probability
to zero before the move and one after the move relative to the omitted $t=-1$
period; in particular, $\hat{\rho}_0\overset{p}{\longrightarrow}1$.

\setcounter{figure}{0}
\renewcommand{\figurename}{Appendix Figure}
\renewcommand\thefigure{\thesection\arabic{figure}}  
\setcounter{table}{0}
\renewcommand{\tablename}{Appendix Table}
\renewcommand\thetable{\thesection\arabic{table}}     
\clearpage
\section{Empirical Appendix}
\subsection{Data Processing Details}\label{sec:data_processing}

In this section, we describe key data processing steps used in the main analysis, model training, and validation exercises.

\subsubsection{Tweet-Linked Article Data}
\label{sec:tweet-linked-article-data}

Our first cleaned article dataset is built from tweets posted by politicians that link to news articles. We begin from a tweet collection containing \NTweetsCollectedMillions\ million tweets and approximately \NPoliticianUsernames\ politician accounts. From these tweets, we recover more than \NTweetedUniqueURLsMillionsFloor{} million unique tweeted URLs. We exclude likely non-news urls including ``youtu.be,'' ``pscp.tv,'' ``facebook.com,''
``instagram.com,'' ``patreon.com,'' and ``google.com.'' Next, we draw a scraping sample of \NSampledURLs\ URLs from this expanded URL pool. Among these sampled URLs, \NArticlesNonEmptyRows\ produce non-empty article text, corresponding to \NArticlesNonEmptyUniqueURL\ unique URLs with non-empty text.

We then combine the scraped article text with politician party labels and tweet-level sentiment information to construct the tweet-linked training data used in the paper. We then use an LLM (OpenAI GPT-5-mini) to classify cases in which the politician's tweet was explicitly criticizing (rather than endorsing or having unclear support for) the article's content. The LLM was provided with both the text of the politician's tweet and the scraped article text in order to make this determination. This filter drops \SentimentCriticizingAll\% of tweets as criticizing (\SentimentCriticizingDem\% of Democrat tweets, \SentimentCriticizingRep\% of Republican tweets). \SentimentEndorsingAll\% of tweets are classified as endorsing the article's content, while \SentimentUnclearAll\% of tweets are labeled as not clearly endorsing or criticizing, and tweets with both such LLM labels are kept in our sample.%

We also apply an LLM-based cleaning pipeline to the scraped article text. In this pipeline, two models identify and remove non-text spans, such as page-navigation text, captions, advertisements, and other interface elements, while a separate binary classifier flags pages that are not valid news articles at all, such as homepages, paywalls, or other non-article content.\footnote{We use OpenAI GPT-5-mini and Claude Haiku 4.5 to detect non-text spans and OpenAI GPT-4.1-mini to identify invalid articles.} Articles flagged as invalid are dropped, and detected non-text spans are removed before downstream cleaning.

After this text-cleaning stage, we impose a minimum post-cleaning text-length threshold of \MinArticleChars\ characters to exclude very short, malformed, or otherwise uninformative text. We also perform language detection on unique article texts and retain only articles classified as English.

The resulting cleaned tweet-linked sample contains \Nfinaltweets\ article--tweet observations, of which \NfinaltweetsDem\ are linked to Democratic politicians and \NfinaltweetsRep\ are linked to Republican politicians. For the tweet-based RoBERTa exercises, we split this cleaned sample into \TrainSplitPercent\ percent training and \ValidationSplitPercent\ percent validation data using a fixed random seed. To avoid leakage across partitions, we ensure that the same article text does not appear in both the training and validation sets by removing tweeted articles from the validation sample if that article's text appears in the training sample (due to also being referenced in a different tweet).

For the RoBERTa training sample, we retain tweets posted on or after June 1,
2011, removing the sparse early tail in which some training months contain
observations from only one party. Within each retained tweet month $m$, let
$N_{m,D}$ and $N_{m,R}$ denote the number of training observations linked to
Democratic and Republican politicians, respectively, and let
$N_m=N_{m,D}+N_{m,R}$. We assign each observation from party
$p\in\{D,R\}$ weight
\[
w_{m,p}=\frac{N_m}{2N_{m,p}}.
\]
Thus each party receives total weight $N_m/2$ within month $m$, while the
month's total weight remains $N_m$. We compute these weights from the training
sample and apply the same month--party weights to validation observations.

Appendix Figure \ref{fig:tweet_time_series} summarizes the monthly
counts and party shares of article-tweet observations in this sample.

\subsubsection{\citet{Budak2016FairBalanced} Article Data}

Our second cleaned article source is based on the labeled article dataset from \citet{Budak2016FairBalanced}. We merge article text files with the associated hand-coded labels, retain observations with non-missing text and URL, and apply the same invalid-article and non-text-cleaning pipeline described above. That is, we drop pages classified as non-articles and replace raw scraped text with text cleaned of interface and other non-news fragments before constructing the final article text measure. We then apply the same minimum article-length restriction of \MinArticleChars\ characters. From these cleaned records, we construct the article-level slant measures used in our validation exercises.

\subsubsection{ProQuest/TDM Analysis Data}

For the author-, outlet-, and event-study analyses, we construct a separate cleaned dataset from the large ProQuest/TDM article archive. Following the ProQuest workflow documented separately, we first pull and preprocess the raw TDM exports and then organize the article-level archive used in the downstream cleaning steps. We next restrict this archive to observations classified as news articles, thereby excluding non-news content or opinion/editorial content from the analysis sample.  %
We then standardize outlet names, remove observations with missing author information, and expand multi-author bylines into author-level observations. After that, we clean and harmonize author names, including nickname-based first-name standardization, and retain only observations with parsable author identities. The ProQuest query selects the Newspapers source type plus enumerated newspaper publication IDs, excluding the separate Wire Feeds source type; downstream, we remove duplicate author--article pairs and exact-text duplicate articles. We then merge in publication and outlet-group metadata and recover missing identifiers when possible. To focus on substantively meaningful journalistic activity, we then drop authors with \MinAuthorArticlesHardcoded\ or fewer articles and retain only authors with more than \MinAuthorArticlesHardcoded\ articles. After these restrictions, the resulting cleaned all-outlets sample contains \SummaryAllOutletsArticles\ articles from \SummaryAllOutletsOutlets\ outlets. For the event-study, outlet-decomposition, AKM, KSS, BLM, and structural analyses, we first drop outlet spells with fewer than \InitialMinSpellArticlesWord\ articles, reconstruct spells, and retain spells with at least \MonthlyPanelMinSpellArticlesWord\ articles. We then exclude authors with more than \EventStudyMaxMovesCap\ valid moves over their complete cleaned outlet history. Parent-company robustness uses the same eligible outlet-level articles before applying its additional restrictions. We estimate AKM and KSS directly on articles from the selected samples. Event-study windows impose additional consecutive-article requirements. When defining these moves, we do not bridge gaps longer than \EventStudyGapCapMonths\ months and exclude adjacent spell pairs whose boundary articles occur on the same day. This cleaned ProQuest/TDM dataset underlies the author-slant, outlet-slant, journalist-mobility, and event-study analyses in the paper. For the mover event-study, outlet-level-decomposition, BLM, and structural analyses, we remove publication-month means estimated once on the full eligible article universe from article slant before constructing article outcomes and outlet means; the AKM and KSS specifications instead use article slant directly with publication-month fixed effects. Here, the eligible article universe is the cleaned sample after the spell screen and move cap.

We infer journalists' gender from first names and race from surnames.

Appendix Table \ref{tab:appendix_sample_construction} tracks observation counts for our main datasets throughout the data cleaning procedures.

\subsection{Anonymized Slant-Measure Comparison}
\label{sec:slant-measure-comparison}

One potential concern is that our slant measure might load on geography or
other proxies that are correlated with the party of a sharing politician, but
may or may not represent political slant itself. We address this in two ways.
First, Section \ref{sec:slant-external-validation} validates our slant measure
against external measures of slant, showing, for example, that our measure is
more strongly correlated with political slant as perceived by human readers
than are other measures in the literature. Second, Appendix Table
\ref{tab:slant_measure_comparison} reports model diagnostics, key results, and
top-contribution tokens for our baseline measure and for two alternative slant
models trained on geography-anonymized or more aggressively anonymized
politician-tweeted article text and then applied to the same non-anonymized
ProQuest articles. The geography version makes generic replacements for U.S.
state and local identifiers, including place names, local facilities, and
related demonyms, while preserving national, foreign, outlet, and other
non-geographic text. For example, it replaces ``Texas Department of
Transportation'' with ``the state Department of Transportation.'' The
aggressive version additionally replaces personal names, organizations and
publications, broader geographic identifiers, and dates or time periods, while
preserving generic political content and numeric references outside dates. For
both versions, GPT-5-mini is used to identify these terms and make
replacements. The table holds each empirical sample and estimator contract
fixed while replacing the slant outcome and the outcome-dependent quantities
required by that estimator. Qualitative takeaways from these alternatives are
largely similar, though the anonymized measures have lower predictive accuracy,
larger event-study coefficients, and
weakly larger journalist shares in some specifications. We use the
non-anonymized slant measure as our baseline because it is most strongly
correlated with human-perceived slant as measured in Section
\ref{sec:slant-external-validation}, while still showing robustness to these
alternatives given ambiguity around whether the anonymized identifiers
represent or are merely correlated with human-perceived slant.

\subsection{Media Topics}
\label{sec:topics}

Our measure of slant intentionally includes both which topics or stories are covered and how that coverage is framed. To help distinguish these dimensions and to better understand the drivers of slant across articles, outlets, and journalists, including around moves, we also use an LLM (OpenAI GPT-4o-mini) to assign each article to exactly one media topic from an existing media topic taxonomy. We source this media topic taxonomy from The International Press Telecommunications Council (IPTC). To assign articles to a topic from this taxonomy, we provide the LLM with the article text and the \NIPTCMediaTopics{} media topics in IPTC's Level One (``Top-Level'') list, and ask it to assign this article to exactly one topic from this list based on the article's text. We use this same approach to apply topics to articles in our politician-tweeted and ProQuest article samples.

These assigned topics allow us to analyze the frequency of topics across datasets, and the ability of topics to explain observed slant. In Appendix Figure \ref{fig:topic_distributions}, we plot for each media topic the share across reweighted politician article tweets (Panel A) and the share across ProQuest articles (Panel B). We see that while all \NIPTCMediaTopics{} topics are represented in both datasets and their frequency is correlated, the datasets differ in the relative frequency of different topics. The ``politics and government'' media topic is the modal topic among politician-tweeted articles, but is only the fourth most common topic among ProQuest articles.
In Appendix Figure \ref{fig:topic_party_share}, we show by topic the share of articles tweeted by Republican (vs.\ Democratic) politicians in our politician-tweeted articles sample. After reweighting to balance the overall share of Democrat vs.\ Republican tweets, we see that most topics are roughly equally likely to be tweeted by each party. Notable exceptions include the ``environment'' and ``society'' topics, which are more frequently shared by Democratic politicians than by Republican politicians. In Appendix Figure \ref{fig:slant_distribution_by_topic}, we plot the distribution of article-level slant by topic in our ProQuest sample. We see mild variation in mean slant across topics, accounting for \AcrossTopicSlantVariancePct\% of the variation in article-level slant. The remaining \WithinTopicSlantVariancePct\% variation in slant occurs within topics. Appendix Figure \ref{fig:eventstudy_topic} follows the main event-study specification after replacing article slant with the mean slant of the article's topic, constructed using the same publication-month adjustment. The resulting coefficients are close to zero.

\subsection{Allowing for Worker-Firm Interactions}
\label{app-sec-blm}
In this section, we extend the framework developed in Section \ref{subsec-akm} by adopting the approach of \citealt{Bonhomme2019DistributionalFramework}, hereafter BLM. Relative to the AKM specification, this framework relaxes additive separability by allowing for latent heterogeneity across journalists and for nonparametric interactions between journalist types and outlet classes.

In our application, we use all articles in the two spells around each journalist's first move and distinguish between two journalist types and four outlet classes. We use the upstream calendar-adjusted article slant, with its common level restored, and impose one shared journalist-type by outlet-class mean matrix before and after the move. Conditional noise variances remain role-specific. Journalist types are latent and estimated from the data.\footnote{Here, ``conservative'' and ``liberal'' refer to the measured slant of the articles produced.} Outlet classes are constructed in a preliminary step by clustering outlets based on their empirical CDF of the same adjusted article slant.

Figure \ref{appfig-blm-slant} shows the mean level of slant produced by journalists of each type working in outlets of each type. The figure reveals variation in slant along both dimensions: type 1 journalists write articles that are \BLMJournalistTypeGapPoints\ points more conservative than their type 0 counterparts, averaging within-class differences across outlet classes, while the most conservative outlet type (type 3) produces articles about \BLMOutletTypeGapPoints\ points more conservative than the most liberal (type 0), on average across journalist types. Figure \ref{appfig-blm-sorting} shows the distribution of journalist types across outlet types.

We explore what observable worker characteristics correlate with their type distribution. In particular, Figure \ref{appfig-blm-obsevables} shows that registered Republicans are \BLMRepDemTypeOneGapPctPts\ percentage points more likely to be type 1 workers than registered Democrats. Male journalists are also \BLMMaleFemaleTypeOneGapPctPts\ percentage points more likely to be type 1 than female journalists.

To quantify departures from additivity, we project the shared slant mean matrix onto additive journalist and outlet type indicators, and examine how much the actual cell means deviate from the additive fit. The interaction residuals (the difference between actual cell means and their best additive approximation) account for \BLMInteractionVarianceSharePct\% of observed calendar-adjusted article variance.

Figure \ref{appfig-blm-decomposition} illustrates the BLM decomposition. Its denominator is observed calendar-adjusted article variance in this first-move sample; the structural decomposition instead uses model-implied stationary adjusted-slant variance, and AKM/KSS use raw slant with conditional month effects. The upstream adjustment subtracts pooled month means on the eligible article-credit universe, which can remove composition shifts as well as common shocks. It is not equivalent to estimating month effects conditional on BLM types and classes. The decomposition attributes \BLMJournalistVarianceSharePct\% of observed calendar-adjusted article variance to journalist types, \BLMOutletVarianceSharePct\% to outlet classes, and \BLMSortingVarianceSharePct\% to sorting of journalists across outlets.

\subsection{Ownership Event Studies}
\label{app-sec-ownership-event-study}

We also analyze changes in newspaper ownership to better understand the potential role of owners in shaping their newspapers' slant. We use the slant of articles published in other news outlets in an acquiring owner's portfolio as a proxy for that owner's preferred slant. We find no evidence that an acquired outlet's slant shifts towards the average article slant of that new owner's portfolio, which may be consistent with a limited role for owners in determining news slant.

More precisely, we define ownership change events
as adjacent-year changes in the newspaper owner.
The ownership panel links ProQuest TDM publication identifiers to year-specific
owners. Owner values are based on \emph{Editor \& Publisher} group-company
ownership fields, matched to ProQuest publications and supplemented with
manual review for publications with incomplete automatic coverage and to remove false ownership changes.

For each event, we construct the acquiring owner's pre-acquisition
portfolio from other outlets owned by the acquiring owner in the year
before the acquisition, excluding the acquired outlet and same-year
acquisitions. Let $e$ index ownership-change events and $l$ index years
relative to the ownership change. The baseline sample is balanced over
$l\in\{-2,-1,0,1\}$ and omits $l=-1$. Let $\tilde{x}_{el}$ be the
acquired outlet's average slant residualized by calendar-year means
estimated among articles from all other outlets that are not acquired during our sample,
and let $g_{e}$ be the raw pre-period slant gap between
other outlets in the acquiring owner's portfolio and the acquired outlet. 

In one rough analog of our journalist movers' specification, we estimate:
\[
\tilde{x}_{el}=\gamma_{e}+\sum_{k\neq -1}\rho^{A}_{k}g_{e}{\bf 1}_{\{l=k\}}+\epsilon_{el},
\]
weighting by $g_{e}^{2}$. The plotted coefficients are $\rho^{A}_{k}$;
positive post-acquisition estimates indicate the share of the initial slant gap closed by the acquired outlet's post-period slant change.

Two alternative specifications measure the extent to which an outlet's average slant and the average slant in the owner's portfolio change around the acquisition. First, we fix the
same acquiring-owner portfolio membership and define time-varying alignment as $a_{el}^{TV}=-|\bar{x}_{el}-\bar{x}_{P(e)l}|$,
where $P(e)$ denotes this fixed portfolio and $\bar{x}_{P(e)l}$ is
its contemporaneous average slant. Second, we define $a_{el}^{FP}=-|\tilde{x}_{el}-\tilde{x}_{P(e),pre}|$,
where $\tilde{x}_{P(e),pre}$ is the acquiring owner's fixed pre-period
portfolio slant after applying the same calendar-year residualization.
For each alignment outcome, we estimate
\[
a_{el}=\gamma_{e}+\sum_{k\neq -1}\rho^{B}_{k}{\bf 1}_{\{l=k\}}+\epsilon_{el}.
\]
The plotted coefficients are relative to the year before acquisition. All standard
errors are clustered at the event level. The data contain \OwnershipEventCount{} ownership changes. \OwnershipBaselineBalancedEvents{} enter the gap-slope and fixed-pre
alignment baselines after restricting to a balanced panel; \OwnershipTimeVaryingAlignmentEvents{} enter the time-varying
alignment baseline because it requires contemporaneous owner-portfolio slant
coverage, and this value is missing in some periods.

Appendix Figures \ref{fig:ownership_event_study}, \ref{fig:ownership_event_study_variation},
and \ref{fig:ownership_event_study_alignment_variation} visualize these event-study coefficients and their underlying variation. The estimates reported in the variation-plot annotations
are using $l=-1$ as the pre period
and the average of $l=0$ and $l=1$ as the pooled post period. The pooled estimates are
$\OwnershipGapSlopePooledPost$ (s.e. $\OwnershipGapSlopePooledPostSE$)
for the movers-specification slope coefficient, $\OwnershipTimeVaryingAlignmentPooledPost$
(s.e. $\OwnershipTimeVaryingAlignmentPooledPostSE$) for time-varying
alignment, and $\OwnershipFixedPreAlignmentPooledPost$ (s.e.
$\OwnershipFixedPreAlignmentPooledPostSE$) for fixed-pre residualized
alignment. Positive estimates would indicate movement toward the acquiring
owner's portfolio; the estimates therefore do not point toward post-acquisition
convergence in slant towards average slant in the owner's other outlets.

\subsection{Cross-Fit Inference and Role-Equality Diagnostics}
\label{sec:outlet_decomp_inference}

We construct the confidence intervals in Table
\ref{tab:outlet_decomp} using a cluster-level fractional-random-weight
bootstrap \citep{Praestgaard1993ExchangeablyWeightedBootstraps,Xu2020FractionalRandomWeightBootstrap}.
For each draw, we assign an independent mean-one exponential weight to each
journalist cluster, hold that weight fixed across the journalist's
observations and all repeated splits, and re-estimate the cross-fit statistic.
Unlike resampling journalist clusters, the strictly positive weights retain
every mover in the estimation sample and therefore preserve the realized
connected support in every draw. Split assignments, outlet means, and outlet
aggregation weights also remain fixed. The reported 95 percent interval uses
the 2.5th and 97.5th percentiles of the resulting estimates.

The model implies equality between the origin and destination role effects,
up to their common normalization. For each split, $V^{o}$ is the cross-half
covariance between the origin-effect estimates, $V^{d}$ is the analogous
covariance between the destination-effect estimates, and $C$ is the average
of the two cross-half origin--destination covariances. We average these
moments across splits and define
\[
Q=V^{o}+V^{d}-2C.
\]
Equivalently, $Q$ is the cross-half covariance of the centered
origin--destination role gaps. Under role equality, these gaps do not vary
across outlets and $Q=0$. To test this restriction, we fit a model with
common origin and destination effects, apply wild-bootstrap Rademacher
multipliers at the journalist level to its residuals, and re-estimate $Q$.
The reported p-value compares the observed statistic with this restricted-null
distribution.

We also report the descriptive agreement ratio $2C/(V^{o}+V^{d})$. The ratio
equals one under role equality, but need not be bounded by zero and one in
finite samples.

\section{Appendix Exhibits\label{sec:appendix-1}}

\setcounter{figure}{0}
\renewcommand{\figurename}{Appendix Figure}
\renewcommand\thefigure{\thesection\arabic{figure}}  
\setcounter{table}{0}
\renewcommand{\tablename}{Appendix Table}
\renewcommand\thetable{\thesection\arabic{table}}     

\input{appendix_figures.tex}

\clearpage
\input{linkage_summary_statistics_filled}

\pagebreak
\input{party_summary_statistics_filled}

\pagebreak
\input{slant_measure_comparison_table_filled}

\pagebreak
\input{consistent_mover_summary_statistics_filled}

\pagebreak
\input{move_summary_statistics_filled}

\clearpage
\input{appendix_sample_construction_filled}
\clearpage
\input{budak_article_filled}

\pagebreak\input{akm_kss_demo_table_filled}

\begin{landscape}\input{akm_table_filled}

\end{landscape}

\begin{table}[H]
\caption{Full Parameter Estimates\label{tab:full_estimates}}

\begin{centering}
\input{modeling_selection_full_table}
\par\end{centering}
\smallskip{}

{\small Notes: The table reports the full parameter vector estimated by simulated maximum
likelihood. Standard errors are clustered by journalist and are conditional on the first-step
$\widehat{\sigma}_e$. Ideal points and median class preferences $\delta_k=\Lam(b_k)$ are in slant
units; $\tau_\zeta$ is in log-odds units. The stay, tenure, and same-class parameters are in utility units; the destination
intercepts $\tilde\alpha_k$ (with $\tilde\alpha_0=0$) and the distance cost
$\tilde\kappa_d$ are in within-nest units (utility divided by $\lambda$).}{\small\par}
\end{table}

\clearpage
\input{structural_variance_table}

\end{document}

%% file: input/scalars.tex
\newcommand{\OutletDecompEqualityPValueResidualized}{0.209}

\newcommand{\OutletDecompCrossfitOutlets}{154}
\newcommand{\OutletDecompNaturalPluginOutlets}{238}
\newcommand{\JournalistExplainedOutletDecompResidualizedPercent}{14}
\newcommand{\OutletDecompCrossfitCILowResidualizedPercent}{11.3}
\newcommand{\OutletDecompCrossfitCIHighResidualizedPercent}{18.9}
\newcommand{\OutletDecompEqualOutletWeightsResidualizedPercent}{15.7}
\newcommand{\OutletDecompEqualOutletWeightsCILowResidualizedPercent}{12.0}
\newcommand{\OutletDecompEqualOutletWeightsCIHighResidualizedPercent}{20.8}
\newcommand{\OutletDecompMatchedPluginResidualizedPercent}{18.0}
\newcommand{\OutletDecompNaturalPluginResidualizedPercent}{15.5}
\newcommand{\OutletGapRight}{0.25}
\newcommand{\ChangeOnMoveRight}{0.17}
\newcommand{\OutletGapLeft}{-0.26}
\newcommand{\ChangeOnMoveLeftAbs}{0.17}
\newcommand{\OutletPsiDAboveMedian}{0.08}
\newcommand{\OutletPsiOAboveMedian}{0.09}
\newcommand{\DescriptiveEventStudyJournalistEffectsShare}{0.33}
\newcommand{\DescriptiveEventStudyJournalistEffectsSharePct}{33}
\newcommand{\BudakRobertaPartitionAccuracy}{70}
\newcommand{\BudakInterRaterRsq}{0.151}
\newcommand{\BLMJournalistTypeGapPoints}{17}
\newcommand{\BLMOutletTypeGapPoints}{43}
\newcommand{\BLMRepDemTypeOneGapPctPts}{13}
\newcommand{\BLMMaleFemaleTypeOneGapPctPts}{19}
\newcommand{\BLMJournalistVarianceSharePct}{9.0}
\newcommand{\BLMOutletVarianceSharePct}{25.7}
\newcommand{\BLMSortingVarianceSharePct}{2.2}
\newcommand{\BLMInteractionVarianceSharePct}{0.06}
\newcommand{\SelW}{0.40}

\newcommand{\SelIdealGap}{0.10}
\newcommand{\SelLadderSpan}{0.71}

\newcommand{\SelSwingFixed}{0.057}
\newcommand{\SelSwingFixedSD}{0.21}
\newcommand{\SelEqualThirdsFixedSD}{0.04}
\newcommand{\SelSwingFree}{0.081}
\newcommand{\SelSwingFreeSD}{0.29}
\newcommand{\SelRepCompShareFree}{32}
\newcommand{\SelGapMin}{0.39}
\newcommand{\SelGapMax}{0.43}
\newcommand{\SelConsShareDemB}{26}
\newcommand{\SelConsShareRepB}{35}
\newcommand{\AKMArticleWeightedEqualThirdsGapRaw}{0.003}

\newcommand{\AKMArticleWeightedRepMinusDemGapRaw}{0.035}
\newcommand{\AKMArticleWeightedEqualThirdsGapSd}{0.01}

\newcommand{\AKMArticleWeightedRepMinusDemGapSd}{0.12}

\newcommand{\StructuralSlantVarianceJournalist}{0.005396}
\newcommand{\StructuralSlantVarianceJournalistTotalPercent}{6.5}

\newcommand{\StructuralSlantVarianceOutletClass}{0.020103}
\newcommand{\StructuralSlantVarianceOutletClassTotalPercent}{24.3}

\newcommand{\StructuralSlantVarianceOfferShock}{0.013901}
\newcommand{\StructuralSlantVarianceOfferShockTotalPercent}{16.8}

\newcommand{\StructuralSlantVarianceTwiceCovJournalistOutletClass}{0.007707}
\newcommand{\StructuralSlantVarianceTwiceCovJournalistOutletClassTotalPercent}{9.3}

\newcommand{\StructuralSlantVarianceTwiceCovJournalistOfferShock}{0.002611}
\newcommand{\StructuralSlantVarianceTwiceCovJournalistOfferShockTotalPercent}{3.2}

\newcommand{\StructuralSlantVarianceTwiceCovOutletClassOfferShock}{-0.008780}
\newcommand{\StructuralSlantVarianceTwiceCovOutletClassOfferShockTotalPercent}{-10.6}

\newcommand{\StructuralSlantVarianceArticleNoise}{0.041768}
\newcommand{\StructuralSlantVarianceArticleNoiseTotalPercent}{50.5}
\newcommand{\StructuralSlantVarianceSystematic}{0.040938}
\newcommand{\StructuralSlantVarianceSystematicTotalPercent}{49.5}
\newcommand{\StructuralSlantVarianceTotal}{0.082706}

\newcommand{\StructuralSlantVarianceJournalistWithSelectionTotalPercentRounded}{10}
\newcommand{\SummaryAllOutletsArticles}{9,005,189}
\newcommand{\SummaryAllOutletsArticlesMillionsFloor}{9}
\newcommand{\SummaryAllOutletsAuthors}{33,571}

\newcommand{\SummaryAllOutletsOutlets}{279}
\newcommand{\SummaryAllOutletsOutletsCeilHundreds}{300}

\newcommand{\KSSJournalistExplainedAKM}{8}
\newcommand{\KSSOutletExplainedAKM}{20}
\newcommand{\KSSCovarianceExplainedAKM}{7}

\newcommand{\KSSJournalistExplainedAKMColOne}{16}
\newcommand{\KSSOutletExplainedAKMColOne}{19}
\newcommand{\KSSCovarianceExplainedAKMColOne}{8}
\newcommand{\KSSCorrelationAKMColOne}{0.245}
\newcommand{\KSSJournalistExplainedAKMColEight}{15}
\newcommand{\KSSJournalistExplainedAKMColNine}{13}
\newcommand{\KSSCorrelationTopAKM}{0.085}

\newcommand{\OwnershipEventCount}{39}
\newcommand{\OwnershipBaselineBalancedEvents}{31}
\newcommand{\OwnershipTimeVaryingAlignmentEvents}{16}
\newcommand{\OwnershipGapSlopePooledPost}{-0.009}
\newcommand{\OwnershipGapSlopePooledPostSE}{0.039}
\newcommand{\OwnershipTimeVaryingAlignmentPooledPost}{-0.028}
\newcommand{\OwnershipTimeVaryingAlignmentPooledPostSE}{0.015}
\newcommand{\OwnershipFixedPreAlignmentPooledPost}{-0.020}
\newcommand{\OwnershipFixedPreAlignmentPooledPostSE}{0.008}
\newcommand{\NMatchedJournalists}{10,670}
\newcommand{\JournalistLTwoMatchedShare}{32}
\newcommand{\JournalistDemShare}{51}
\newcommand{\JournalistRepShare}{21}
\newcommand{\JournalistNonPartisanShare}{27}

\newcommand{\SentimentEndorsingAll}{92}
\newcommand{\SentimentUnclearAll}{5}
\newcommand{\SentimentCriticizingAll}{3}

\newcommand{\SentimentCriticizingDem}{2}

\newcommand{\SentimentCriticizingRep}{4}

\newcommand{\Nfinaltweets}{182,538}

\newcommand{\NfinaltweetsDem}{128,105}
\newcommand{\NfinaltweetsRep}{54,433}

\newcommand{\BudakSampleArticlesRoundedDownTenThousand}{10,000}
\newcommand{\BudakSampleOutlets}{15}
\newcommand{\NIPTCMediaTopics}{17}
\newcommand{\NTweetsCollectedMillions}{4.4}

\newcommand{\NPoliticianUsernames}{1,800}

\newcommand{\NTweetedUniqueURLsMillionsFloor}{1}
\newcommand{\NSampledURLs}{600,000}
\newcommand{\NArticlesNonEmptyRows}{348,821}
\newcommand{\NArticlesNonEmptyUniqueURL}{328,966}
\newcommand{\TweetClassificationValidationBalancedAccuracy}{82}
\newcommand{\TweetClassificationValidationROCAUC}{0.90}

\newcommand{\ConnectedSampleJournalists}{30,504}
\newcommand{\ConnectedSampleMovers}{4,091}
\newcommand{\ConnectedSampleOutlets}{266}
\newcommand{\ConnectedSampleObservations}{7,065,499}
\newcommand{\AcrossTopicSlantVariancePct}{9.2}
\newcommand{\WithinTopicSlantVariancePct}{90.8}
\newcommand{\OutletStratificationMixedBalancedAccuracyPct}{83.1}

\newcommand{\OutletStratificationHeldoutBalancedAccuracyPct}{76.9}

\newcommand{\OutletStratificationHeldoutMinusMixedBalancedAccuracyGapWelchPValue}{0.139}
\newcommand{\OutletStratificationMixedROCAUC}{0.93}

\newcommand{\OutletStratificationHeldoutROCAUC}{0.89}

\newcommand{\OutletStratificationHeldoutMinusMixedROCAUCGapWelchPValue}{0.054}

\newcommand{\RoBERTaEffectiveBatchSize}{128}
\newcommand{\RoBERTaAttentionDropoutPercent}{20}
\newcommand{\RoBERTaHiddenDropoutPercent}{20}
\newcommand{\RoBERTaLearningRate}{\ensuremath{0.5\times10^{-5}}}
\newcommand{\RoBERTaWarmupSteps}{1000}
\newcommand{\RoBERTaTrainingEpochs}{25}
\newcommand{\TrainSplitPercent}{70}
\newcommand{\ValidationSplitPercent}{30}

\newcommand{\RoBERTaTrainingLoss}{0.262}
\newcommand{\RoBERTaValidationLoss}{0.449}

\newcommand{\EventStudytzeroTfive}{77}
\newcommand{\EventStudyPostMoveLevelTfive}{0.77}

\newcommand{\EventStudyBetaApproxTfive}{0.29}

\newcommand{\InitialMinSpellArticlesWord}{five}

\newcommand{\EventStudyMinSpellArticlesWord}{five}
\newcommand{\EventStudyMaxMovesCap}{10}
\newcommand{\EventStudyGapCapMonths}{12}

\newcommand{\MonthlyPanelMinSpellArticlesWord}{five}
\newcommand{\MinArticleChars}{1,000}
\newcommand{\MinAuthorArticlesHardcoded}{10}
\newcommand{\BERTVocabSizeHardcoded}{30,000}
\newcommand{\BERTPretrainingWordsBillionsHardcoded}{3.3}
\newcommand{\BERTBaseLayersHardcoded}{12}
\newcommand{\BERTBaseHiddenSizeHardcoded}{768}
\newcommand{\BERTBaseAttentionHeadsHardcoded}{12}
\newcommand{\BERTBaseParamsMillionsHardcoded}{110}
\newcommand{\BERTLargeParamsMillionsHardcoded}{340}
\newcommand{\JournalistDemShareWeaverHardcoded}{28.1}
\newcommand{\JournalistRepShareWeaverHardcoded}{7.1}
\newcommand{\JournalistDemShareWeaverHardcodedfloor}{28}
\newcommand{\JournalistRepShareWeaverHardcodedfloor}{7}
\newcommand{\JournalistRepShareWeaverSeventyoneHardcoded}{25.7}
\newcommand{\JournalistDemShareWeaverSeventyoneHardcoded}{35.5}
\newcommand{\JournalistVotingtoClintonSharePovich}{89}
\newcommand{\ConsumerExplainedGSMediaBias}{20}

%% file: figures.tex
\begin{figure}[H]
\caption{Partisan Distribution of Journalists\label{fig:weaver}}

\begin{centering}
\emph{Panel A: Surveyed Journalists' Self-Reported Partisan Identification
(\citealt{Weaver2019AmericanJournalistDigitalAge})}
\par\end{centering}
\begin{centering}
\includegraphics[scale=0.75]{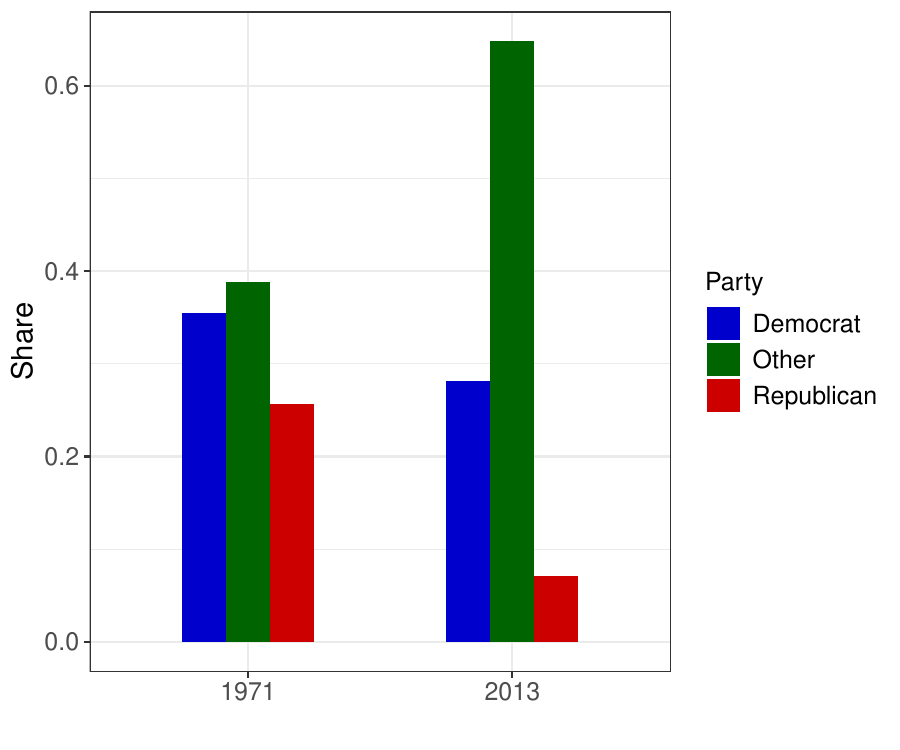}
\par\end{centering}
\begin{centering}
\emph{Panel B: Surveyed Washington Correspondents' Self-Reported Vote
Choice in 1992 (\citealt{Povich1996PartnersAdversaries})}
\par\end{centering}
\begin{centering}
\includegraphics[scale=0.75]{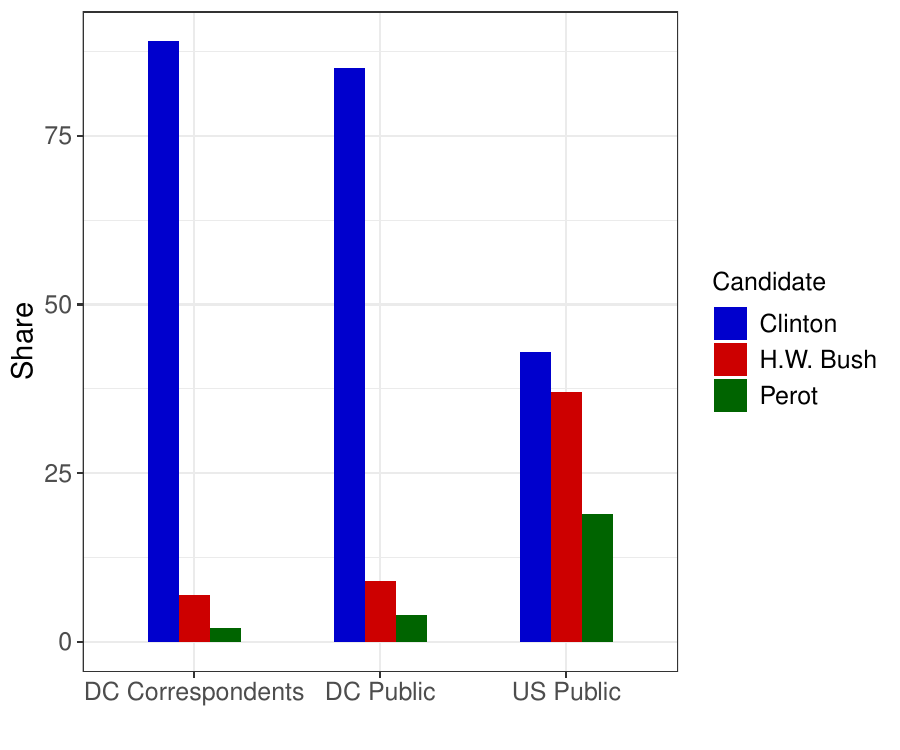}
\par\end{centering}
Notes: Panel A reports the share of surveyed journalists in 1971 and
2013 that self-reported as a Democrat (blue), a Republican (red),
or as non-partisan or a third party (green). Data is taken from \citet{Weaver2019AmericanJournalistDigitalAge}. Panel B reports the self-reported vote choices of Washington
Correspondents from \citet{Povich1996PartnersAdversaries}, as well as information on vote
choices for the District of Columbia and the US population as a whole.
\end{figure}

\begin{figure}[H]
\caption{Comparing to Human Perceived Slant\label{fig:human_compare_rank}}
\smallskip{}

\begin{centering}
\emph{Panel A: Non-parametric Rank Correlation}
\par\end{centering}
\begin{centering}
\includegraphics[scale=0.65]{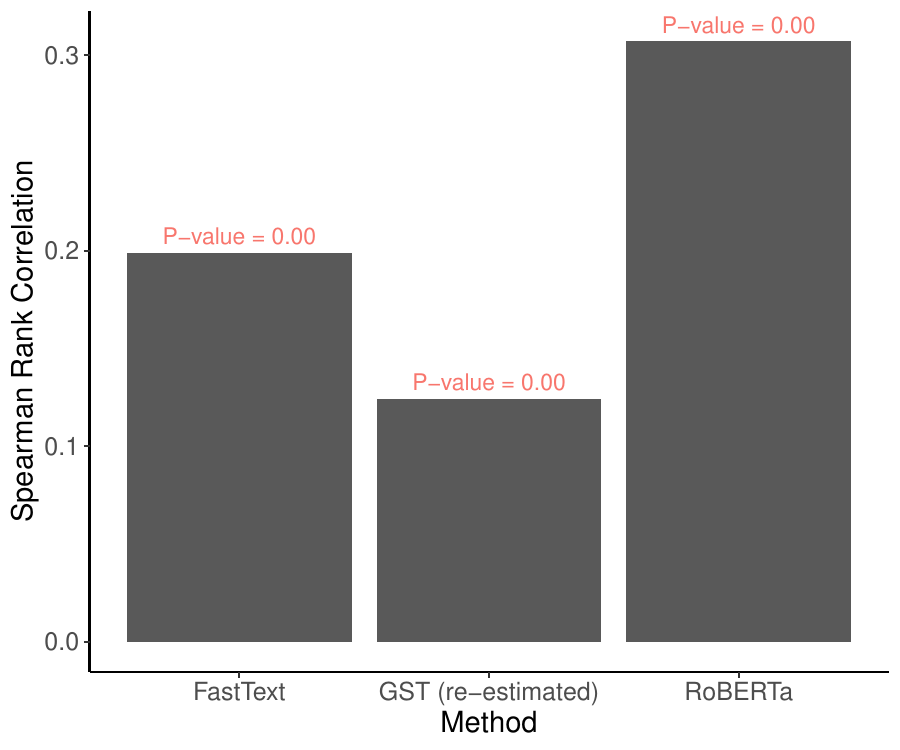}
\par\end{centering}
\smallskip{}

\begin{centering}
\emph{Panel B: Accuracy among Clearly Partisan Articles}
\par\end{centering}
\begin{centering}
\includegraphics[scale=0.65]{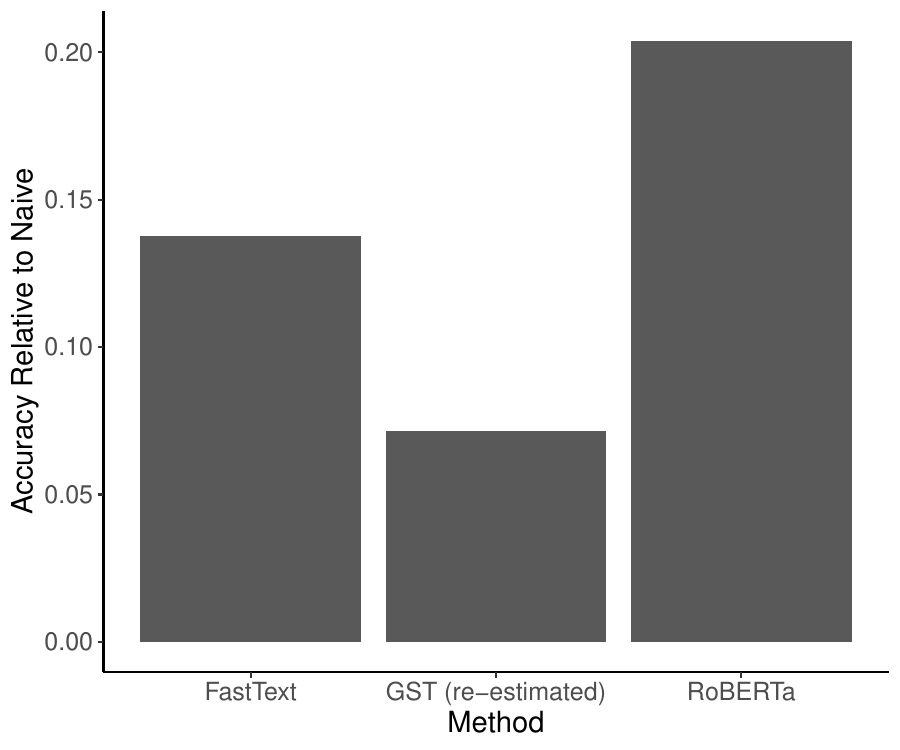}\smallskip{}
\par\end{centering}
Notes: Panel A plots the rank correlation of the \citet{Budak2016FairBalanced}
perceived slant measure for each article as defined in Appendix Table
\ref{tab:budak_et_al_article} versus the model-based slant measures
using all articles. GST uses the re-estimated Republican posterior;
its saved log-odds preserve ranks when probabilities round to one.
Panel B plots the accuracy of each model-based
method in partitioning the dataset in half between more liberal and
more conservative articles relative to a naive algorithm. In Panel
B, the sample is restricted to articles with a non-zero blinded-coder
rating. Each measure and the human rating are partitioned at their
respective sample medians; the figure reports accuracy minus 0.5.
\end{figure}

\begin{figure}[H]
\caption{Average Outlet Slant and Journalist Partisanship \label{fig:outlet_slant_partisanship}}
\smallskip{}

\begin{centering}
\includegraphics[scale=0.55]{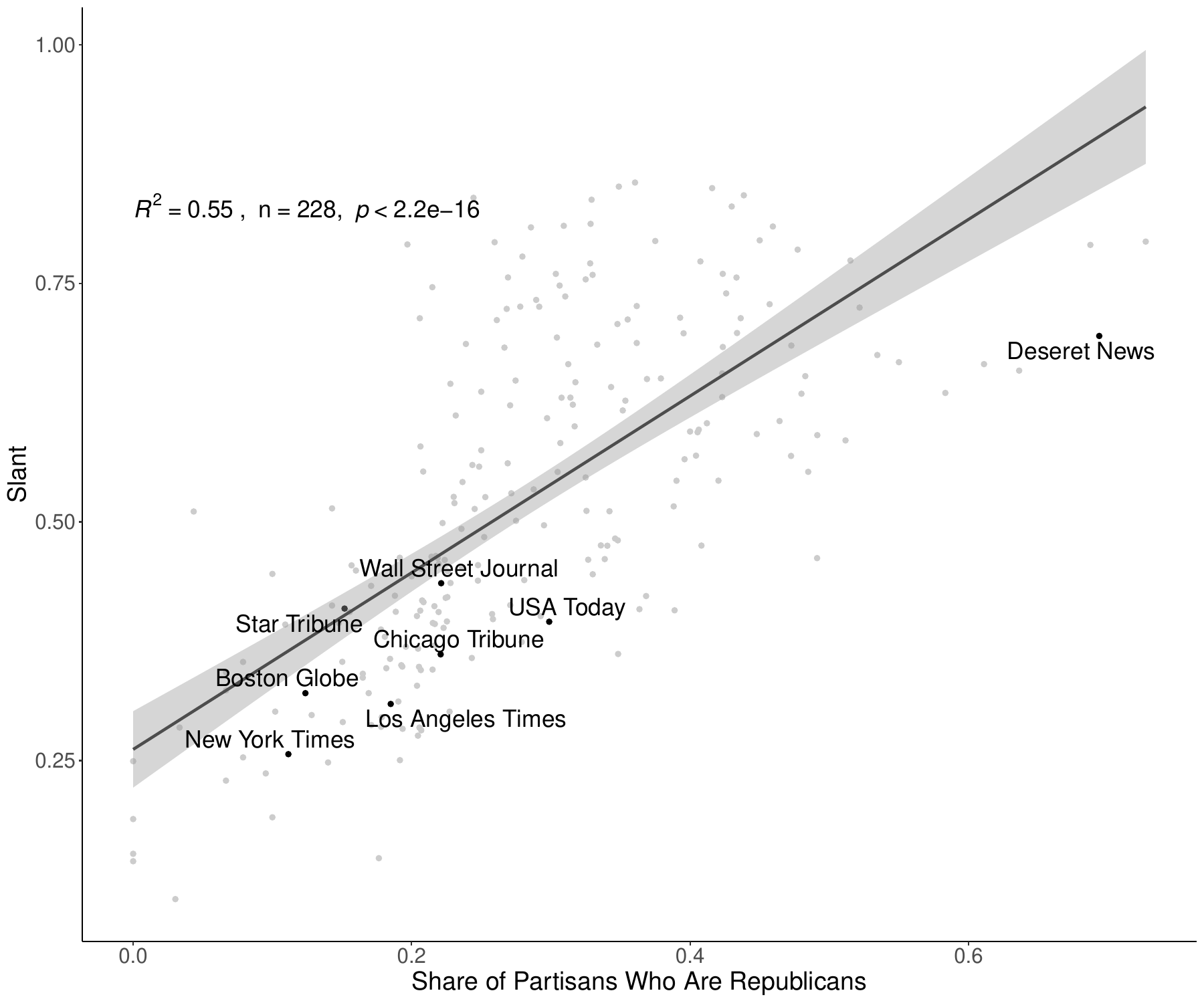}
\par\end{centering}
Notes: Figure plots the relationship between the average estimated
slant at an outlet (y-axis) with the share of matched partisan journalists
who are Republicans (x-axis). Slant ranges from 0 to 1, with larger values
indicating content estimated to be more conservative.
\end{figure}

\begin{figure}[H]
\caption{Authors at \emph{The New York Times} versus \emph{The Wall Street
Journal\label{fig:nyt_vs_wsj}}}
\smallskip{}

\begin{centering}
\emph{Panel A: All Authors}
\par\end{centering}
\begin{centering}
\includegraphics[scale=0.7]{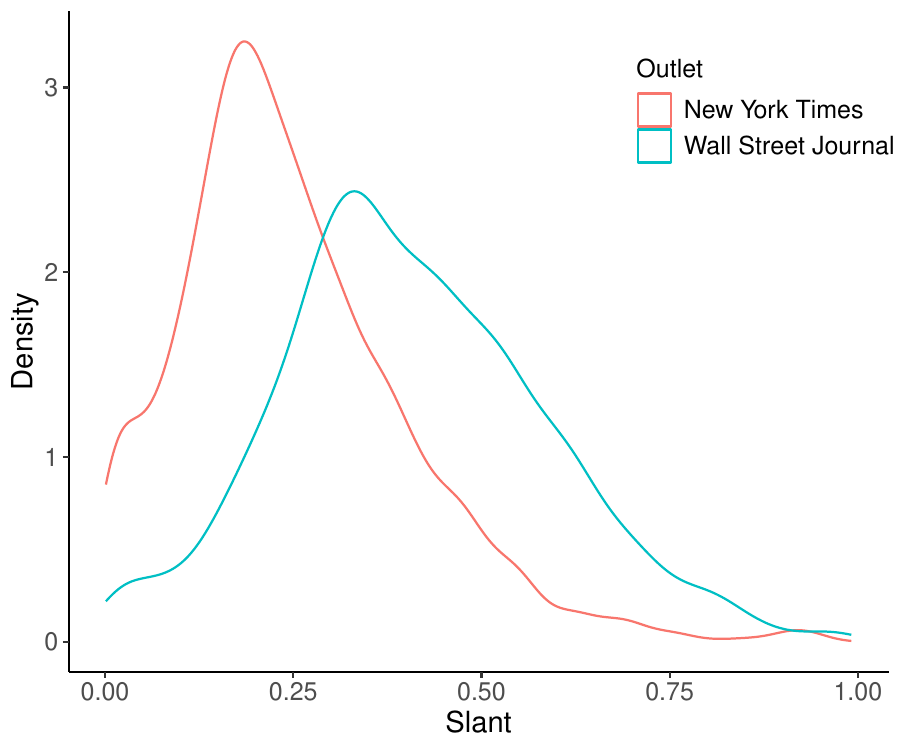}
\par\end{centering}
\smallskip{}

\begin{centering}
\emph{Panel B: Movers Between NYT and WSJ}
\par\end{centering}
\begin{centering}
\includegraphics[scale=0.7]{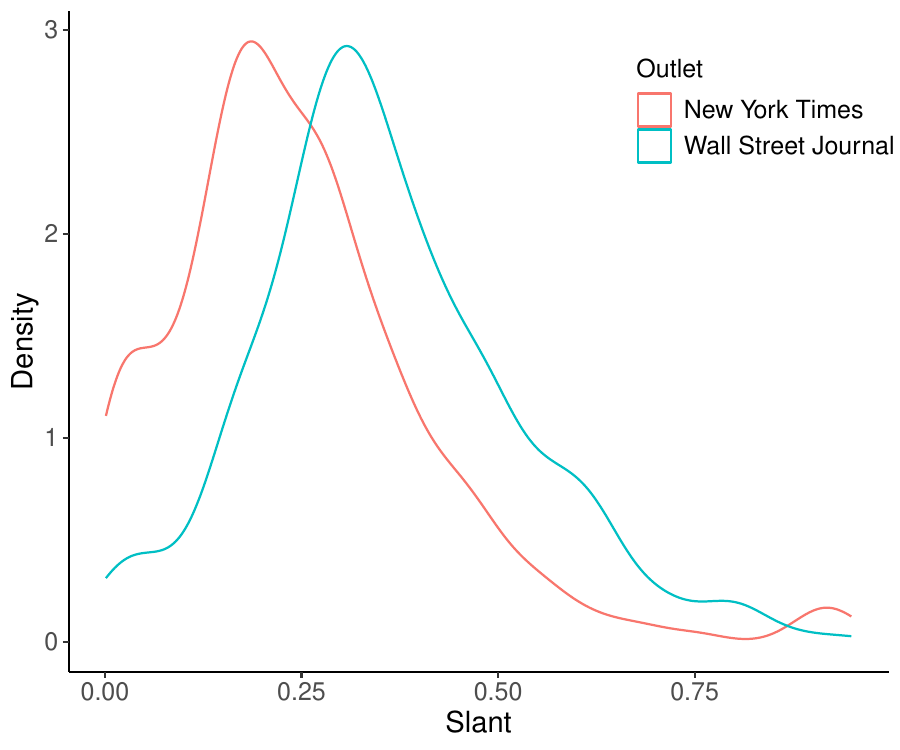}\smallskip{}
\par\end{centering}
Notes: Figure plots the distribution, across authors, of the average
slant of each author's article as measured by our RoBERTa model of
slant for each outlet. Panel A includes all authors who appear at
a given outlet. Panel B restricts attention to authors who have written
at both \emph{The New York Times} and \emph{The Wall Street Journal}.
Slant ranges from 0 to 1, with larger values indicating content estimated
to be more conservative.
\end{figure}

\begin{figure}[H]
\caption{Trends in Slant\label{fig:slant_trends}}
\smallskip{}

\begin{centering}
\includegraphics[scale=0.99]{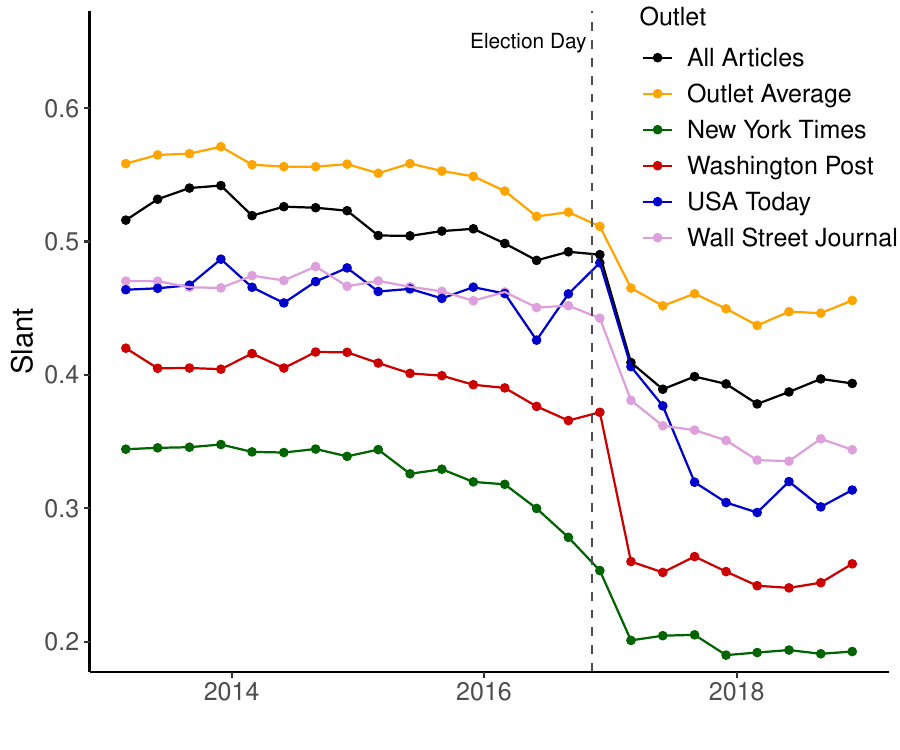}
\par\end{centering}
\smallskip{}

Notes: Figure plots the trends in average estimated slant over time, aggregated
by quarter. This average is calculated across all articles (black) as well as for selected outlets (various
colors) in our main analysis sample. We also plot an outlet-weighted (rather than article-weighted) average of slant across all outlets (yellow). The dashed vertical line is Election
Day 2016. Slant ranges from 0 to 1, with larger values indicating
content estimated to be more conservative.
\end{figure}

\begin{figure}[H]
\caption{Distribution of Origin-Destination Outlet Slant Scores in Event Study
Samples\label{fig:movedist}}

\begin{centering}
\emph{Panel A: $T=5$}
\par\end{centering}
\begin{centering}
\includegraphics[scale=0.72]{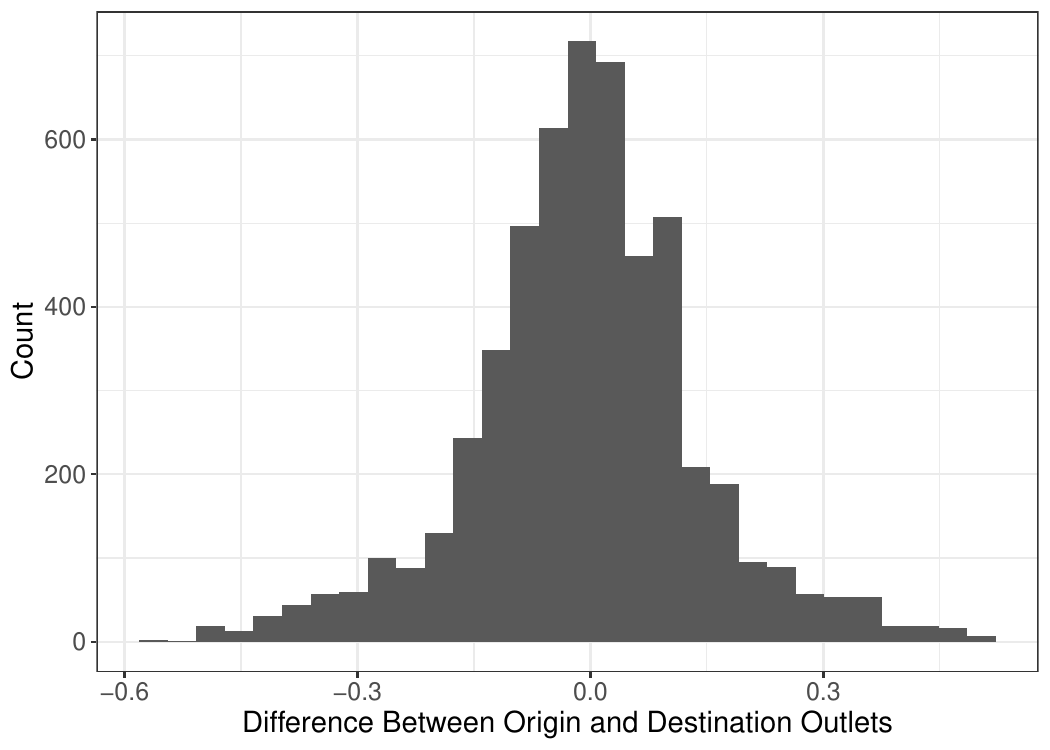}
\par\end{centering}
\begin{centering}
\emph{Panel B: $T=10$}
\par\end{centering}
\begin{centering}
\includegraphics[scale=0.72]{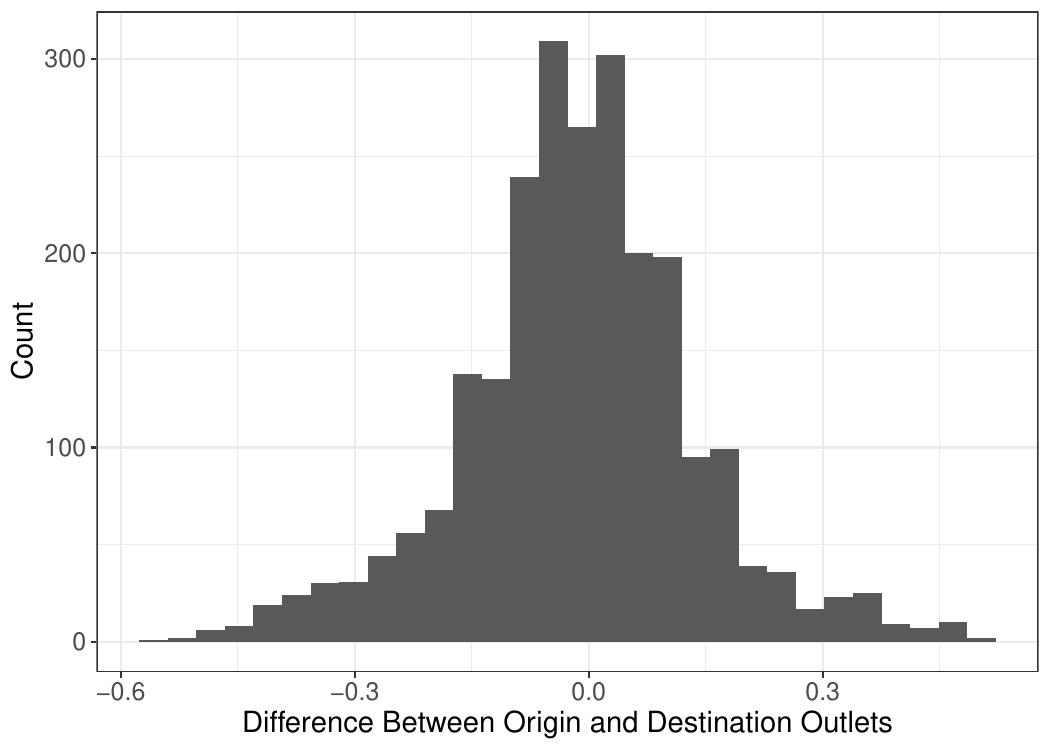}
\par\end{centering}
Notes: Figure plots the distribution of moves
$\bar{x}_{d(q)}^{(-i)}-\bar{x}_{o(q)}^{(-i)}$
for consistent movers in the baseline $T=5$ estimation sample (Panel A) and
the $T=10$ horizon robustness sample (Panel B). Outlet means use the full-period
author--article panel after publication-month residualization, excluding every
article involving the moving journalist and all associated coauthor-expanded rows.
\end{figure}

\begin{figure}[H]
\caption{Event Study Changes in Slant Around Author Moves\label{fig:eventstudy}}

\begin{centering}
\emph{Panel A: $T=5$}
\par\end{centering}
\begin{centering}
\includegraphics[scale=0.688]{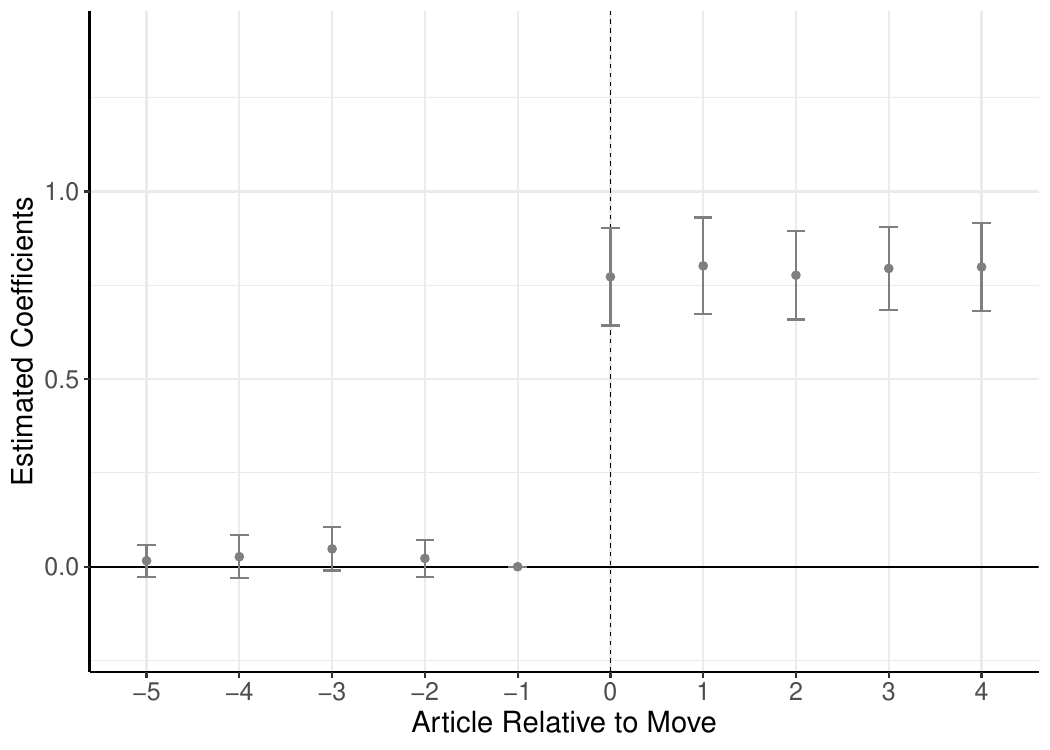}
\par\end{centering}
\begin{centering}
\emph{Panel B: $T=10$}
\par\end{centering}
\begin{centering}
\includegraphics[scale=0.688]{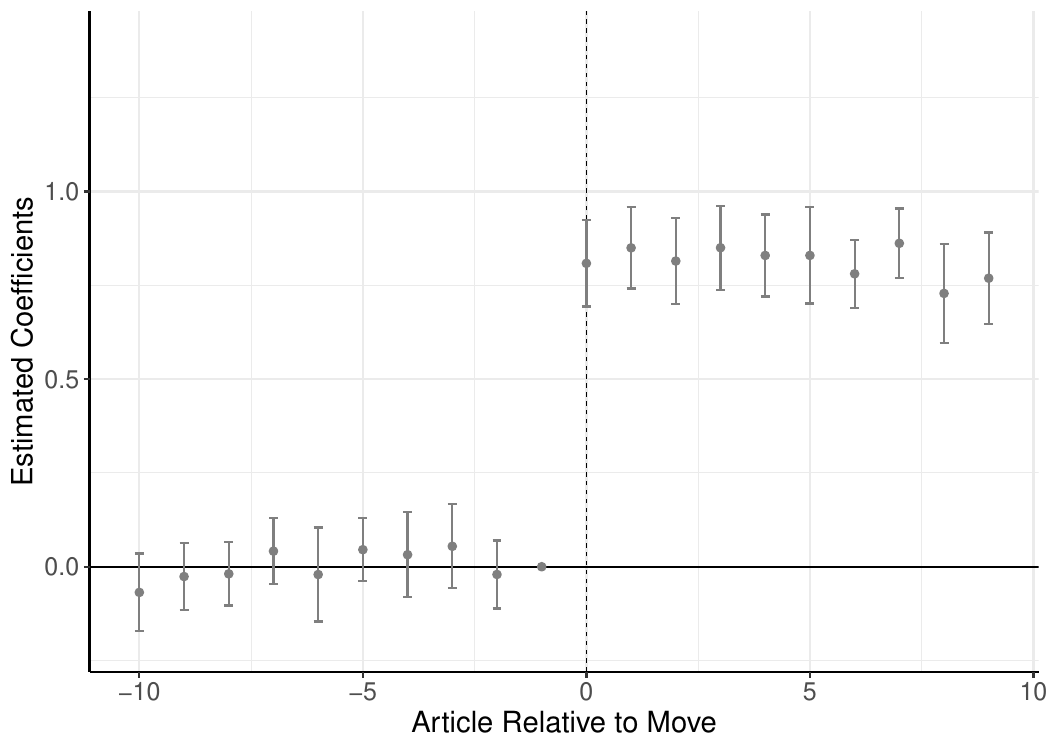}
\par\end{centering}
Notes: Figure plots the estimated event study coefficients $\hat{\rho}_{l}$
for the baseline $T=5$ estimation sample (Panel A) and the $T=10$ horizon robustness
sample (Panel B). Event study coefficients estimate the moving author's
change in slant relative to their slant at $t=-1$, expressed as a
fraction of the full-period author--article-weighted leave-focal-author-out
origin--destination gap after article slant is residualized by publication-month
means estimated once on the full eligible article universe. Regressions weight
observations by the squared leave-out gap and include move-pair
fixed effects.
Standard errors are two-way clustered at the origin- and destination-outlet
levels.
\end{figure}

\begin{figure}[H]
\caption{Exploring Potential Asymmetry\label{fig:asymmetry}}

\begin{centering}
\emph{Panel A: Event Study by Direction of Move}
\par\end{centering}
\begin{centering}
\includegraphics[scale=0.514]{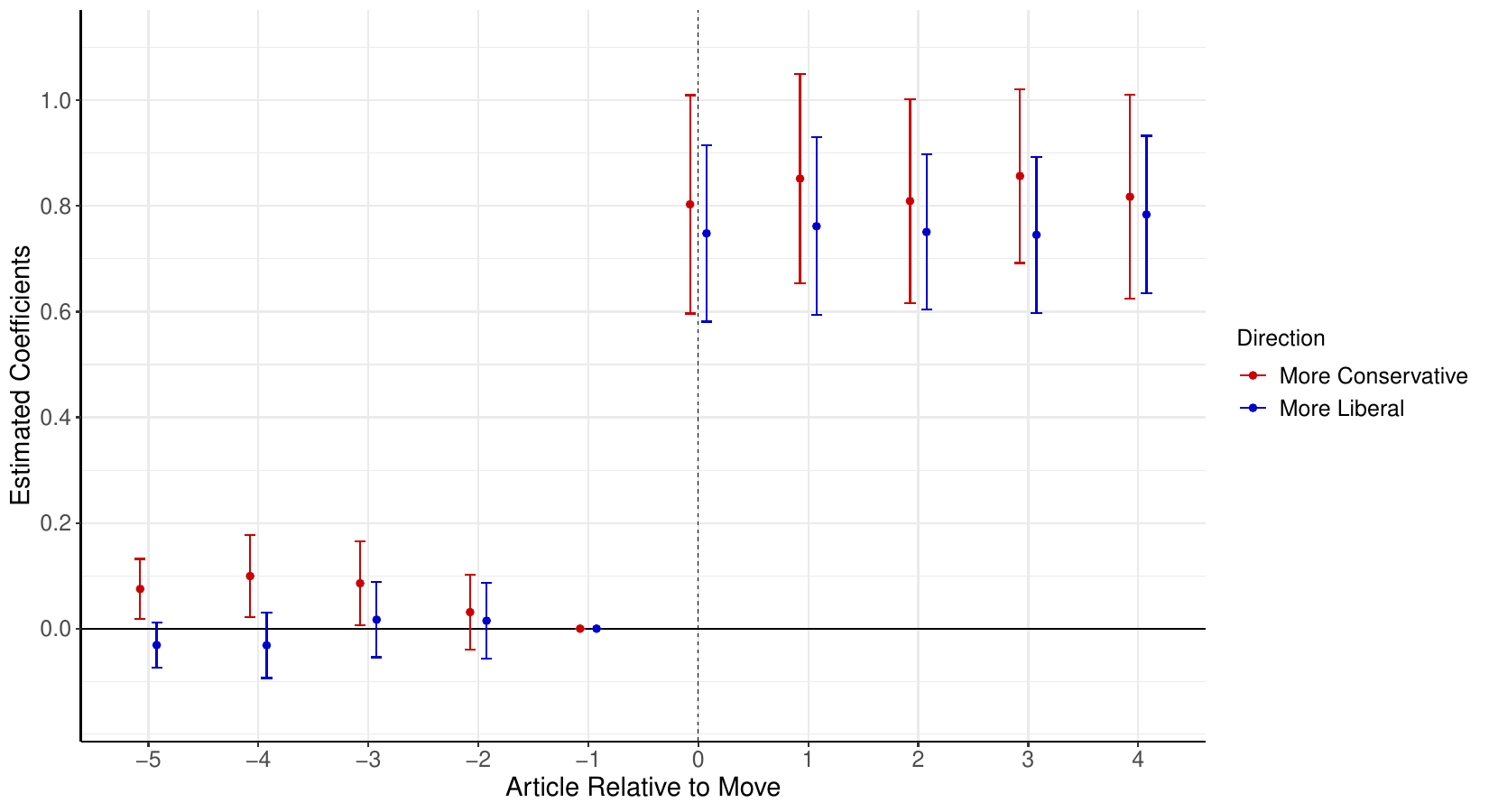}
\par\end{centering}
\begin{centering}
\emph{Panel B: Predicting the Change in Journalist Slant with the
Change in Outlet Slant}
\par\end{centering}
\begin{centering}
\includegraphics[scale=0.514]{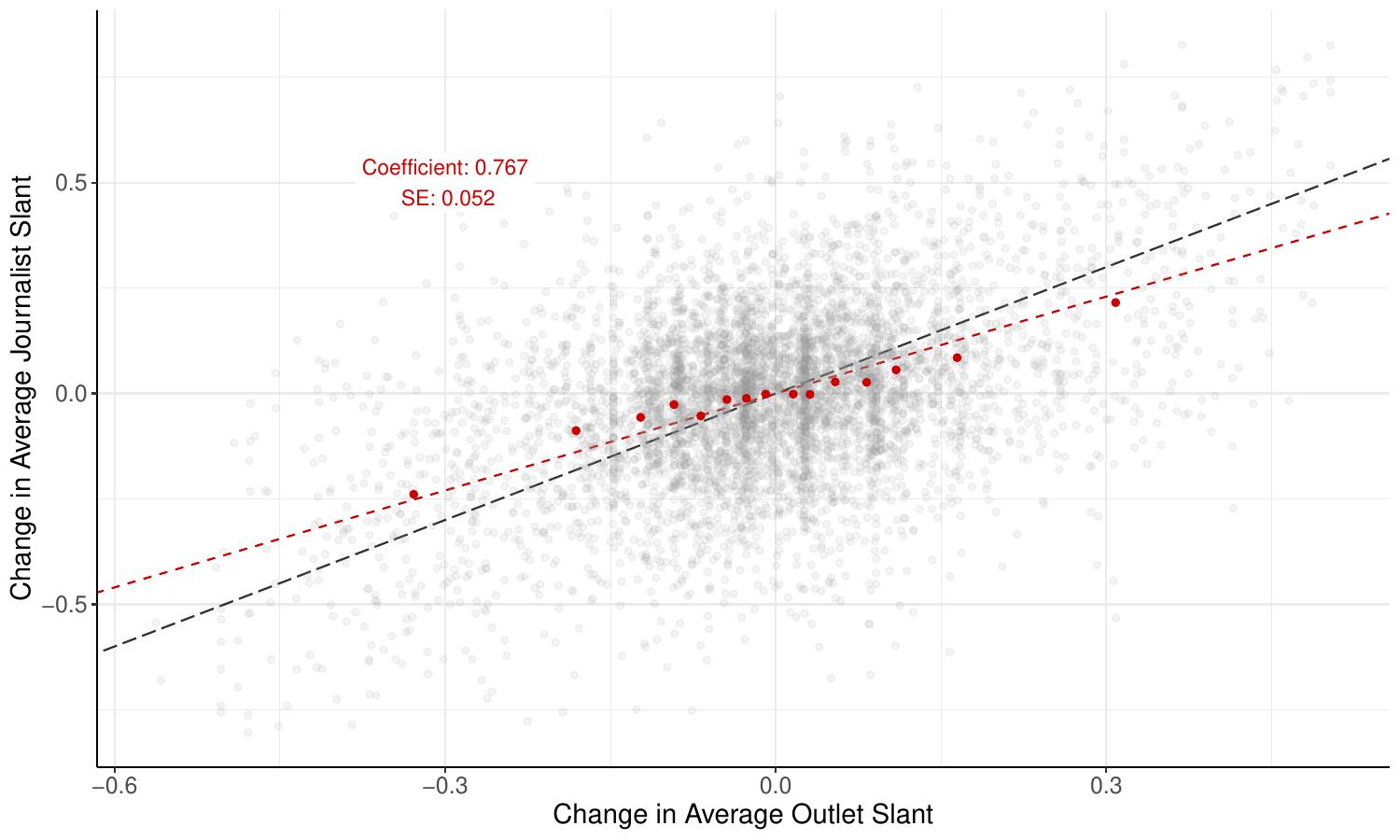}
\par\end{centering}
Notes: Panel A plots the event study coefficients for the baseline $T=5$
estimation sample when the model is estimated separately among the
set of movers with $\bar{x}_{d(q)}^{(-i)}-\bar{x}_{o(q)}^{(-i)}>0$ in red and the set
with $\bar{x}_{d(q)}^{(-i)}-\bar{x}_{o(q)}^{(-i)}<0$ in blue. Panel B plots the mover-level scatter
plot for the same $T=5$ estimation sample with $\bar{x}_{d(q)}^{(-i)}-\bar{x}_{o(q)}^{(-i)}$
on the x-axis and $\frac{1}{T}\sum_{l\geq0}x_{il}-\frac{1}{T}\sum_{l<0}x_{il}$
on the y-axis. In Panel B, the red points show unweighted means within 15
equal-count bins. The fitted line is constrained through the origin and weights observations by
$(\bar{x}_{d(q)}^{(-i)}-\bar{x}_{o(q)}^{(-i)})^{2}$.
Both panels use the common publication-month adjustment; Panel A includes move-pair fixed effects. Panel A standard
errors and the standard error reported for the Panel B slope are two-way
clustered at the origin- and destination-outlet levels.
\end{figure}

\begin{figure}[H]
\caption{Heterogeneity in Journalist AKM Fixed Effects\label{fig:akm_hetero}}

\begin{centering}
\emph{Panel A: Party}
\par\end{centering}
\begin{centering}
\includegraphics[scale=0.4]{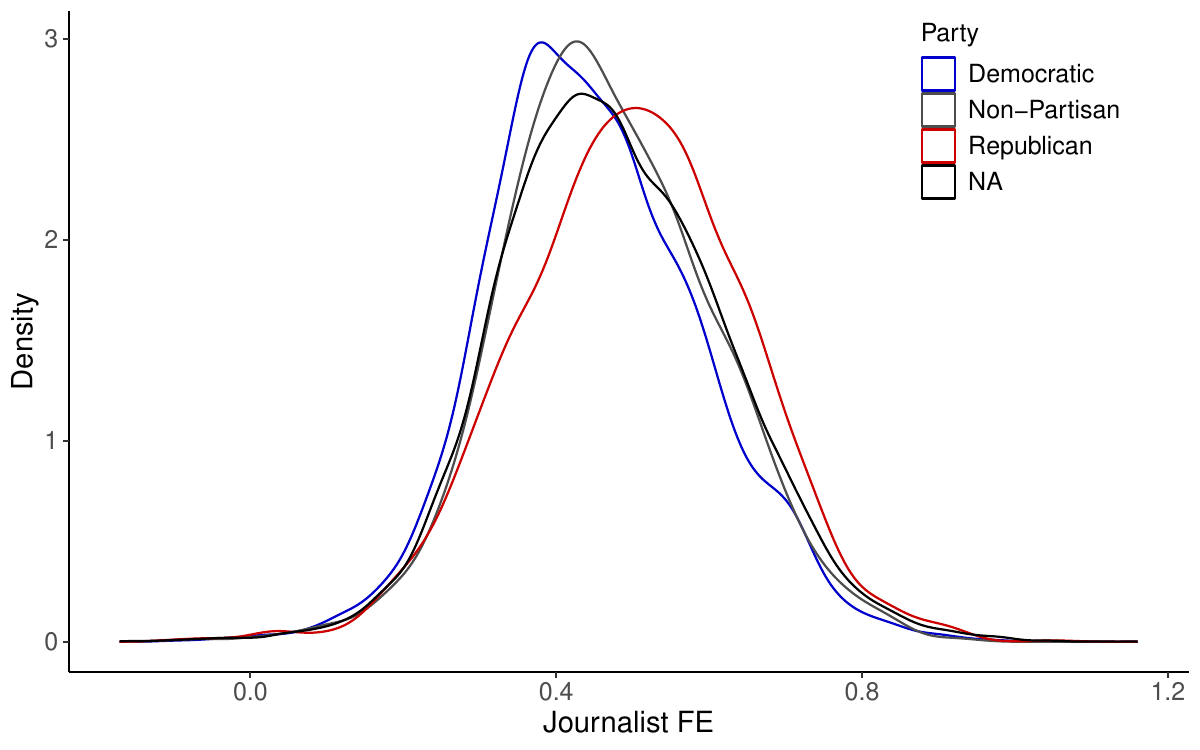}
\par\end{centering}
\begin{centering}
\emph{Panel B: Gender}
\par\end{centering}
\begin{centering}
\includegraphics[scale=0.4]{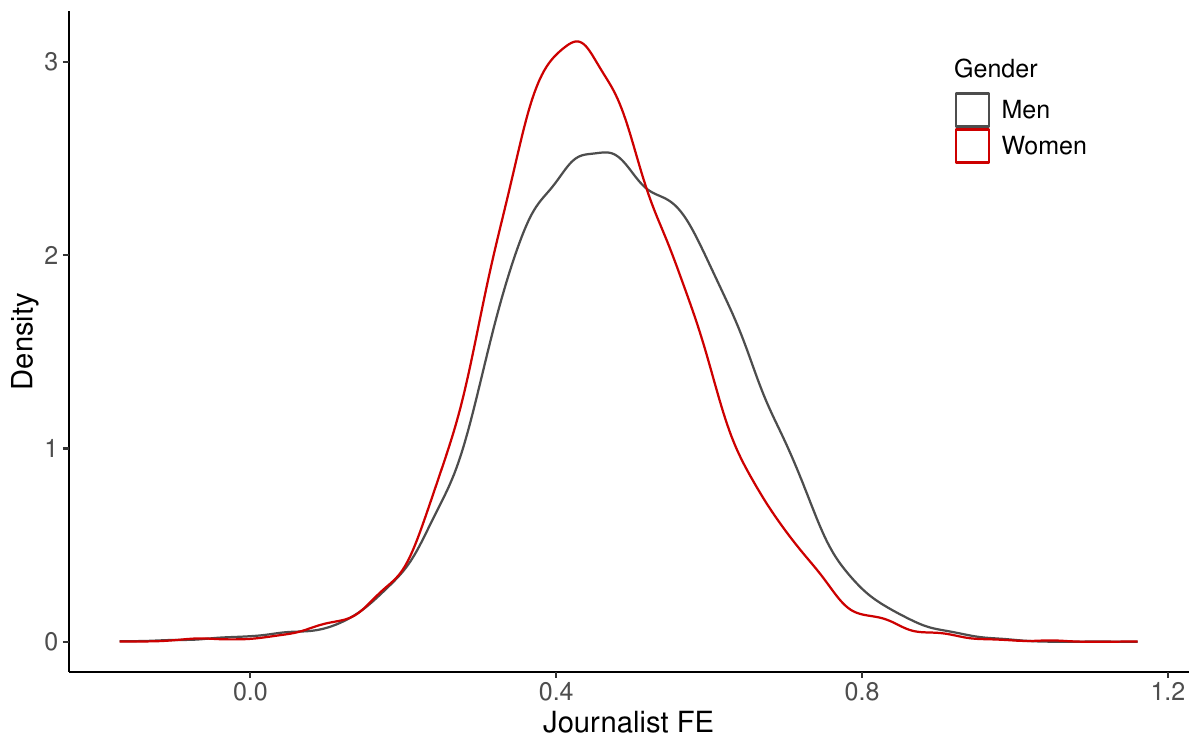}
\par\end{centering}
\begin{centering}
\emph{Panel C: Race}
\par\end{centering}
\begin{centering}
\includegraphics[scale=0.4]{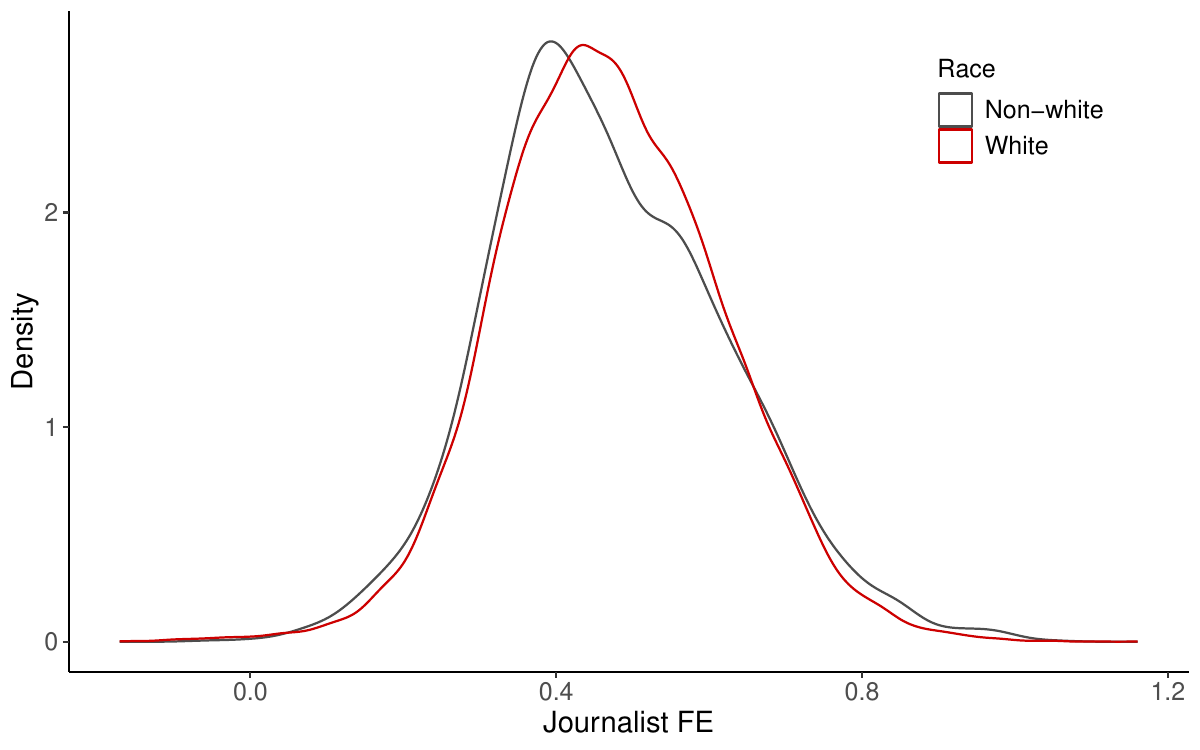}
\par\end{centering}
Notes: Each panel plots the distribution in estimated AKM journalist
fixed effects across various sub-groups of journalists. Panel A examines
heterogeneity across journalists by their partisan affiliation. The NA
group includes journalists whose party
affiliation is not mapped to the Democratic, Republican, or non-partisan
categories by the L2 matching and recoding procedure. Panel B examines
heterogeneity by gender, as predicted by the author's name. Panel C
examines heterogeneity by race, as predicted by the author's name.
\end{figure}

\begin{figure}[H]
\caption{Decomposition Event Study\label{fig:descriptive_event_study}}

\begin{centering}
\emph{Panel A: Below-Median to Above-Median Movers}
\par\end{centering}
\begin{centering}
\includegraphics[scale=0.55]{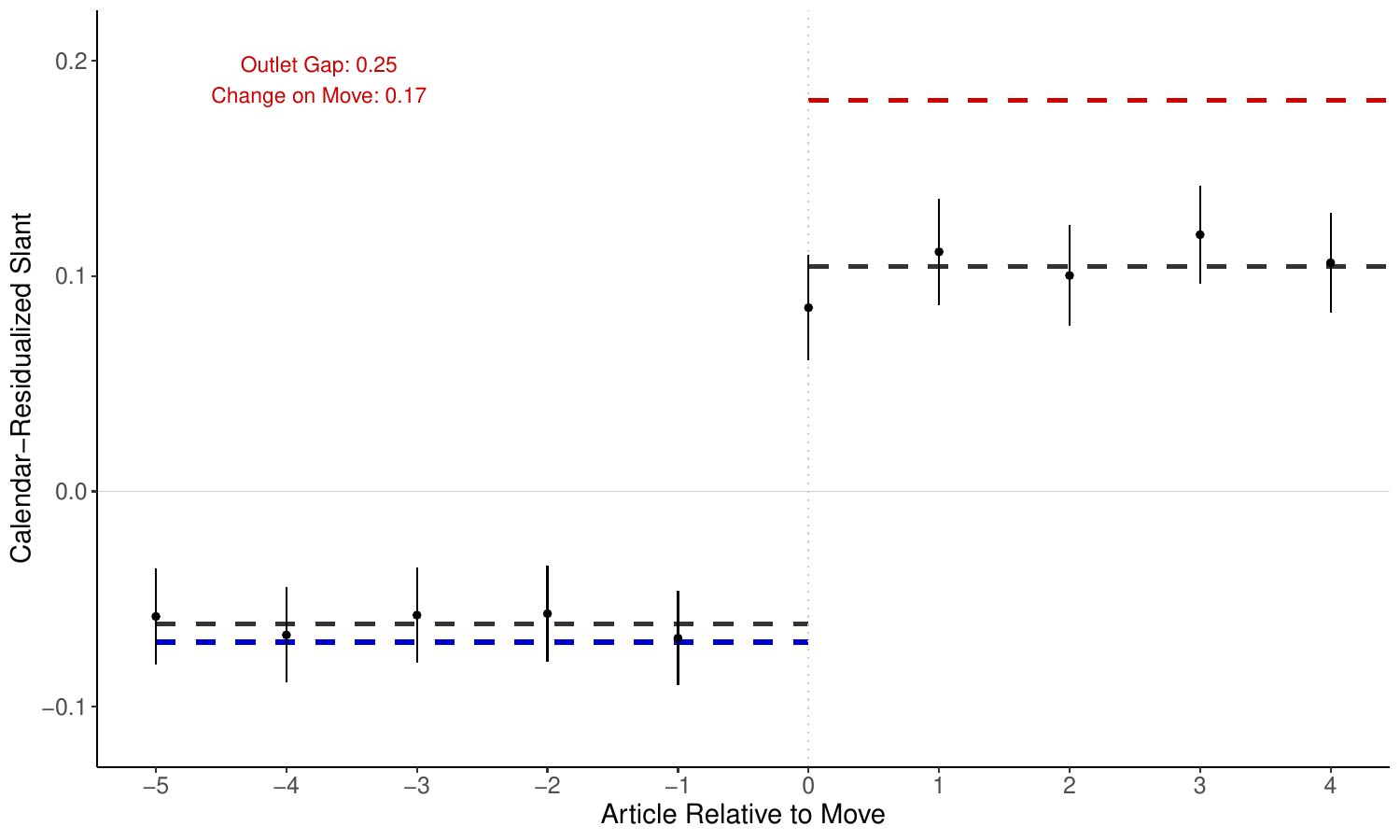}
\par\end{centering}
\begin{centering}
\emph{Panel B: Above-Median to Below-Median Movers}
\par\end{centering}
\begin{centering}
\includegraphics[scale=0.55]{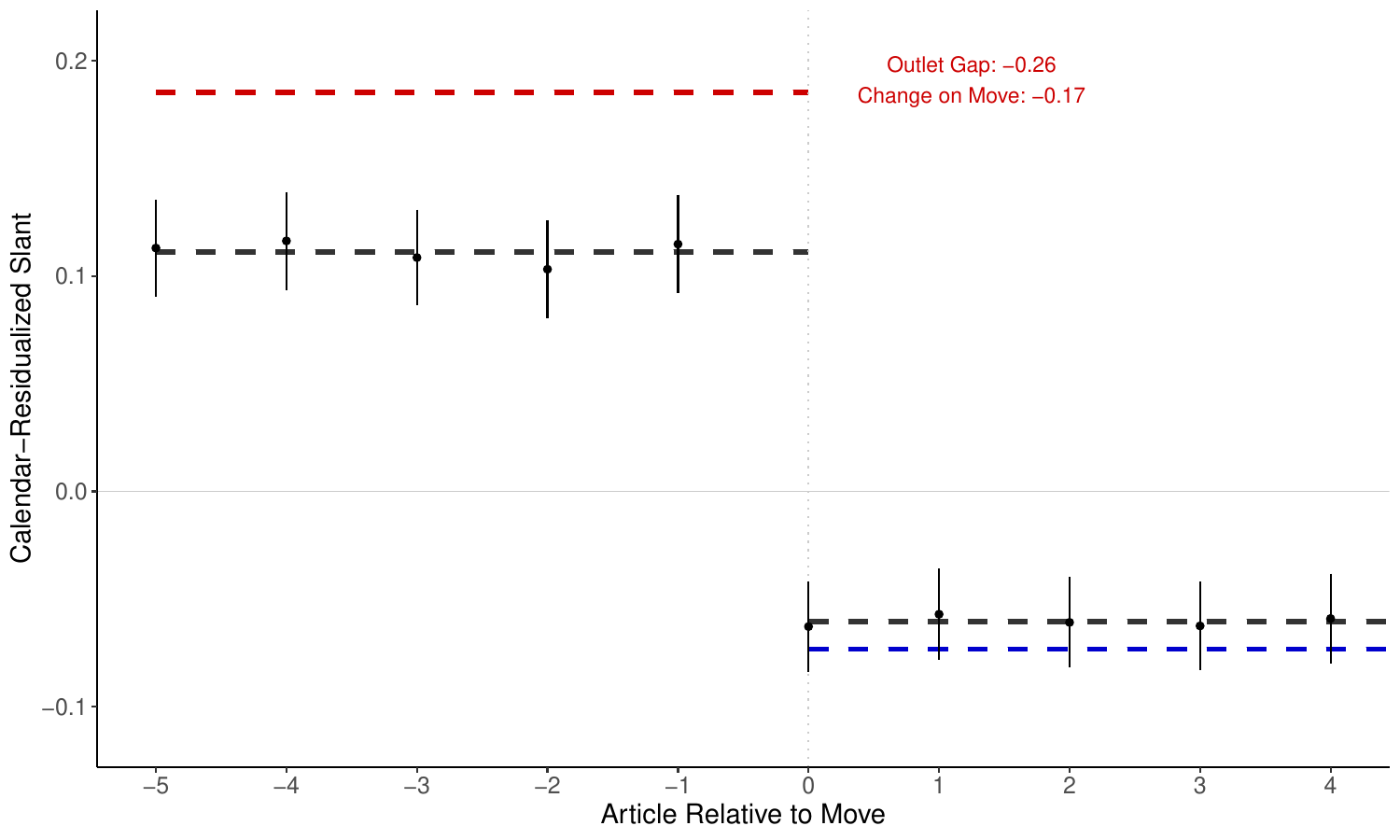}
\par\end{centering}
Notes: Panel A plots the average values of $\bar{x}_{j'}$, $\bar{x}_{j}$,
$x_{ij'}$, and $x_{ij}$ for the set of movers who move from below-median
to above-median outlets using the baseline $T=5$ event window. Outlet classifications and all plotted slant measures
are constructed after removing publication-month means. The dashed blue (red)
line is the average $\bar{x}_{j}$ for below-median (above-median) outlets. The
grey points and associated confidence intervals are the average article slant for the $t$th article relative to the
move for each journalist in the corresponding set. The dashed grey
lines are the average across the grey points before and after the
move respectively. Panel B is the same as Panel A, except using the
set of movers from above-median to below-median outlets.
\end{figure}

\begin{figure}[H]
\caption{Counterfactual Slant Changes\label{fig:cf_contributions}}

\begin{centering}
\emph{Panel A: Outlet sizes fixed}
\par\end{centering}
\begin{centering}
\includegraphics[width=.9\linewidth]{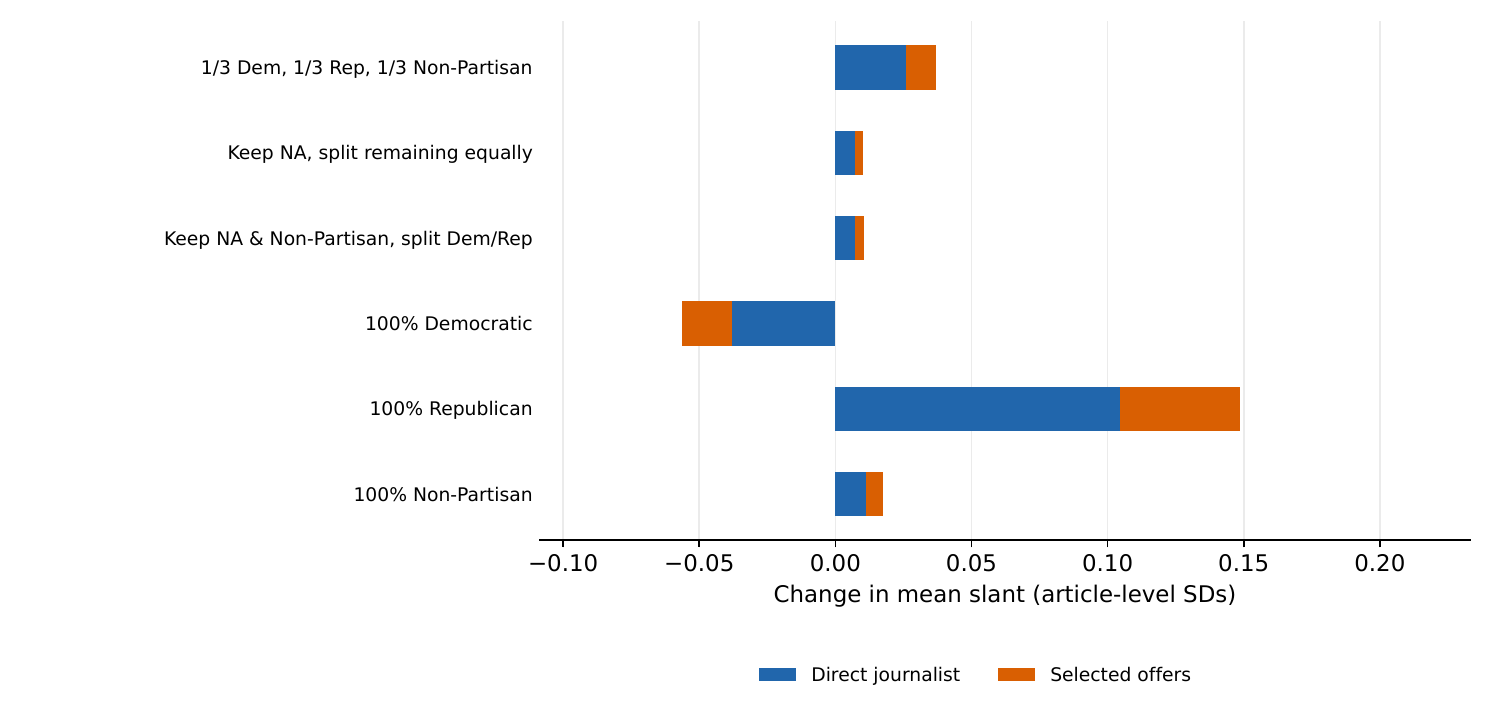}
\par\end{centering}
\begin{centering}
\emph{Panel B: Outlet sizes free}
\par\end{centering}
\begin{centering}
\includegraphics[width=.9\linewidth]{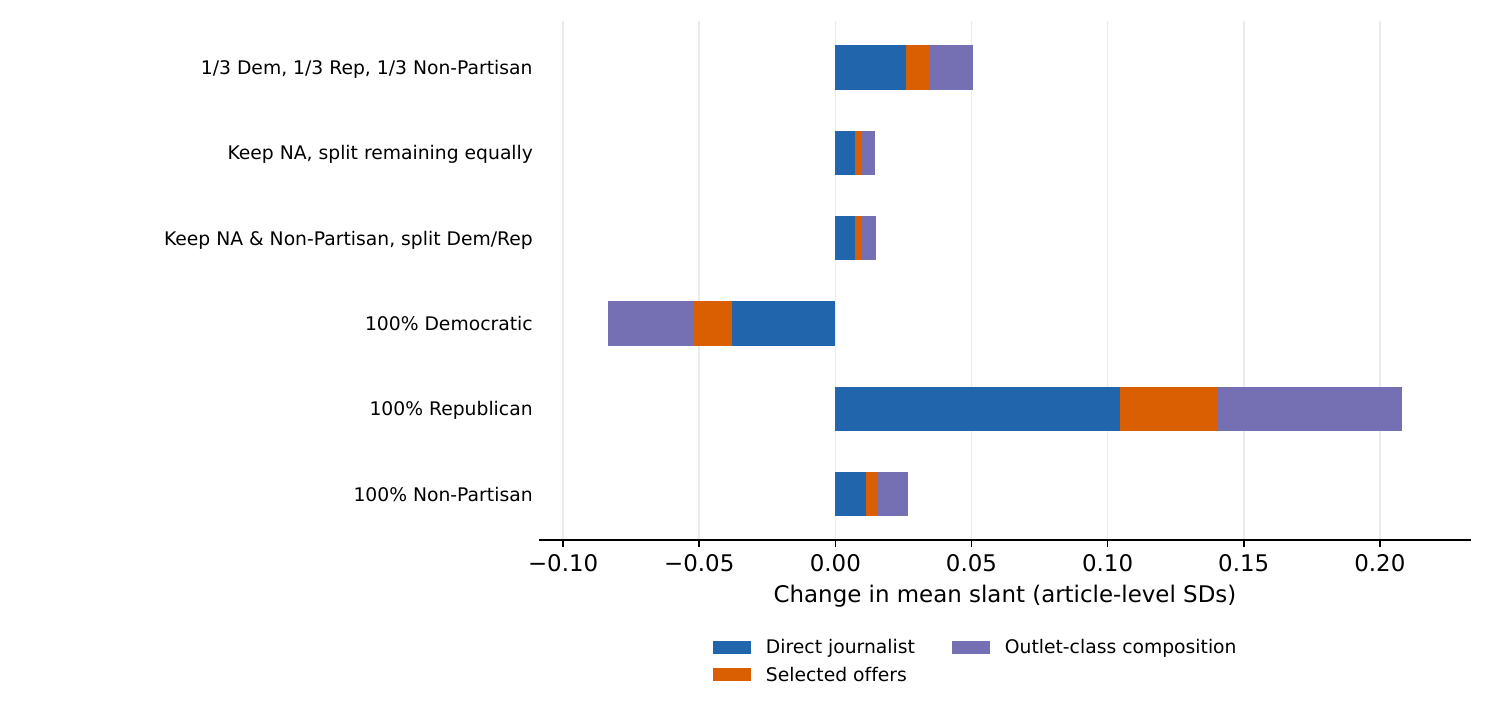}
\par\end{centering}
Notes: Change in mean slant relative to the observed composition, in article-level standard
deviations. Bars decompose each scenario's total change into the direct journalist effect and
selected offers; in Panel B, where class employment shares adjust freely, a third component
captures the change in outlet-class composition.
\end{figure}

%% file: descriptive_tables_filled.tex
{\tiny{}

\begin{table}[H]

{\tiny\caption{Summary Statistics\label{tab:summary}}

}{\tiny\par}

\begin{centering}

{\small{}%

\begin{tabular}{lcccccc}

\hline 

 & \multicolumn{6}{c}{{\small Variable}}\tabularnewline

\cline{2-7}

 & {\small Articles} & {\small Authors} & {\small Slant} & {\small Slant (Single outlet)} & {\small Multiple outlets} & {\small Single outlet}\tabularnewline

\hline 

{\small\emph{Boston Globe}} & {\small217,479} & {\small2,025} & {\small0.32} & {\small0.28} & {\small1,585} & {\small440}\tabularnewline

{\small\emph{Chicago Tribune}} & {\small236,470} & {\small5,607} & {\small0.36} & {\small0.38} & {\small5,300} & {\small307}\tabularnewline

{\small\emph{Los Angeles Times}} & {\small194,002} & {\small2,475} & {\small0.31} & {\small0.23} & {\small2,373} & {\small102}\tabularnewline

{\small\emph{New York Times}} & {\small288,844} & {\small2,990} & {\small0.26} & {\small0.25} & {\small2,440} & {\small550}\tabularnewline

{\small\emph{USA Today}} & {\small107,852} & {\small2,663} & {\small0.40} & {\small0.59} & {\small2,646} & {\small17}\tabularnewline

{\small\emph{Wall Street Journal}} & {\small470,103} & {\small2,927} & {\small0.44} & {\small0.46} & {\small1,751} & {\small1,176}\tabularnewline

{\small\emph{Washington Post}} & {\small357,805} & {\small3,033} & {\small0.35} & {\small0.35} & {\small2,790} & {\small243}\tabularnewline

\hline 

{\small All Outlets} & {\small9,005,189} & {\small33,571} & {\small0.47} & {\small0.49} & {\small21,323} & {\small12,248}\tabularnewline

\hline 

\end{tabular}}{\small\par}

\par\end{centering}

{\small Notes: Table shows summary statistics for our cleaned sample of ProQuest data. Single- and multiple-outlet counts classify publication histories in this broad sample; they do not impose the valid-move definition used in the mobility analyses.}{\small\par}

\end{table}

}{\tiny\par}

%% file: akm_kss_table_filled.tex
{\tiny{}

\begin{table}[H]

{\tiny\caption{AKM Decompositions\label{tab:akm_kss_table}}

}{\tiny\par}

\begin{centering}

{\small{}%

\begin{tabular}{lcccccccccc}

\hline 

 & {\small (1)} & {\small (2)} & {\small (3)} & {\small (4)} & {\small (5)} & {\small (6)} & {\small (7)} & {\small (8)} & {\small (9)} & {\small (10)}\tabularnewline

\hline 

{\small Sample:} & {\small All} & {\small Dropping} & {\small Dropping} & {\small Max 2} & {\small Drop 1 Mo.} & {\small Top 20} & {\small$T=5$} & {\small$T=5$} & {\small$T=10$} & {\small Parent}\tabularnewline

 &  & {\small Stayers (Full)} & {\small Stayers (Large)} & {\small Outlets} & {\small Moves} & {\small Outlets} & {\small Full} & {\small Large} & {\small Full} & {\small Company}\tabularnewline

\hline 

 &  &  &  &  &  &  &  &  &  & \tabularnewline

\multicolumn{11}{c}{{\small\emph{Panel A: Plug-in Estimator}}}\tabularnewline

\hline 

\multicolumn{11}{c}{{\small Share of Variance}{\scriptsize , $x_{ijt}$}}\tabularnewline

\hline 

{\small Journalist, $\text{var}(\gamma_{i})$} & {\small0.194} & {\small0.167} & {\small0.185} & {\small0.202} & {\small0.194} & {\small0.266} & {\small0.236} & {\small0.273} & {\small0.216} & {\small0.352}\tabularnewline

{\small Outlet, $\text{var}(\alpha_{j})$} & {\small0.217} & {\small0.224} & {\small0.263} & {\small0.215} & {\small0.222} & {\small0.094} & {\small0.186} & {\small0.241} & {\small0.209} & {\small0.082}\tabularnewline

{\small Sorting, $2\times\text{cov}(\gamma_{i},\alpha_{j})$} & {\small0.038} & {\small0.040} & {\small-0.001} & {\small0.030} & {\small0.031} & {\small-0.001} & {\small0.057} & {\small-0.020} & {\small0.028} & {\small0.033}\tabularnewline

\hline 

 & \multicolumn{10}{c}{{\small Correlation}}\tabularnewline

\hline 

{\small Sorting, $\text{cor}(\gamma_{i},\alpha_{j})$} & {\small0.092} & {\small0.103} & {\small-0.002} & {\small0.073} & {\small0.074} & {\small-0.004} & {\small0.136} & {\small-0.039} & {\small0.067} & {\small0.096}\tabularnewline

\hline 

 &  &  &  &  &  &  &  &  &  & \tabularnewline

\multicolumn{11}{c}{{\small\emph{Panel B: Leave-Out Estimator (\citealt{Kline2020LeaveOutVariance})}}}\tabularnewline

\hline 

\multicolumn{11}{c}{{\small Share of Variance}{\scriptsize , $x_{ijt}$}}\tabularnewline

\hline 

{\small Journalist, $\text{var}(\gamma_{i})$} & {\small0.155} & {\small0.085} & {\small0.080} & {\small0.155} & {\small0.152} & {\small0.245} & {\small0.143} & {\small0.148} & {\small0.129} & {\small0.325}\tabularnewline

{\small Outlet, $\text{var}(\alpha_{j})$} & {\small0.192} & {\small0.201} & {\small0.230} & {\small0.182} & {\small0.194} & {\small0.081} & {\small0.164} & {\small0.188} & {\small0.183} & {\small0.069}\tabularnewline

{\small Sorting, $2\times\text{cov}(\gamma_{i},\alpha_{j})$} & {\small0.085} & {\small0.075} & {\small0.072} & {\small0.095} & {\small0.086} & {\small0.024} & {\small0.084} & {\small0.040} & {\small0.063} & {\small0.059}\tabularnewline

\hline 

\multicolumn{11}{c}{{\small Correlation}}\tabularnewline

\hline 

{\small Sorting, $\text{cor}(\gamma_{i},\alpha_{j})$} & {\small0.245} & {\small0.286} & {\small0.266} & {\small0.284} & {\small0.249} & {\small0.085} & {\small0.273} & {\small0.120} & {\small0.205} & {\small0.197}\tabularnewline

\hline 

\end{tabular}}{\small\par}

\par\end{centering}

{\small Notes: Table reports AKM variance decomposition exercises using the plug-in estimator (Panel A) and the \citet{Kline2020LeaveOutVariance} leave-out estimator (Panel B) with various samples. Both panels use the leave-out connected set of outlets. All columns use raw article slant, the raw outcome-variance denominator, and publication-month fixed effects. Column (1) uses the full sample. Column (2) drops journalists observed at only one retained outlet. Column (3) retains only origin and destination spells around moves whose absolute calendar-residualized leave-focal-author-out outlet gap exceeds 0.1, counting each author--article observation once. Column (4) drops journalists that appear in more than two outlets. Column (5) restricts to tenures of longer than one month in duration. Column (6) restricts to the top 20 outlets according to weekend circulation. Column (7) uses the $T=5$ event study sample. Column (8) is the same as Column (7) except that it restricts to moves whose absolute calendar-residualized leave-focal-author-out outlet gap exceeds 0.1. Column (9) uses the $T=10$ event study sample. Column (10) treats all outlets owned by a single ``Parent Company'' as the same outlet and restricts attention to outlets for which we observe the parent company.}{\small\par}

\end{table}

}{\tiny\par}

%% file: outlet_decomp_table_filled.tex
{\tiny{}

\begin{table}[H]

{\tiny\caption{Outlet-level Journalist Effects\label{tab:outlet_decomp}}

}{\tiny\par}

\begin{centering}

{\small{}%

\setlength{\tabcolsep}{1.8pt}%

\begin{tabular}{lcccccccc}

\hline 

 & {\small (1)} & {\small (2)} & {\small (3)} & {\small (4)} & {\small (5)} & {\small (6)} & {\small (7)} & {\small (8)}\tabularnewline

\hline 

{\small Sample:} & {\small All} & {\small 12+} & {\small Max 1} & {\small Drop 1 Mo.} & {\small Top 20} & $T=5$ & $T=10$ & {\small Parent}\tabularnewline

 &  & {\small Months} & {\small Move} & {\small Moves} & {\small Outlets} & {\small Sample} & {\small Sample} & {\small Company}\tabularnewline

\hline 

 & \multicolumn{8}{c}{{\small Share of Variance, $\bar{x}_{j}$}}\tabularnewline

\hline 

{\small Cross-fit, article weights} & {\small0.143} & {\small0.126} & {\small0.201} & {\small0.144} & {\small0.008} & {\small0.165} & {\small0.103} & {\small0.247}\tabularnewline

{\small \textit{95\% CI}} & {\small\textit{[0.113,}} & {\small\textit{[0.089,}} & {\small\textit{[0.125,}} & {\small\textit{[0.084,}} & {\small\textit{[-0.146,}} & {\small\textit{[0.122,}} & {\small\textit{[0.022,}} & {\small\textit{[0.160,}}\tabularnewline

 & {\small\textit{0.189]}} & {\small\textit{0.178]}} & {\small\textit{0.293]}} & {\small\textit{0.208]}} & {\small\textit{0.175]}} & {\small\textit{0.224]}} & {\small\textit{0.201]}} & {\small\textit{0.345]}}\tabularnewline

{\small Cross-fit, equal outlet weights} & {\small0.157} & {\small0.127} & {\small0.176} & {\small0.155} & {\small-0.077} & {\small0.177} & {\small0.056} & {\small0.350}\tabularnewline

{\small \textit{95\% CI}} & {\small\textit{[0.120,}} & {\small\textit{[0.084,}} & {\small\textit{[0.104,}} & {\small\textit{[0.100,}} & {\small\textit{[-0.293,}} & {\small\textit{[0.124,}} & {\small\textit{[-0.031,}} & {\small\textit{[0.256,}}\tabularnewline

 & {\small\textit{0.208]}} & {\small\textit{0.188]}} & {\small\textit{0.255]}} & {\small\textit{0.219]}} & {\small\textit{0.156]}} & {\small\textit{0.244]}} & {\small\textit{0.150]}} & {\small\textit{0.461]}}\tabularnewline

{\small Plug-in, cross-fit support} & {\small0.180} & {\small0.161} & {\small0.193} & {\small0.174} & {\small0.089} & {\small0.191} & {\small0.171} & {\small0.264}\tabularnewline

{\small Plug-in, natural support} & {\small0.155} & {\small0.151} & {\small0.117} & {\small0.117} & {\small0.044} & {\small0.168} & {\small0.094} & {\small0.334}\tabularnewline

\hline 

 & \multicolumn{8}{c}{{\small Role-Equality Diagnostics}}\tabularnewline

\hline 

{\small P-value, equal role effects} & {\small0.209} & {\small0.055} & {\small0.209} & {\small0.026} & {\small0.286} & {\small0.298} & {\small0.069} & {\small0.305}\tabularnewline

{\small Agreement ratio, $2C/(V^{o}+V^{d})$} & {\small0.956} & {\small0.897} & {\small0.891} & {\small0.798} & {\small0.209} & {\small0.961} & {\small0.643} & {\small0.977}\tabularnewline

\hline 

 & \multicolumn{8}{c}{{\small Sample Size}}\tabularnewline

\hline 

{\small Moves} & {\small4,700} & {\small3,681} & {\small821} & {\small1,798} & {\small573} & {\small4,700} & {\small1,256} & {\small1,091}\tabularnewline

{\small Journalists} & {\small2,633} & {\small1,977} & {\small821} & {\small1,426} & {\small433} & {\small2,633} & {\small956} & {\small832}\tabularnewline

{\small Outlets, cross-fit support} & {\small154} & {\small134} & {\small48} & {\small89} & {\small15} & {\small154} & {\small65} & {\small20}\tabularnewline

{\small Outlets, natural plug-in support} & {\small238} & {\small233} & {\small194} & {\small220} & {\small17} & {\small238} & {\small205} & {\small32}\tabularnewline

\hline 

\end{tabular}}{\small\par}

\par\end{centering}

Notes: Table reports the share of variance in average outlet slant attributed to outlet-level journalist effects. Articles are residualized relative to calendar-month fixed effects before calculating both journalist-spell means and outlet-level means. The baseline averages symmetric cross-half covariances from five repeated journalist-disjoint splits and uses outlets identified in every split half. Except in the equal-outlet-weight row, outlets are weighted by their full-sample number of articles. The plug-in rows use, respectively, the cross-fit support and the larger connected support available without split completeness. Confidence intervals use a journalist-cluster weighted bootstrap. The role-equality rows compare the model's origin and destination effects; p-values use a restricted-null journalist-cluster wild bootstrap, and Appendix \ref{sec:outlet_decomp_inference} defines the test and agreement ratio. Move and journalist counts and the cross-fit outlet count use the cross-fit support; the final row reports the natural plug-in support. In Columns (1)--(5) and (8), moves pair strictly forward adjacent article spells with at least \EventStudyMinSpellArticlesWord\ articles, after the common preliminary \InitialMinSpellArticlesWord-article screen and reconstruction; gaps longer than \EventStudyGapCapMonths\ months are not bridged. All columns exclude journalists above \EventStudyMaxMovesCap\ valid moves in the complete cleaned outlet history. The preferred baseline uses full-spell means. Columns (1)--(8) report, respectively, the full sample; journalists observed in at least 12 months; journalists with at most one retained move; spells longer than one month; the top 20 outlets by weekend circulation; the $T=5$ and $T=10$ event-study samples; and parent-company groups.

\end{table}

}{\tiny\par}

%% file: modeling_selection_key_table.tex
\begin{tabular}{lcc}
\toprule
 & Estimate & SE \\
\midrule
Journalist contract weight, $\beta/(\beta+1)$ & 0.403 & 0.019 \\
Slant preference, $\beta$ & 0.674 & 0.053 \\
Mean ideal point: Democratic & 0.537 & 0.027 \\
Mean ideal point: Republican & 0.636 & 0.028 \\
Mean ideal point: Non-Partisan & 0.571 & 0.028 \\
Mean ideal point: Unknown party & 0.562 & 0.017 \\
\bottomrule
\end{tabular}

%% file: appendix_figures.tex
\begin{figure}[H]
\caption{Politician Op-Eds\label{fig:opeds}}
\smallskip{}

\smallskip{}

\begin{centering}
\emph{Panel A: The New York Times}
\par\end{centering}
\begin{centering}
\includegraphics[scale=0.4]{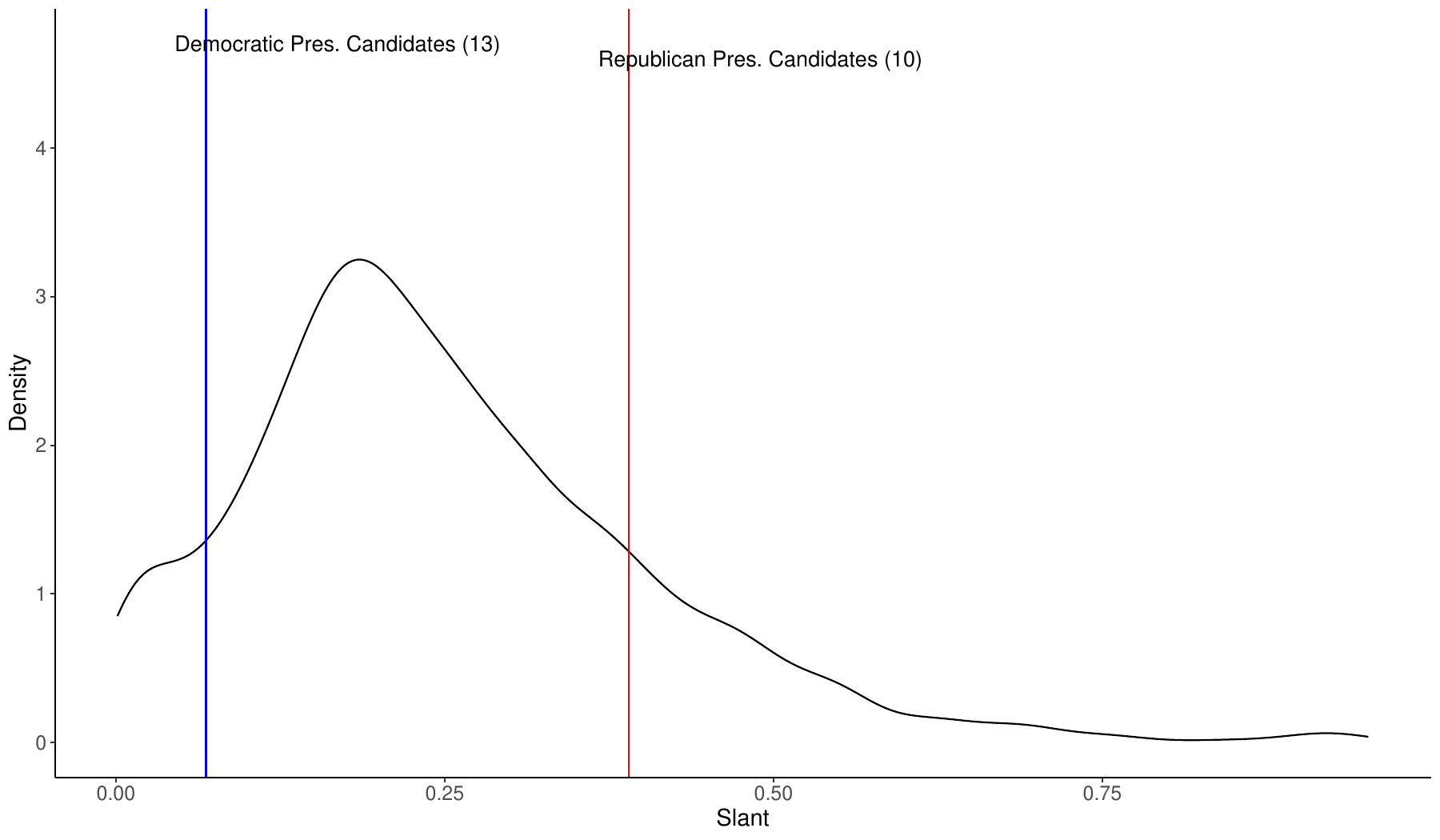}\smallskip{}
\par\end{centering}
\begin{centering}
\emph{Panel B: The Wall Street Journal}
\par\end{centering}
\begin{centering}
\includegraphics[scale=0.4]{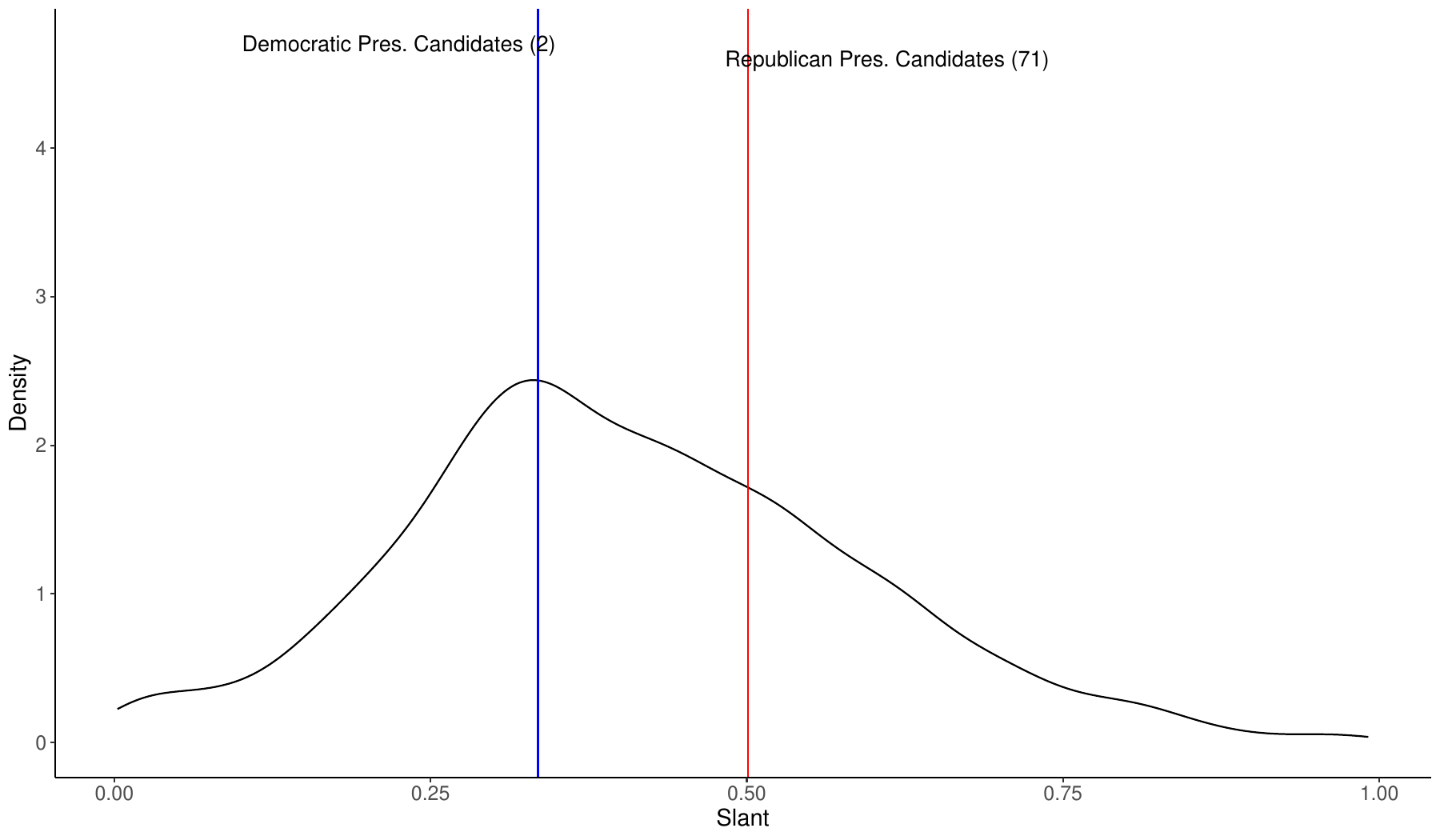}
\par\end{centering}
\smallskip{}

Notes: Panel A plots the distribution, across authors, of the average
slant of each author's article as measured by our RoBERTa model of
slant for \emph{The New York Times} for our main analysis dataset.
The vertical lines display the average slant for major Democratic
presidential candidates (blue) and major Republican presidential candidates
(red) in 2016 across articles written by those authors in \emph{The
New York Times}. Panel B is the same as Panel A, except for \emph{The
Wall Street Journal. }The number of articles in each set is indicated
in parentheses. The politician articles include articles outside the
main analysis dataset.
\end{figure}

\begin{figure}[H]
\caption{Comparison to Other Measures of Slant or Consumer Partisanship\label{fig:gs_bk_comparison}}
\smallskip{}

\begin{centering}
\emph{Panel A: Comparison to Berkman-Klein}
\par\end{centering}
\begin{centering}
\includegraphics[scale=0.34]{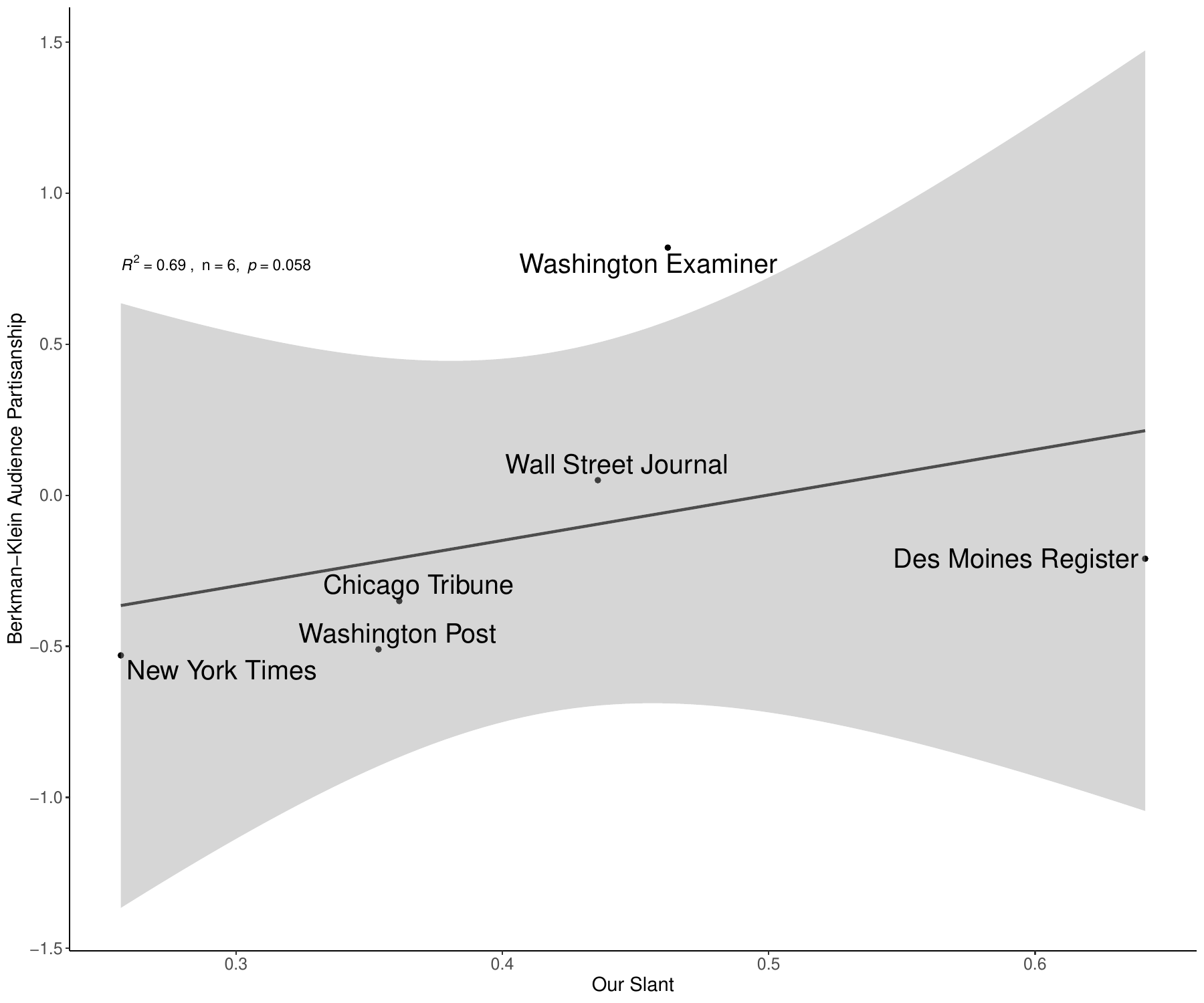}\smallskip{}
\par\end{centering}
\begin{centering}
\emph{Panel B: Comparison to Gentzkow and Shapiro (2010)}
\par\end{centering}
\begin{centering}
\includegraphics[scale=0.34]{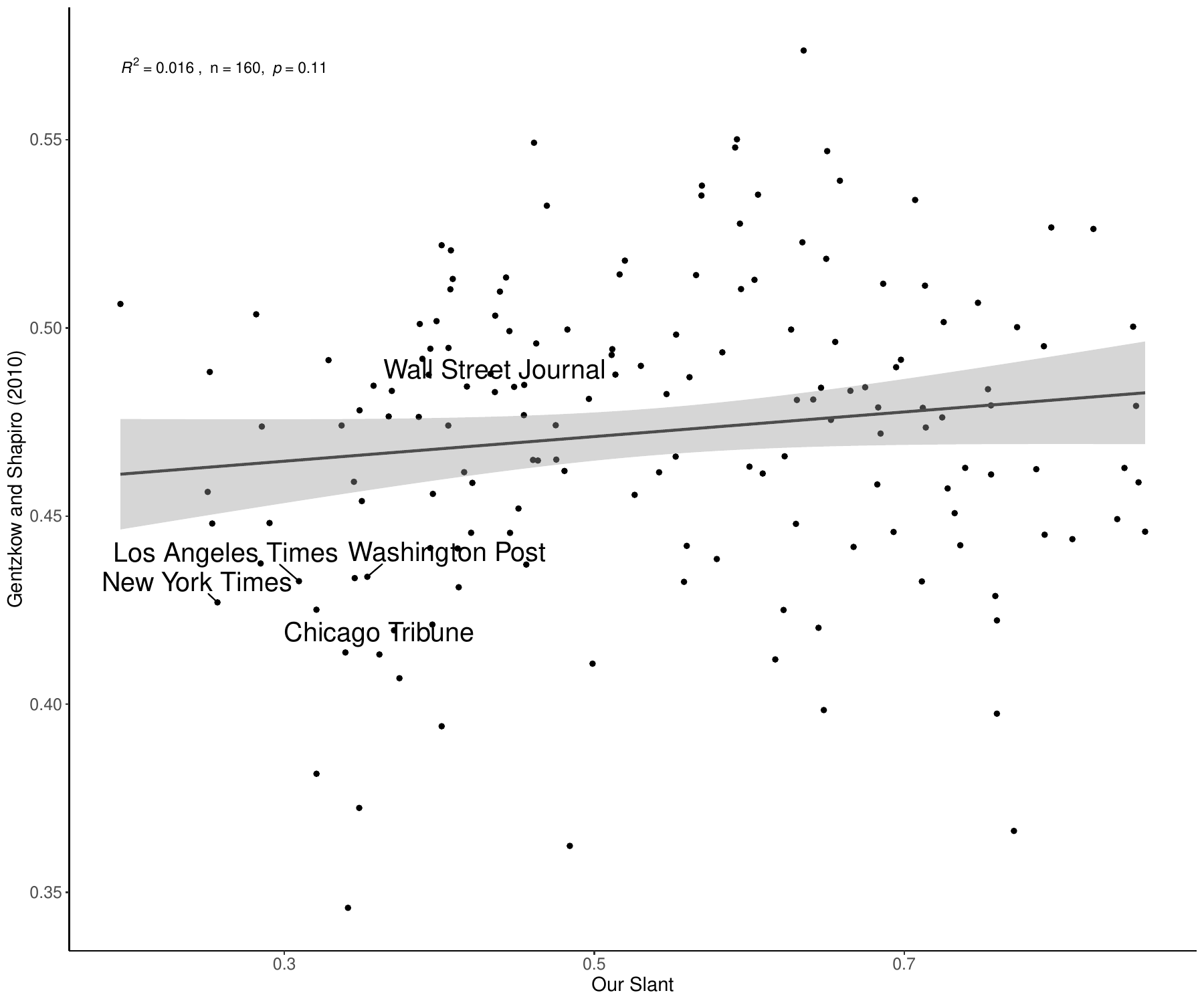}
\par\end{centering}
Notes: Panel A plots the comparison of our measure of slant (x-axis)
to a measure of audience partisanship from the Berkman-Klein study
on the 2016 election (y-axis). Panel B plots the outlet-level slant
as measured in \citet{Gentzkow2010WhatDrivesMediaBias} on the x-axis, and the
average slant across articles as measured by our RoBERTa model of
slant. Each point represents a single outlet. We match outlets between
the ProQuest data and the \citet{Gentzkow2010WhatDrivesMediaBias} data using a
fuzzy string matching algorithm. We exclude one outlier with a negative
slant score in the \citet{Gentzkow2010WhatDrivesMediaBias} data.
\end{figure}

\begin{figure}[H]
\caption{Comparing \citet{Gentzkow2010WhatDrivesMediaBias} to Berkman-Klein Partisanship\label{fig:gs_vs_bk}}
\smallskip{}

\begin{centering}
\includegraphics[scale=0.6]{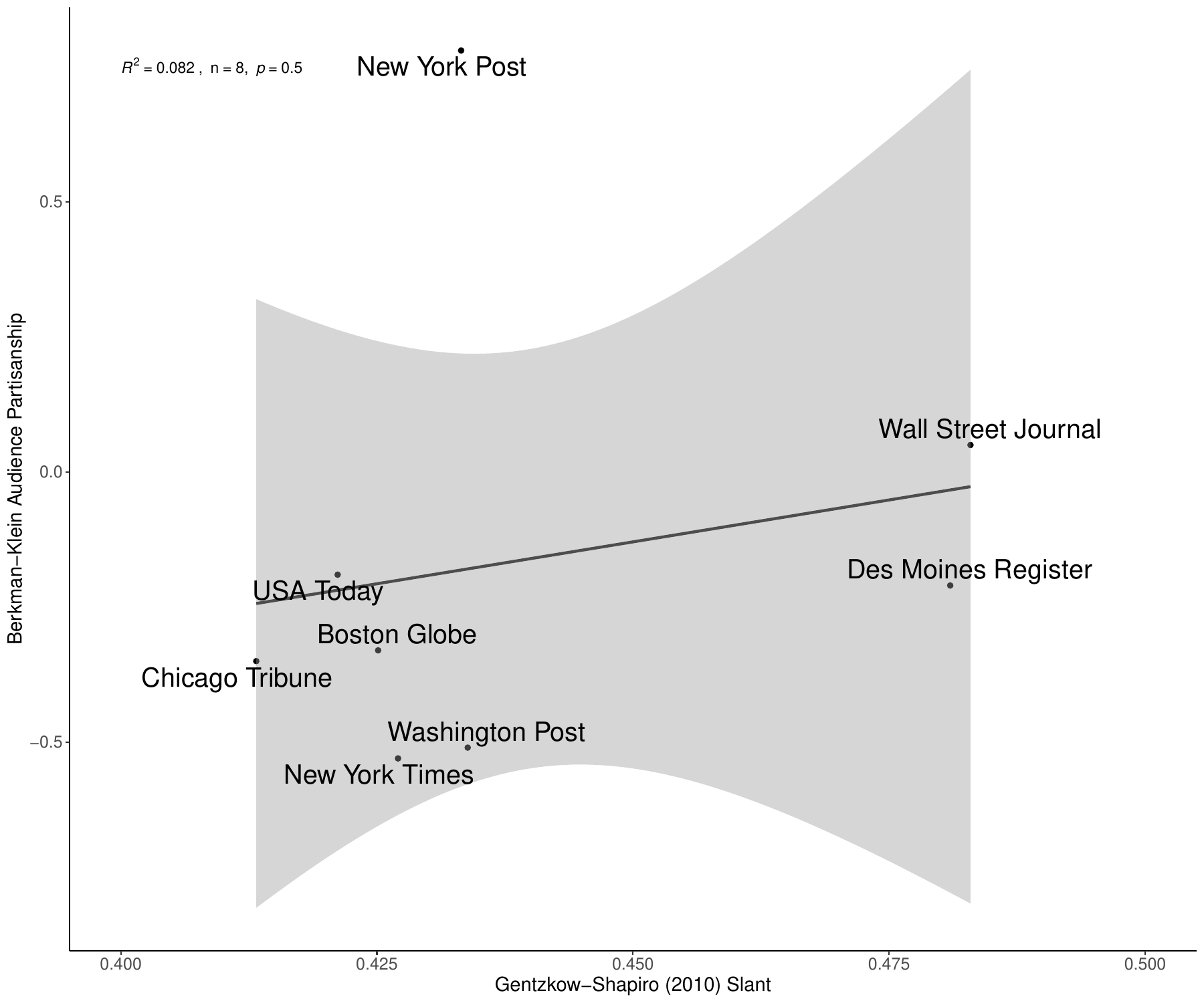}\smallskip{}
\par\end{centering}
Notes: Figure plots the outlet-level slant as measured in \citet{Gentzkow2010WhatDrivesMediaBias} on the x-axis and a measure of audience partisanship
from the Berkman-Klein 2016 election study on the y-axis. Each point
represents a single outlet. We exclude one outlier with a negative
slant score in the \citet{Gentzkow2010WhatDrivesMediaBias} data.
\end{figure}

\begin{figure}[H]
\caption{Authors at \emph{The New York Times} and \emph{The Wall Street Journal
}--- 2013 vs. 2018\emph{\label{fig:nyt_vs_wsj_13_18}}}
\smallskip{}

\begin{centering}
\emph{Panel A: New York Times}
\par\end{centering}
\begin{centering}
\includegraphics[scale=0.7]{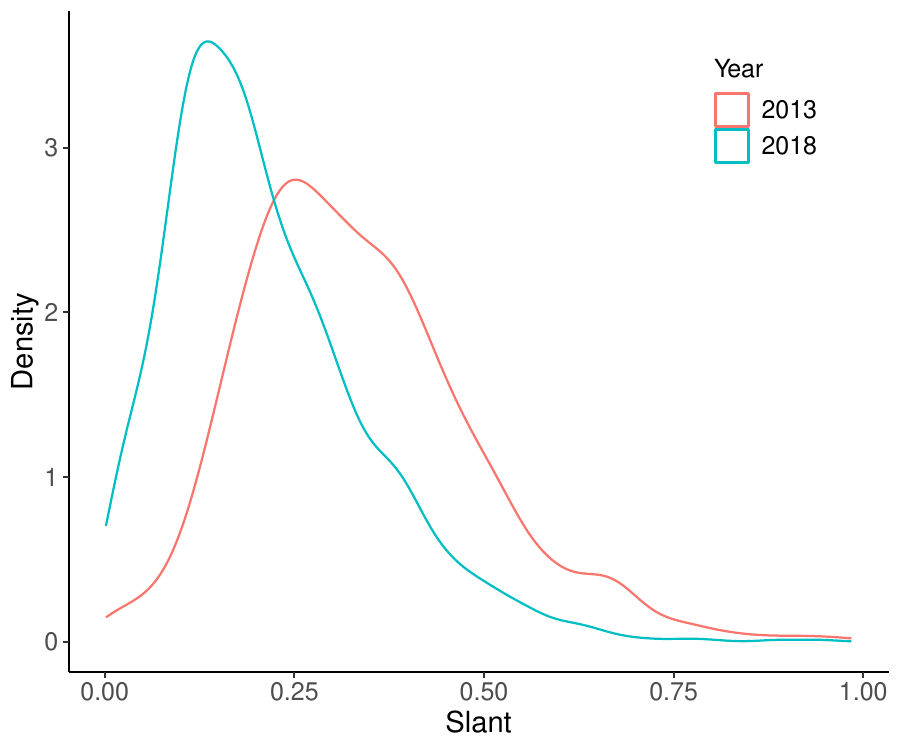}
\par\end{centering}
\smallskip{}

\begin{centering}
\emph{Panel B: Wall Street Journal}
\par\end{centering}
\begin{centering}
\includegraphics[scale=0.7]{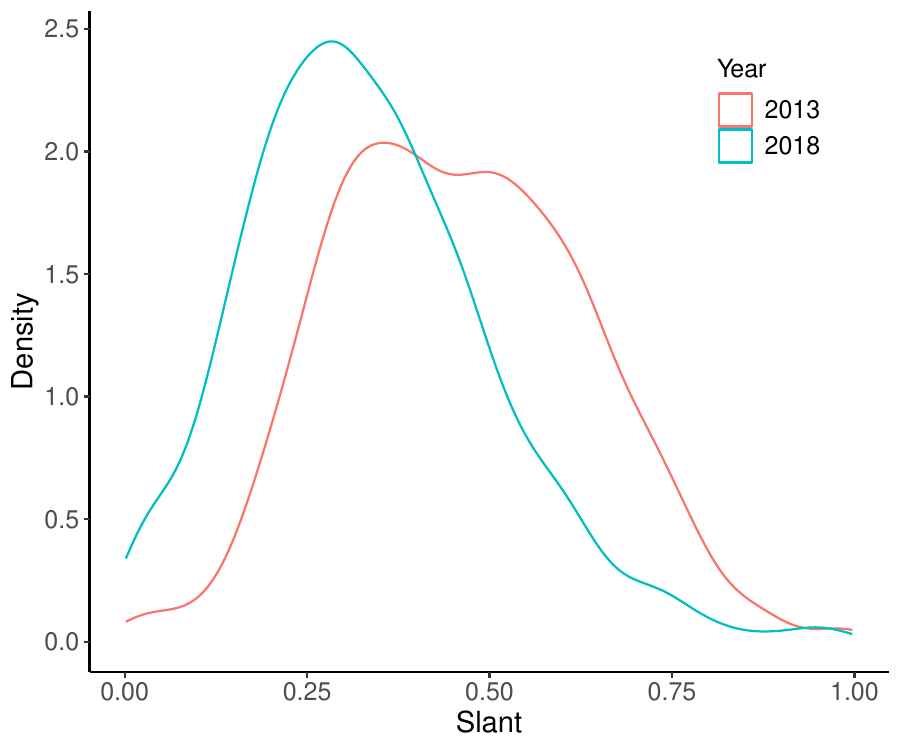}\smallskip{}
\par\end{centering}
Notes: Figure plots the distribution, across authors, of the average
slant of each author's article as measured by our RoBERTa model of
slant for each outlet. Panel A includes all authors who appear in
\emph{The New York Times} for 2013 and 2018 separately. Panel B is
the same as Panel A, except for \emph{The Wall Street Journal}.
\end{figure}

\begin{figure}[H]
\caption{Horizon Robustness for Pooled Pre/Post Changes in Journalist Slant\label{fig:role_of_consistency}}

\begin{centering}
\cprotect\emph{Panel A: $T=5$ Event-Study Sample}
\par\end{centering}
\begin{centering}
\includegraphics[scale=0.5]{input/predicting_slant_changes_T5}
\par\end{centering}
\begin{centering}
\cprotect\emph{Panel B: $T=10$ Event-Study Sample}
\par\end{centering}
\begin{centering}
\includegraphics[scale=0.5]{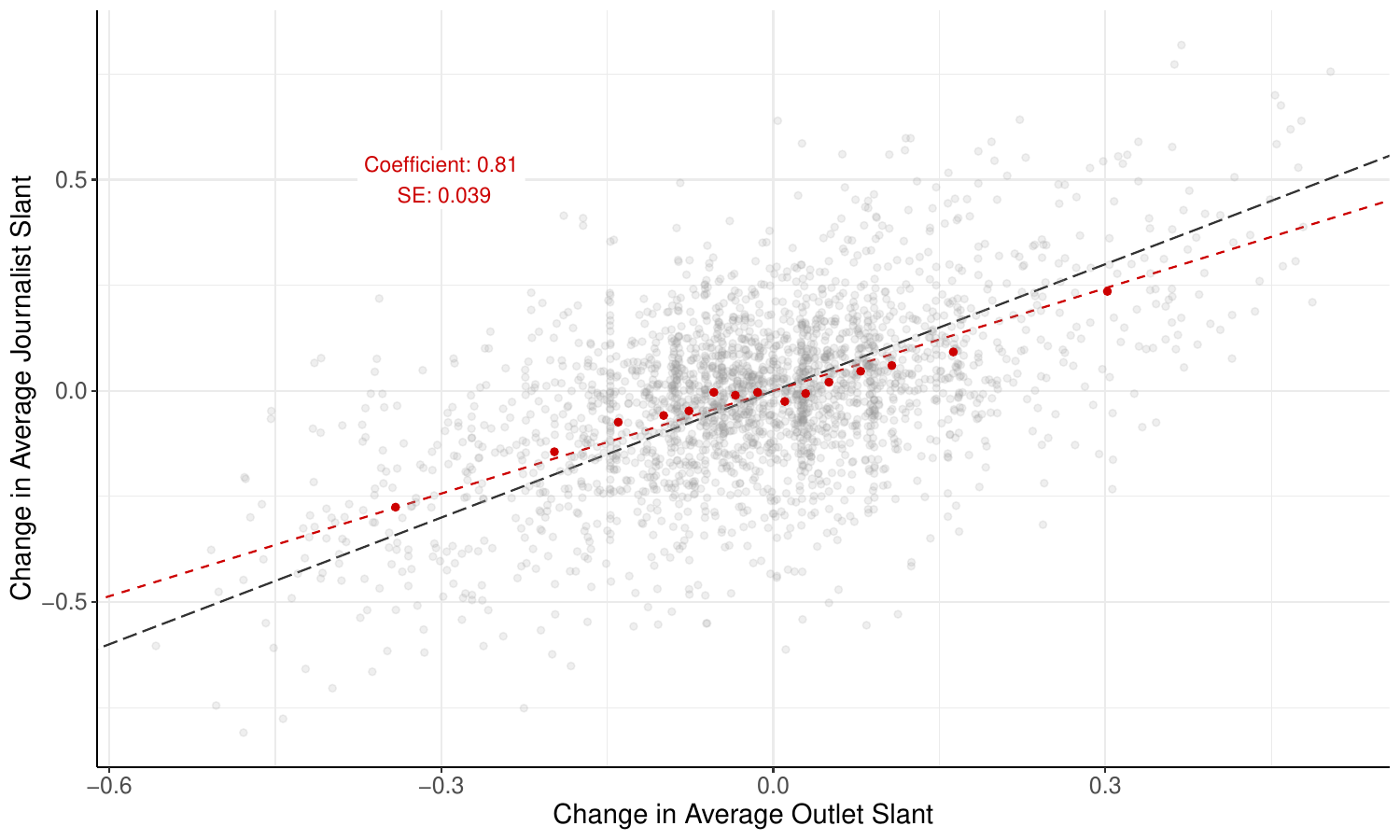}
\par\end{centering}
Notes: Each panel plots the mover-level change in mean journalist slant between
the post- and pre-move windows against the leave-focal-author-out outlet gap.
Both variables are constructed after removing publication-month means. The samples
match the corresponding event studies. Panel A uses the baseline $T=5$ window
and Panel B uses $T=10$ as a horizon robustness check. The red points show unweighted means
within 15 equal-count bins. The fitted line is constrained
through the origin and weights observations by the squared outlet gap;
standard errors are two-way clustered at the origin- and destination-outlet
levels.
\end{figure}

\begin{figure}[H]
\begin{centering}
\caption{Comparing to Human Perceived Slant --- Distributions\label{fig:budak_distributions}}
\par\end{centering}
\begin{centering}
\begin{centering}
\subfloat[RoBERTa]{\label{sfig:a}\includegraphics[scale=.4]{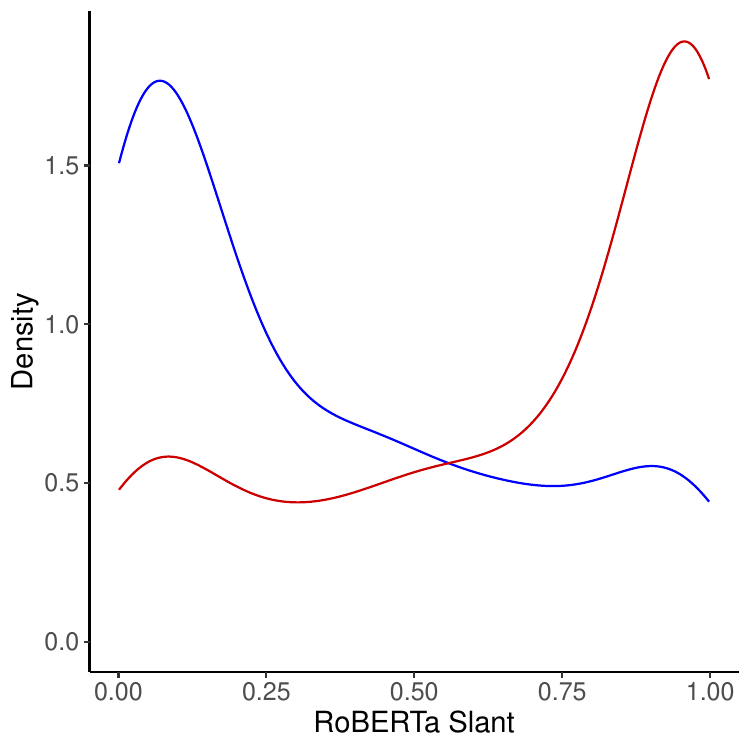}} \hspace{1mm}
\subfloat[FastText]{\label{sfig:b}\includegraphics[scale=.4]{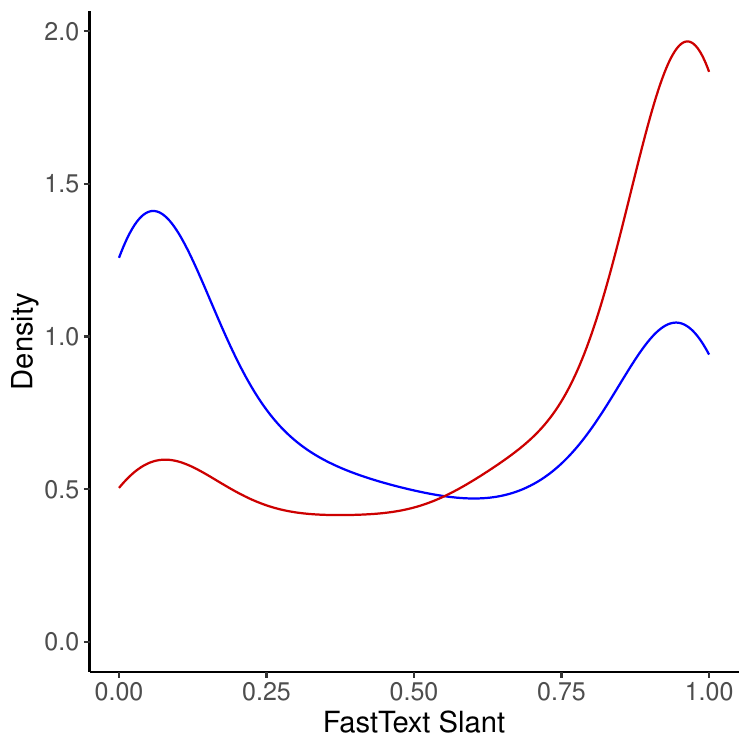}} \hspace{1mm}
\subfloat[GST (re-estimated)]{\label{sfig:c}\includegraphics[scale=.4]{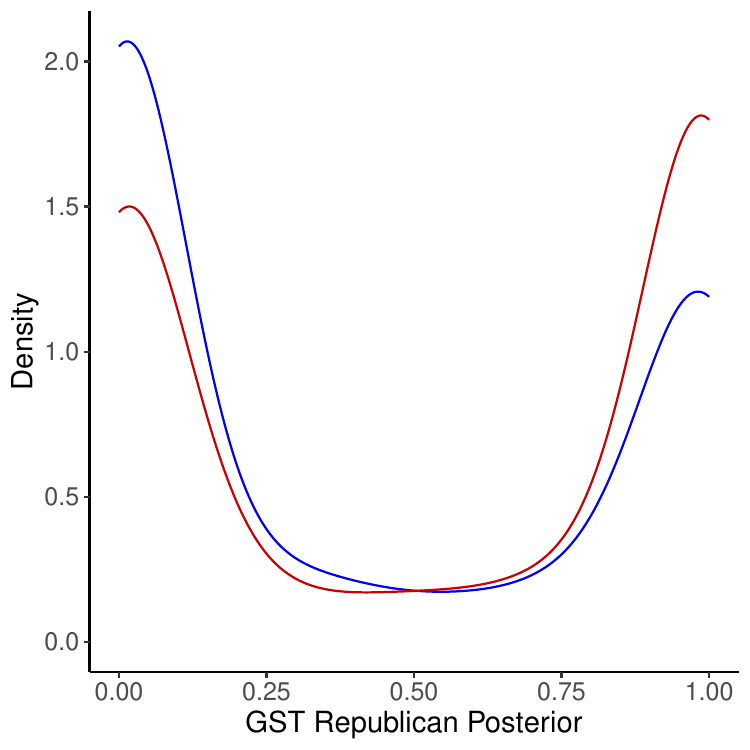}} \hspace{1mm}
\end{centering}
\par\end{centering}
\begin{centering}
\cprotect\emph{\smallskip{}
}
\par\end{centering}
\begin{centering}
\cprotect\emph{\smallskip{}
}
\par\end{centering}
Notes: Figure plots the distribution of the text-based measure separately
for articles coded as leaning Republican (red) and articles coded
as leaning Democratic (blue) by the MTurk coders in \citet{Budak2016FairBalanced}.
The sample is restricted to articles with a non-zero blinded-coder rating.
GST is the re-estimated Republican posterior probability. The density
bandwidth is 0.1 for each measure; curves may extend beyond the score support.
\end{figure}

\clearpage
\begin{figure}[H]
\caption{Validation Balanced Accuracy by Tweet Month\label{fig:monthly_oos_accuracy}}

\begin{centering}
\includegraphics[width=.86\textwidth,height=.65\textheight,keepaspectratio]{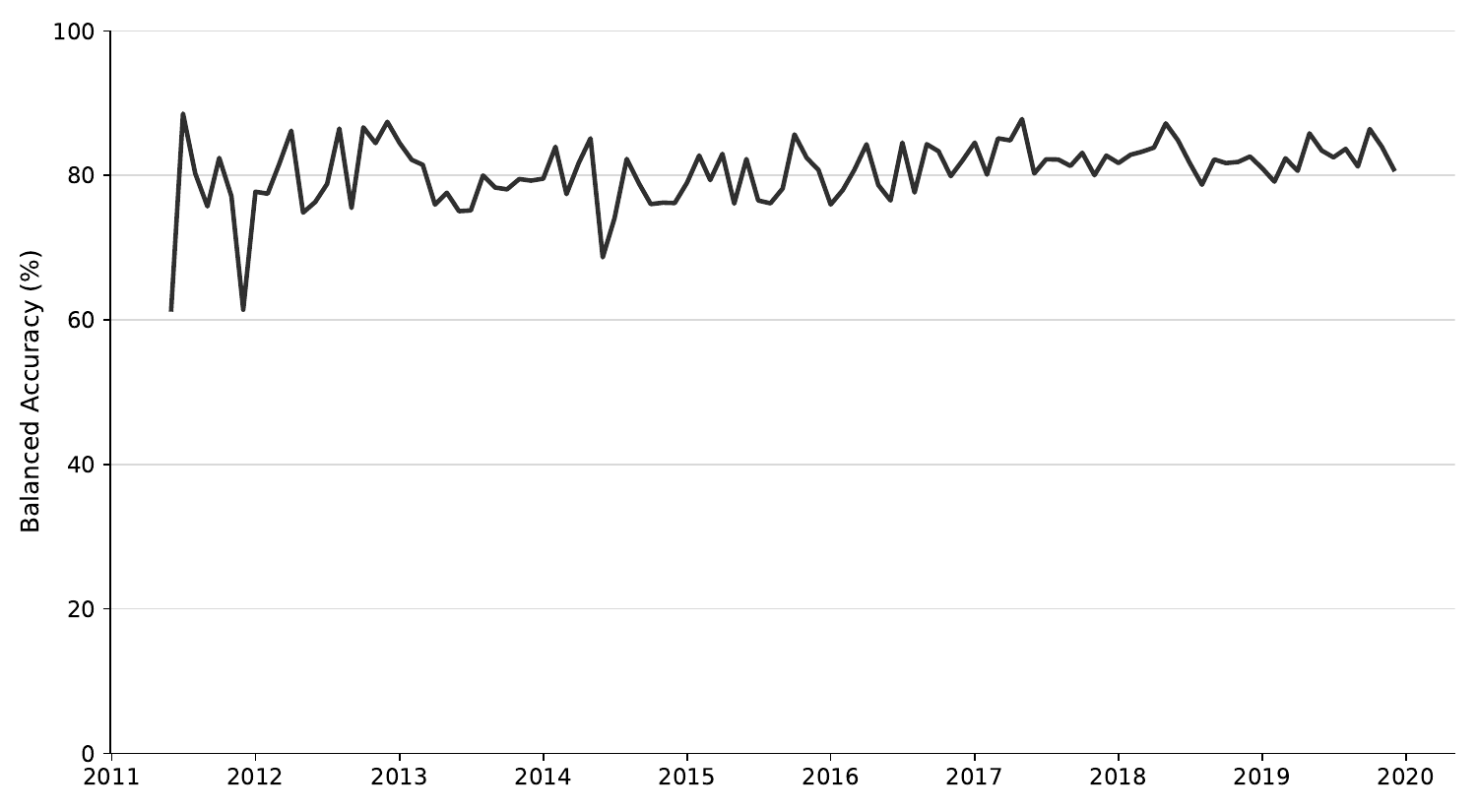}
\par\end{centering}
Notes: Figure plots balanced accuracy in the validation sample, calculated
separately by the month in which the article was tweeted.
\end{figure}

\clearpage
\begin{figure}[H]
\caption{Out-of-Sample Validation for Article-Stratified vs. Outlet-Stratified Splits\label{fig:outlet_stratification}}

\begin{centering}
\emph{Panel A: Balanced Accuracy}
\par\end{centering}
\begin{centering}
\includegraphics[width=.64\textwidth,height=.31\textheight,keepaspectratio]{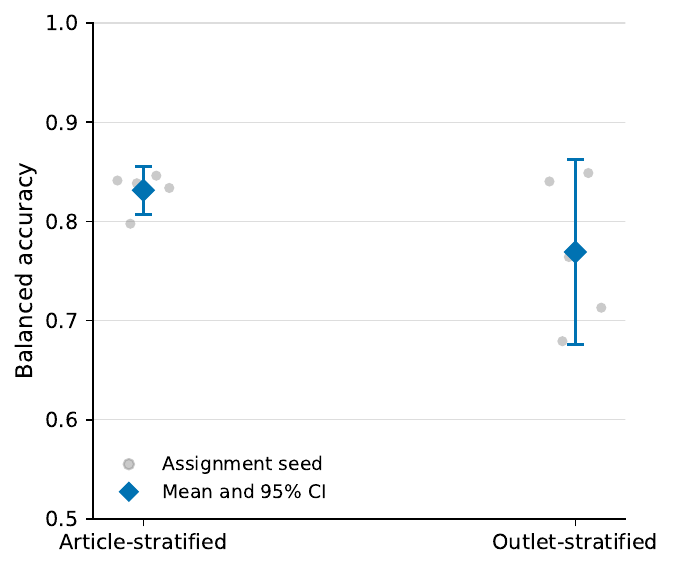}
\par\end{centering}
\begin{centering}
\emph{Panel B: ROC-AUC}
\par\end{centering}
\begin{centering}
\includegraphics[width=.64\textwidth,height=.31\textheight,keepaspectratio]{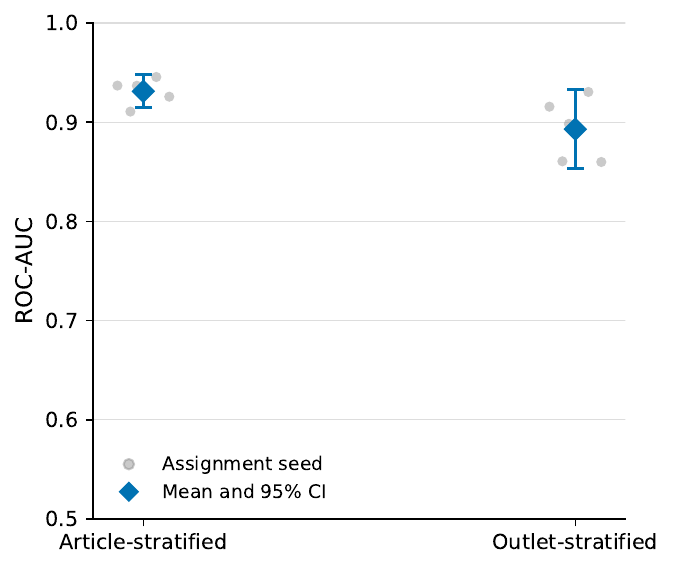}
\par\end{centering}
{\small Notes: Both designs use the same sample of politician-tweeted articles
from June 1, 2011 onward, excluding missing or unknown outlets and identical
texts attributed to multiple outlets. Article-stratified splits assign whole groups of
identical article texts; outlet-stratified splits assign whole outlets. Neither
design allows identical texts in both training and validation. Dots show five
random split assignments per design, diamonds show means, and whiskers show
two-sided 95 percent confidence intervals across assignments. The training specification
and model initialization seed are held fixed. Two-sided unpaired t-tests
(allowing unequal variances) of
the differences between the outlet- and article-stratified means give
$p=\OutletStratificationHeldoutMinusMixedBalancedAccuracyGapWelchPValue$ for
balanced accuracy and $p=\OutletStratificationHeldoutMinusMixedROCAUCGapWelchPValue$
for ROC-AUC.\par}
\end{figure}

\clearpage
\begin{figure}[H]
\caption{Heterogeneity in Journalist AKM Fixed Effects --- WSJ and NYT\label{fig:akm_hetero_wsj_nyt}}

\begin{centering}
\emph{Panel A: New York Times vs. Wall Street Journal}
\par\end{centering}
\begin{centering}
\includegraphics[scale=0.43]{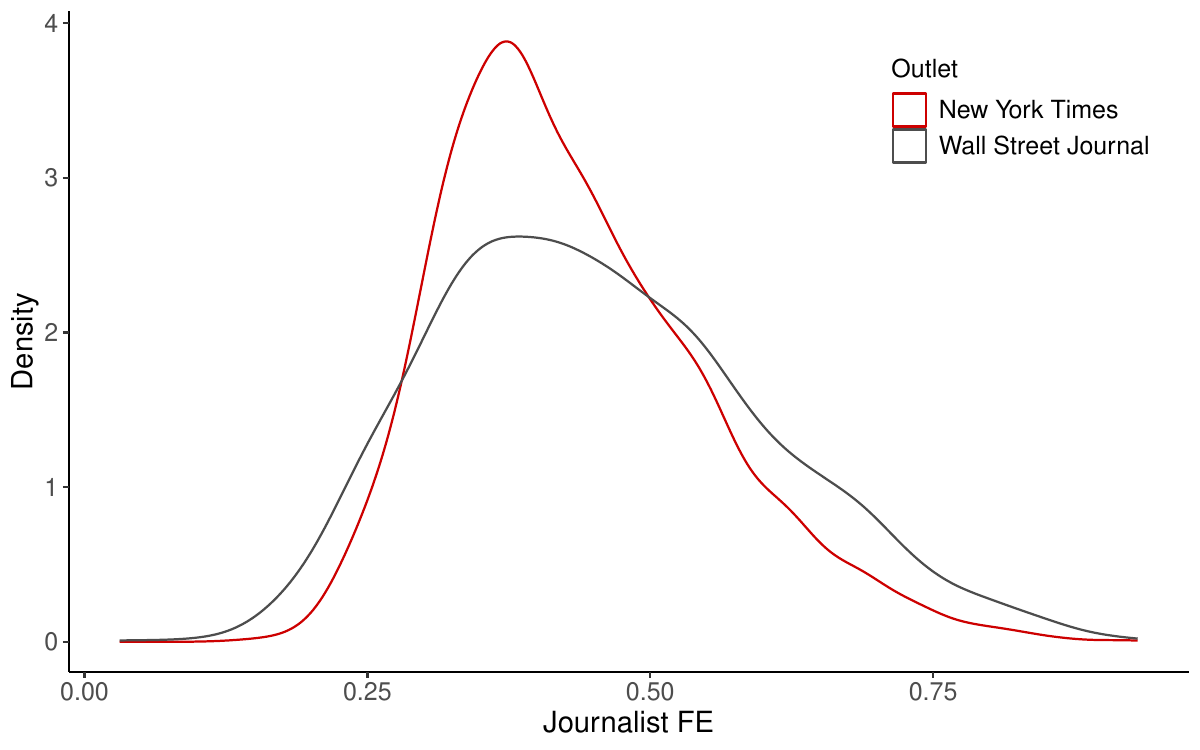}
\par\end{centering}
\begin{centering}
\emph{Panel B: New York Times -- By Party}
\par\end{centering}
\begin{centering}
\includegraphics[scale=0.43]{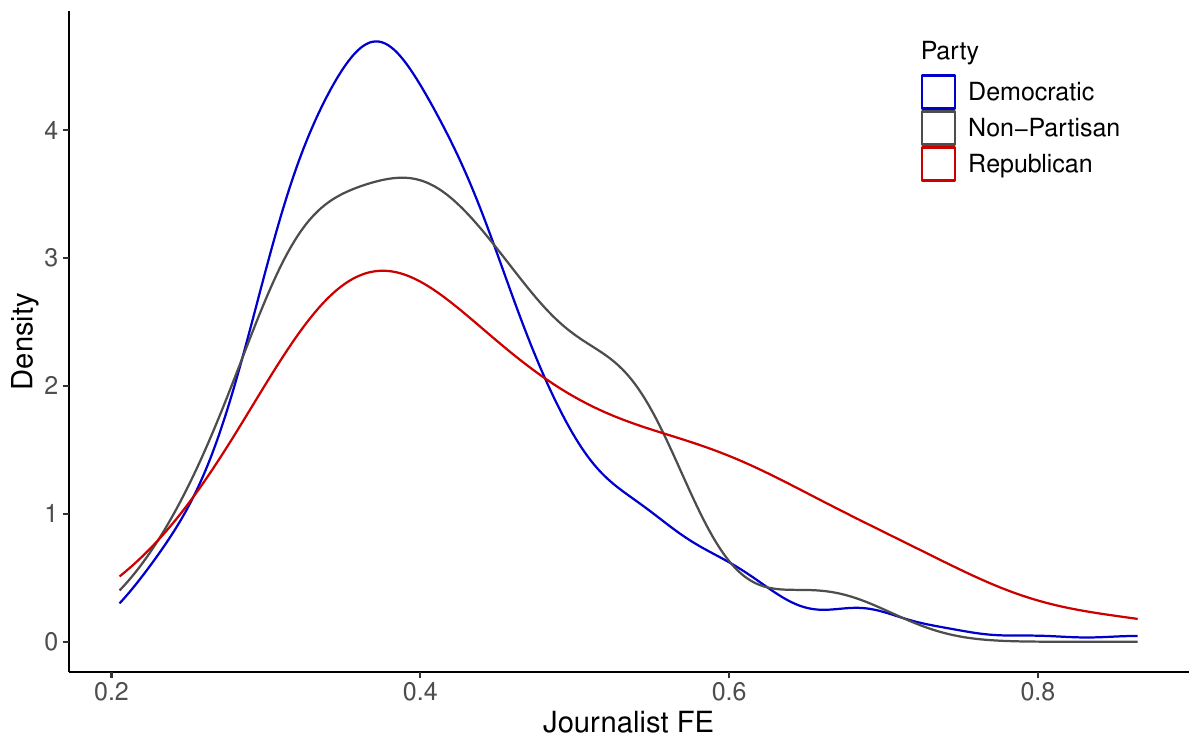}
\par\end{centering}
\begin{centering}
\emph{Panel C: Wall Street Journal -- By Party}
\par\end{centering}
\begin{centering}
\includegraphics[scale=0.43]{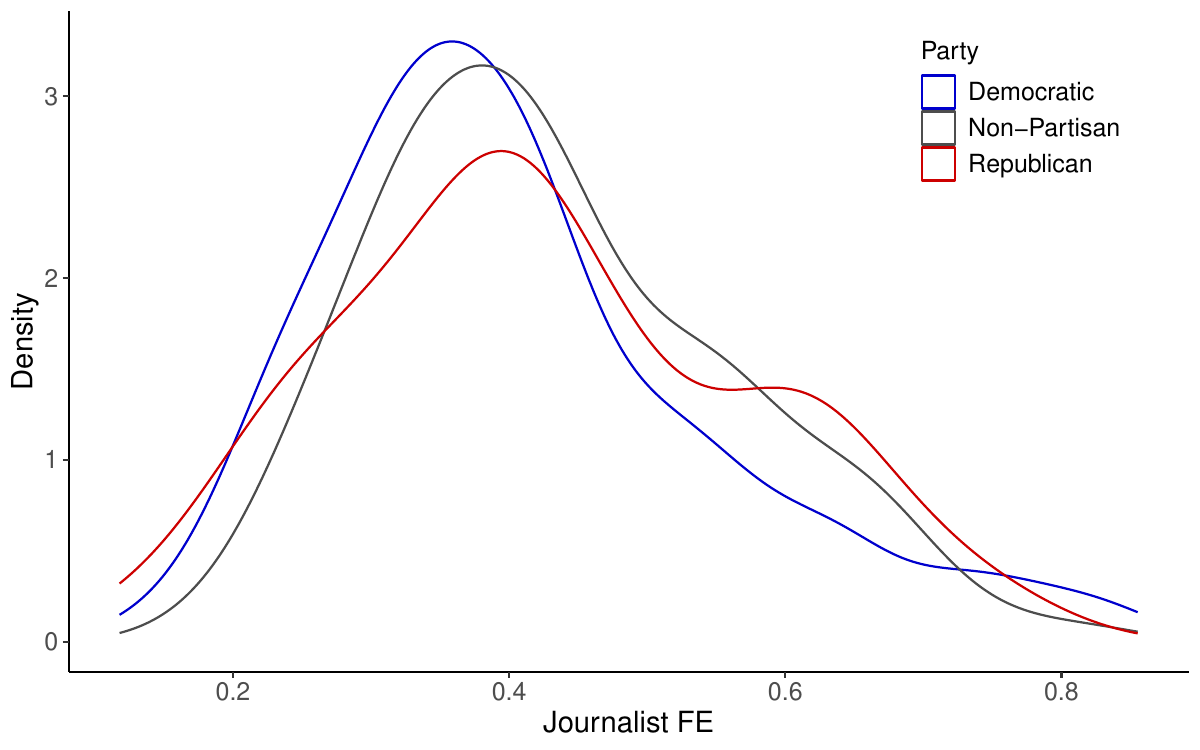}
\par\end{centering}
Notes: Each panel plots the distribution in estimated AKM journalist
fixed effects across various sub-groups of journalists. Panel A examines
heterogeneity across journalists at \emph{The New York Times }and
\emph{The Wall Street Journal}. Panel B examines heterogeneity at
\emph{The New York Times} by partisan affiliation. Panel C examines
heterogeneity at \emph{The Wall Street Journal} by party affiliation.
\end{figure}

\begin{figure}[H]
\caption{Topic Distributions in Politician-Tweeted and ProQuest Article Samples\label{fig:topic_distributions}}
\begin{centering}
\emph{Panel A: Politician-Tweeted Articles}
\par\end{centering}
\begin{centering}
\includegraphics[width=.88\textwidth]{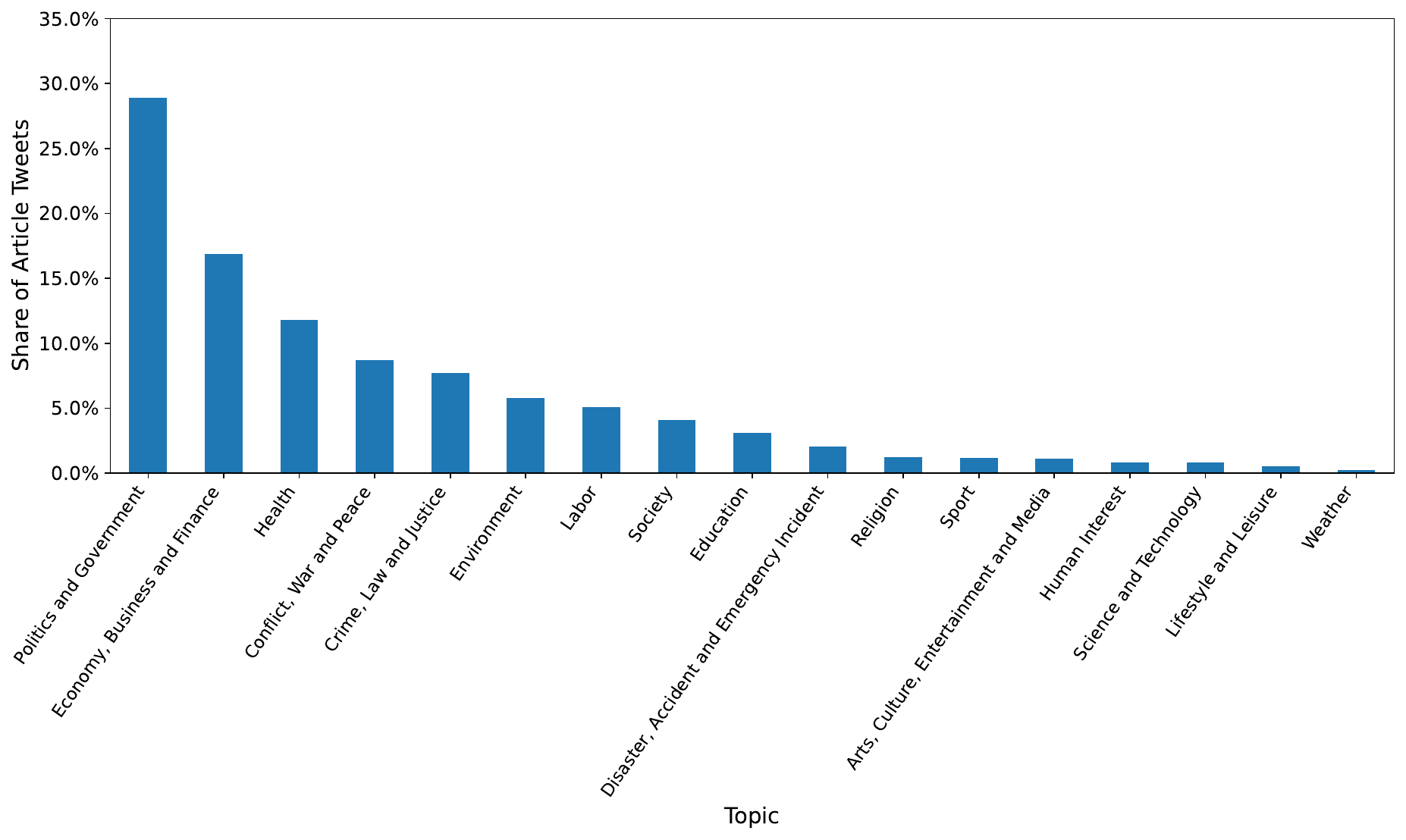}
\par\end{centering}
\begin{centering}
\emph{Panel B: ProQuest Articles}
\par\end{centering}
\begin{centering}
\includegraphics[width=.88\textwidth]{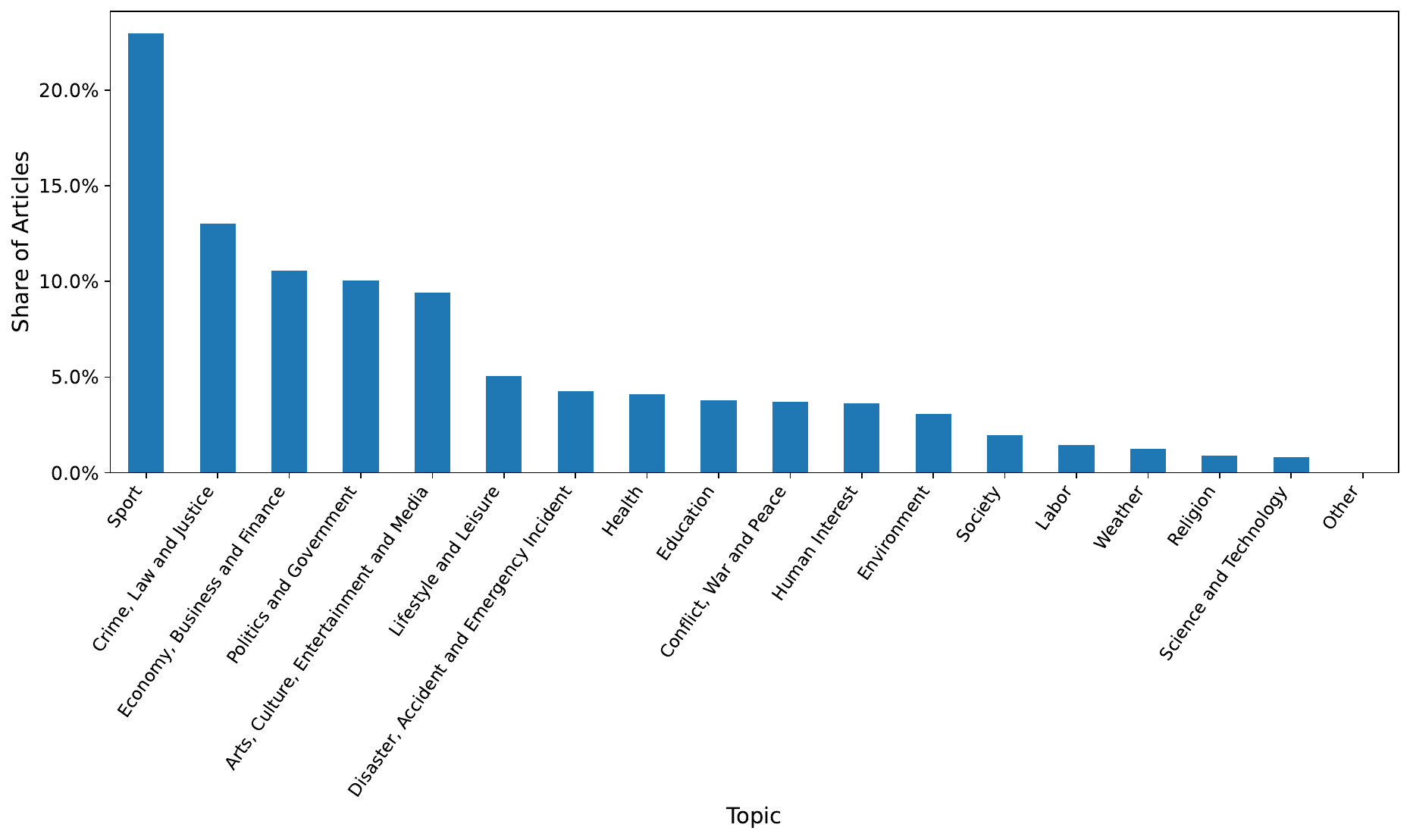}
\par\end{centering}
Notes: Panel A reports the share of reweighted politician article
tweets in each IPTC topic, after reweighting article tweets to balance
the overall Democratic and Republican shares. Panel B reports the share
of ProQuest articles in each IPTC topic in the article sample used for
the topic-slant analysis.
\end{figure}

\begin{figure}[H]
\caption{Republican Share by Topic in Politician-Tweeted Articles\label{fig:topic_party_share}}

\begin{centering}
\includegraphics[width=.95\textwidth]{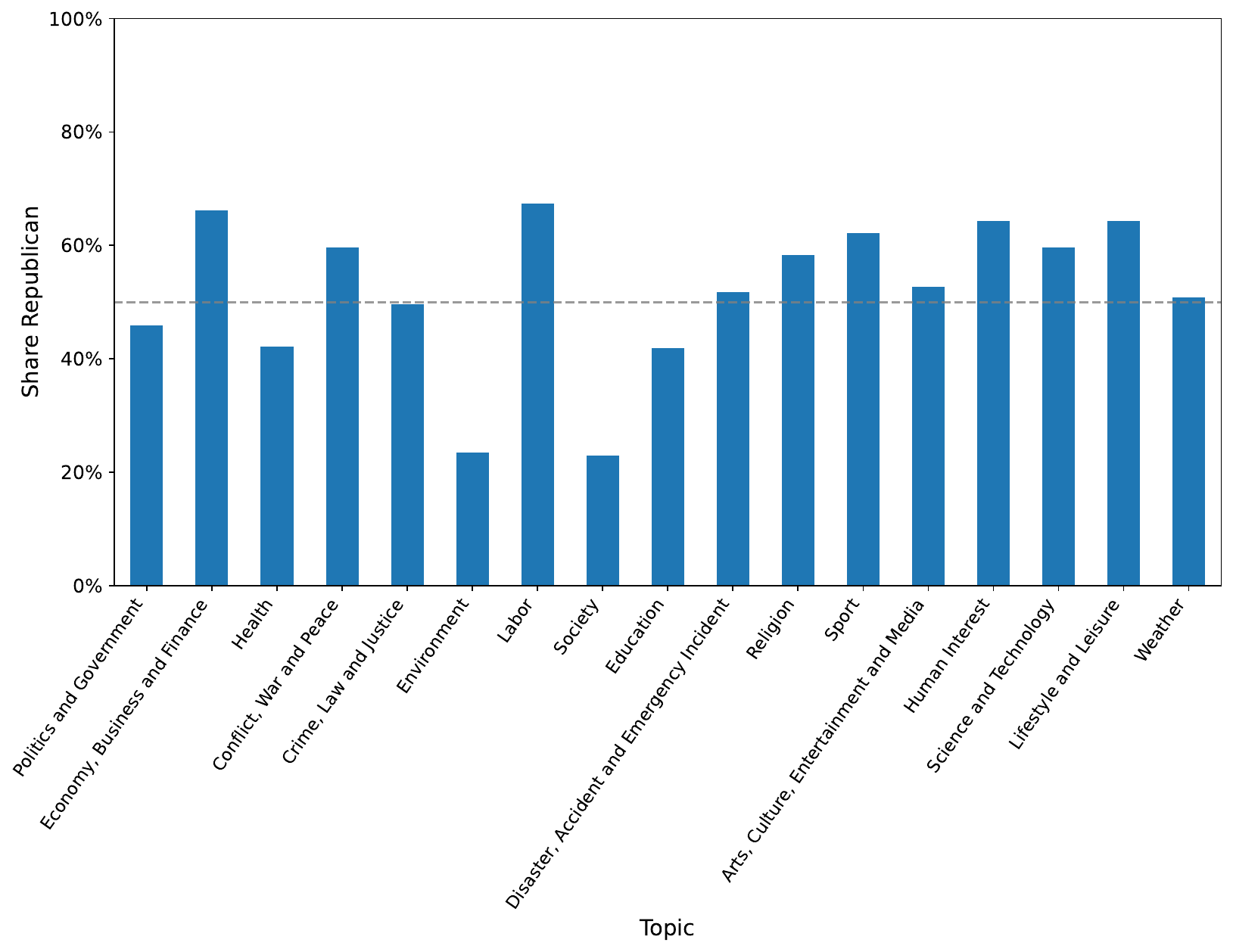}
\par\end{centering}
Notes: Figure reports the weighted share of politician-tweeted articles
within each IPTC topic that were tweeted by Republican politicians,
using article-tweet weights that balance the overall Democratic and
Republican shares.
\end{figure}

\begin{figure}[H]
\caption{Politician-Tweeted Article Time Series\label{fig:tweet_time_series}}

\begin{centering}
\emph{Panel A: Article-Tweet Counts by Party}
\par\end{centering}
\begin{centering}
\includegraphics[width=.88\textwidth,height=.34\textheight,keepaspectratio]{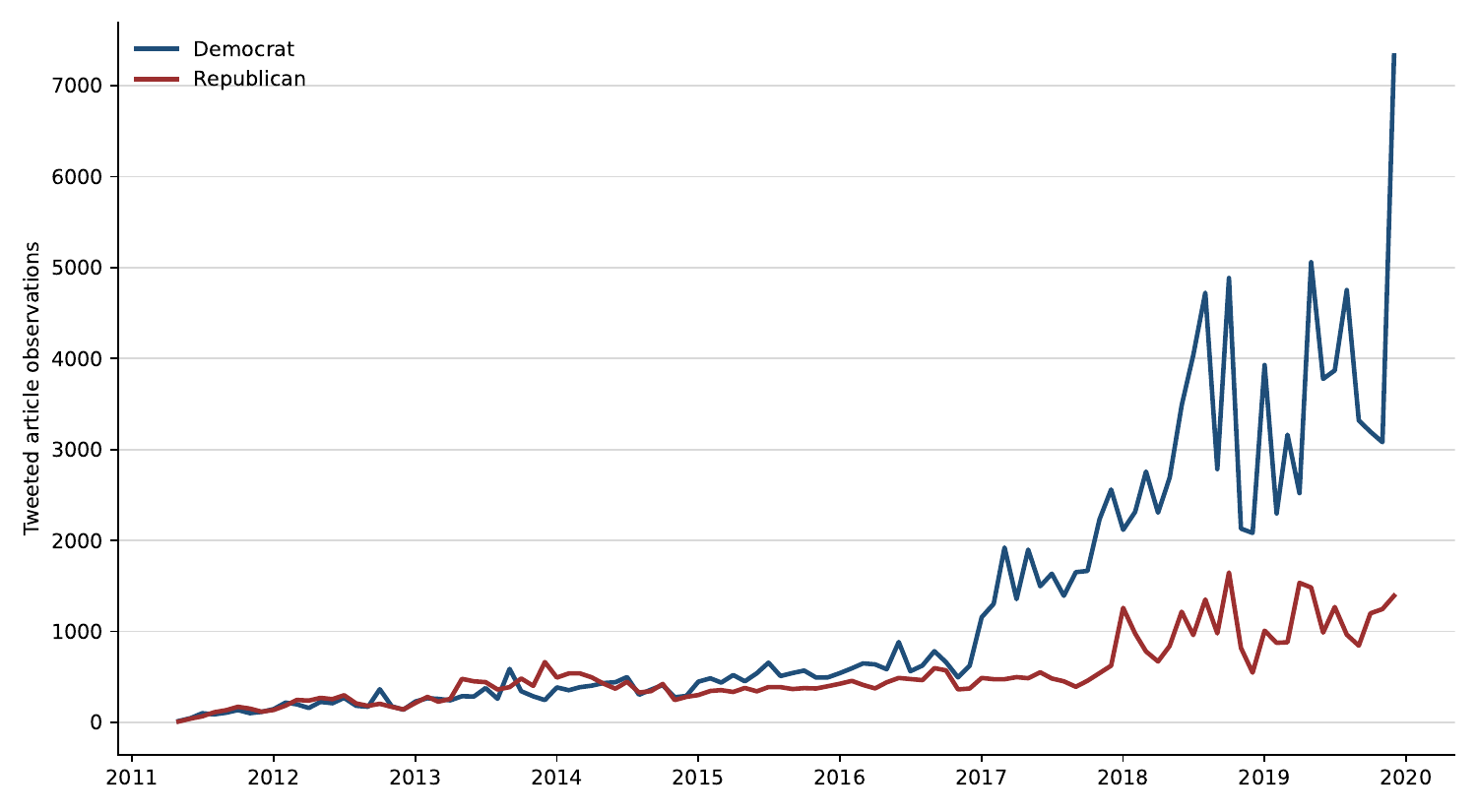}
\par\end{centering}
\begin{centering}
\emph{Panel B: Article-Tweet Party Shares}
\par\end{centering}
\begin{centering}
\includegraphics[width=.88\textwidth,height=.34\textheight,keepaspectratio]{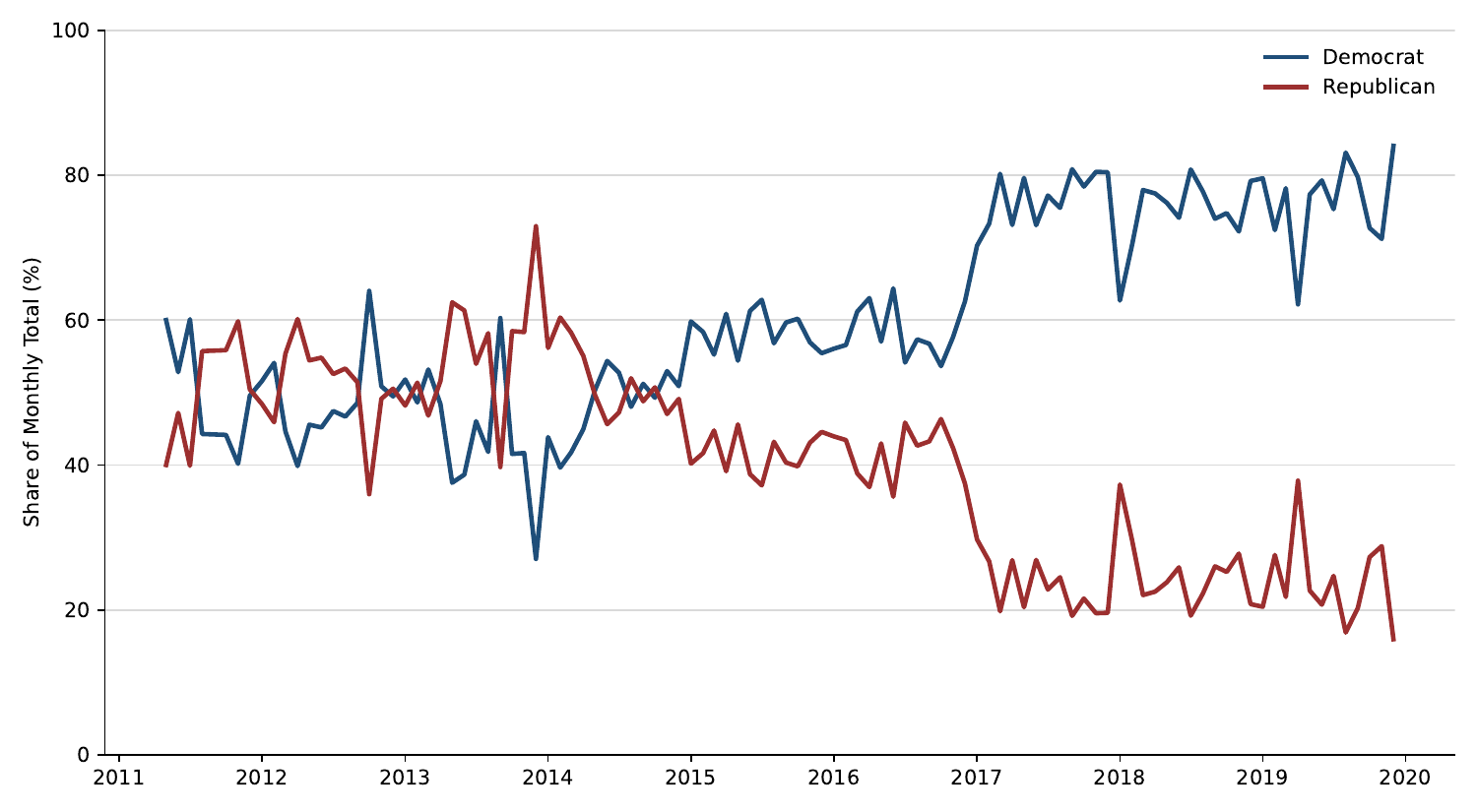}
\par\end{centering}
Notes: Figure summarizes the cleaned politician-tweeted article sample.
Panel A reports article-tweet counts separately by month $\times$
the party of the politician who tweeted the article. Panel B reports
the corresponding party shares among article-tweet observations.
For visualization purposes, both panels restrict to the longest contiguous
monthly span containing at least 25 tweeted-article observations per
month.
\end{figure}

\begin{figure}[H]
\caption{Outlet-Level Sharing by Politician Party in the Tweeted-Article
Sample\label{fig:twitter_tabulate_plot}}

\begin{centering}
\includegraphics[width=\textwidth,height=.78\textheight,keepaspectratio]{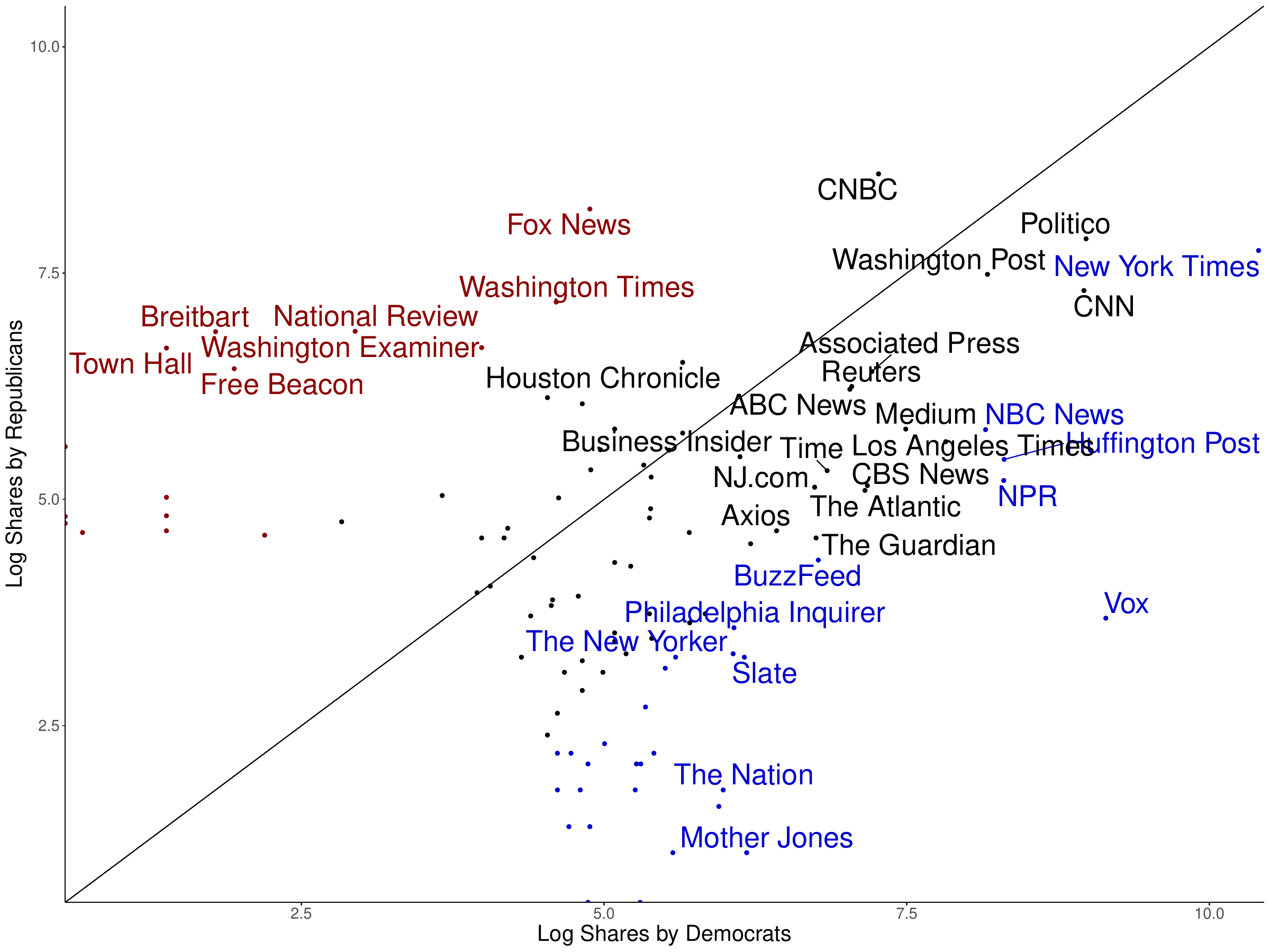}
\par\end{centering}
Notes: Figure plots, by outlet, the log number of article tweets shared
by Democratic politicians on the x-axis and Republican politicians on the
y-axis. Points above the 45-degree line are shared more often by Republican
politicians, while points below the line are shared more often by Democratic
politicians. Labeled outlets are selected high-frequency outlets or outlets
with highly asymmetric sharing by politician party.
\end{figure}

\clearpage
\begin{figure}[H]
\caption{Slant Distributions by Topic in ProQuest\label{fig:slant_distribution_by_topic}}

\begin{centering}
\includegraphics[width=\textwidth,height=.88\textheight,keepaspectratio]{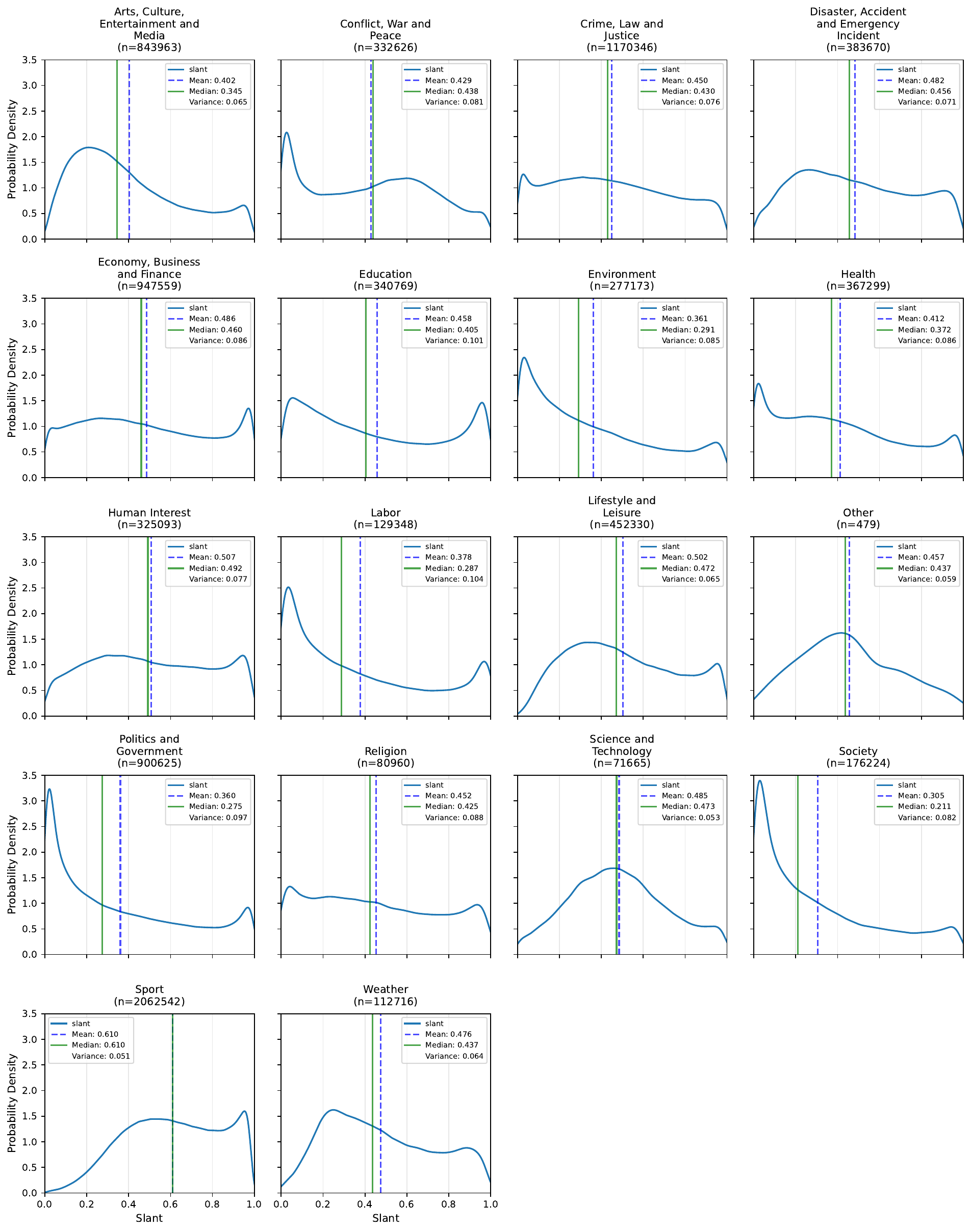}
\par\end{centering}
Notes: Each facet plots the distribution of article slant within a
cleaned topic in the ProQuest sample. The dashed blue line marks the
mean slant within the topic, and the solid green line marks the median.
\end{figure}

\begin{figure}[H]
\caption{Single-Author Event Study Changes in Slant Around Author Moves\label{fig:eventstudy_single_author}}

\begin{centering}
\emph{Panel A: $T=5$}
\par\end{centering}
\begin{centering}
\includegraphics[scale=0.7]{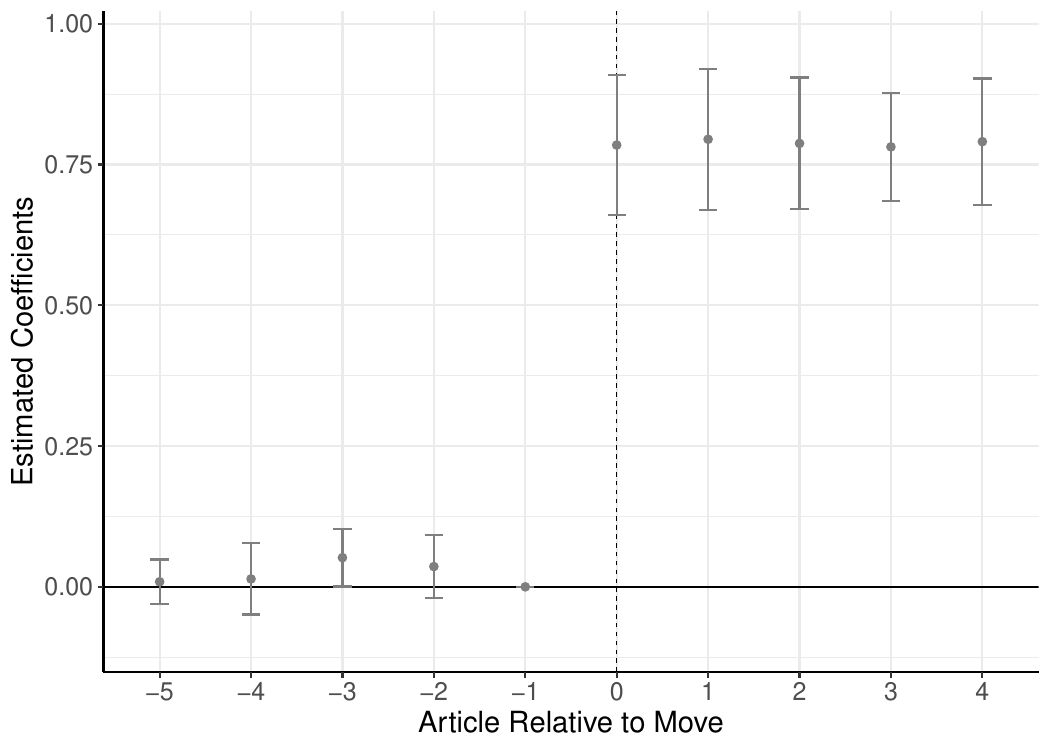}
\par\end{centering}
\begin{centering}
\emph{Panel B: $T=10$}
\par\end{centering}
\begin{centering}
\includegraphics[scale=0.7]{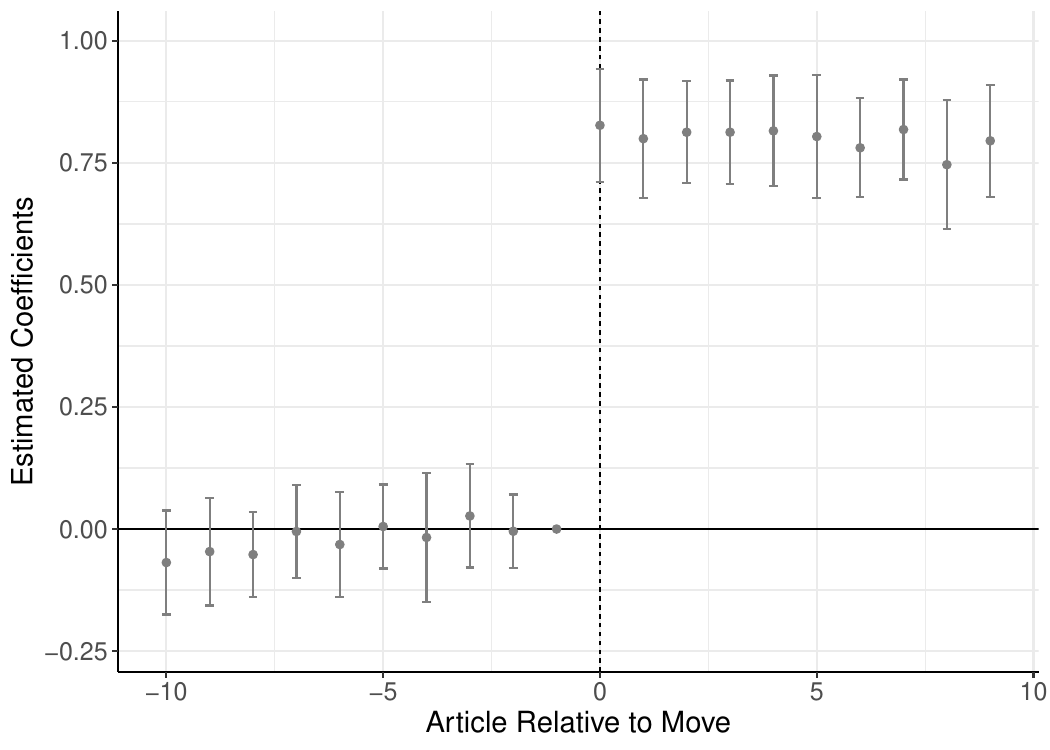}
\par\end{centering}
Notes: Figure follows Figure \ref{fig:eventstudy} in plotting the
estimated mover event-study coefficients, but restricts the event-study
sample to articles with a single author. Panel A uses the baseline $T=5$
window and Panel B uses $T=10$ as a horizon robustness check. The coefficients are normalized and
weighted using the leave-focal-author-out origin-destination outlet slant gap
constructed after removing publication-month means. Standard
errors are two-way clustered at the origin- and
destination-outlet levels.
\end{figure}

\begin{figure}[H]
\caption{Topic-Level Event Study Changes in Slant Around Author Moves\label{fig:eventstudy_topic}}

\begin{centering}
\emph{Panel A: $T=5$}
\par\end{centering}
\begin{centering}
\includegraphics[scale=0.67]{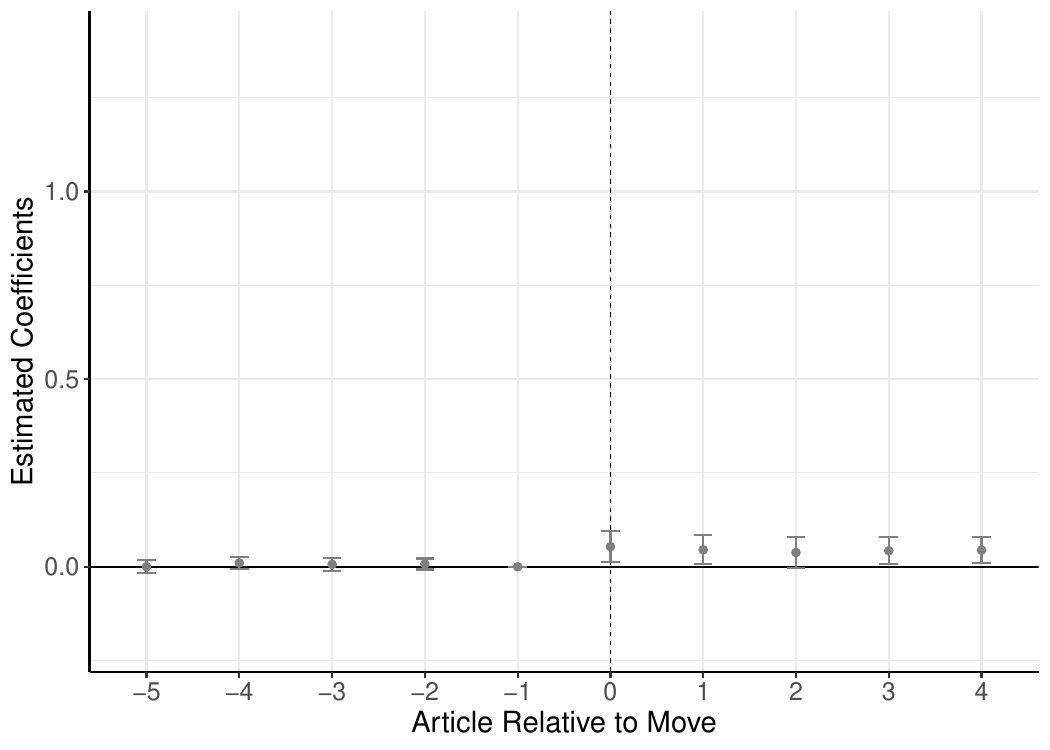}
\par\end{centering}
\begin{centering}
\emph{Panel B: $T=10$}
\par\end{centering}
\begin{centering}
\includegraphics[scale=0.67]{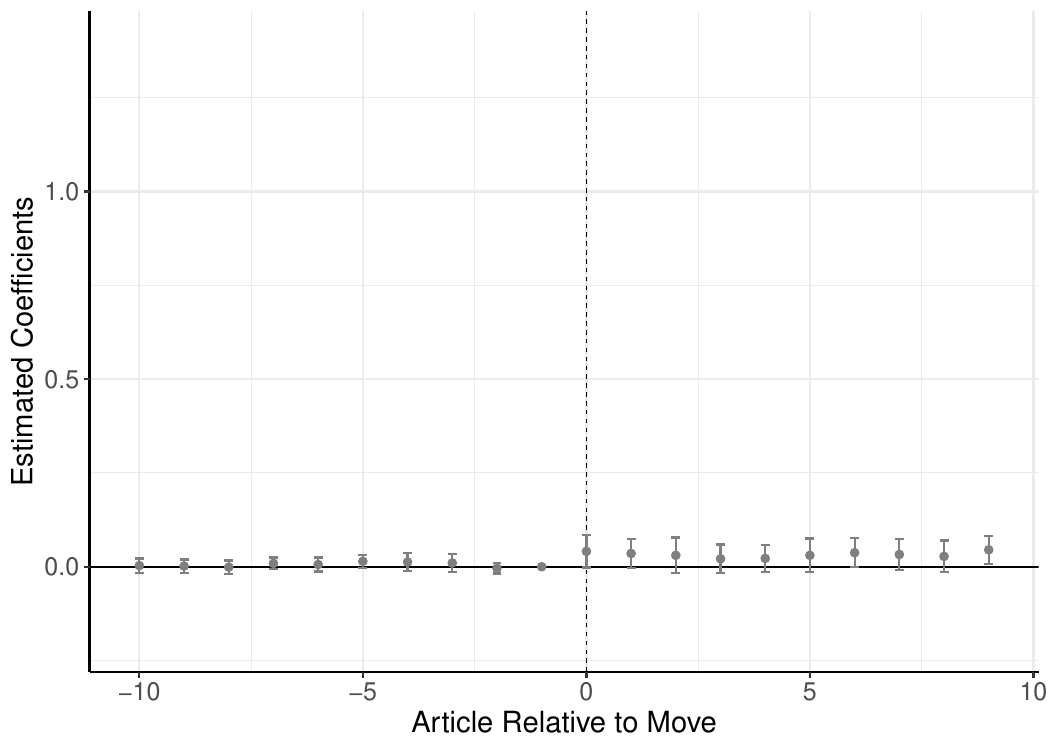}
\par\end{centering}
Notes: Figure follows Figure \ref{fig:eventstudy} in plotting the estimated mover event-study coefficients, but modifies
the outcome variable to be the mean slant for the article's topic, rather than the slant of the individual article itself.
Topic means are calculated across all eligible articles after removing publication-month means. Panel A uses the baseline $T=5$
window and Panel B uses $T=10$ as a horizon robustness check. The coefficients
are normalized and weighted using the leave-focal-author-out origin-destination outlet
slant gap, with move-pair fixed effects. Standard errors are two-way
clustered at the origin- and destination-outlet levels.
\end{figure}

\begin{figure}[H]
\caption{Event Study Changes in Slant Around Author Moves by Party Affiliation\label{fig:eventstudy_party_four_groups}}

\begin{centering}
\includegraphics[width=.9\textwidth,height=.7\textheight,keepaspectratio]{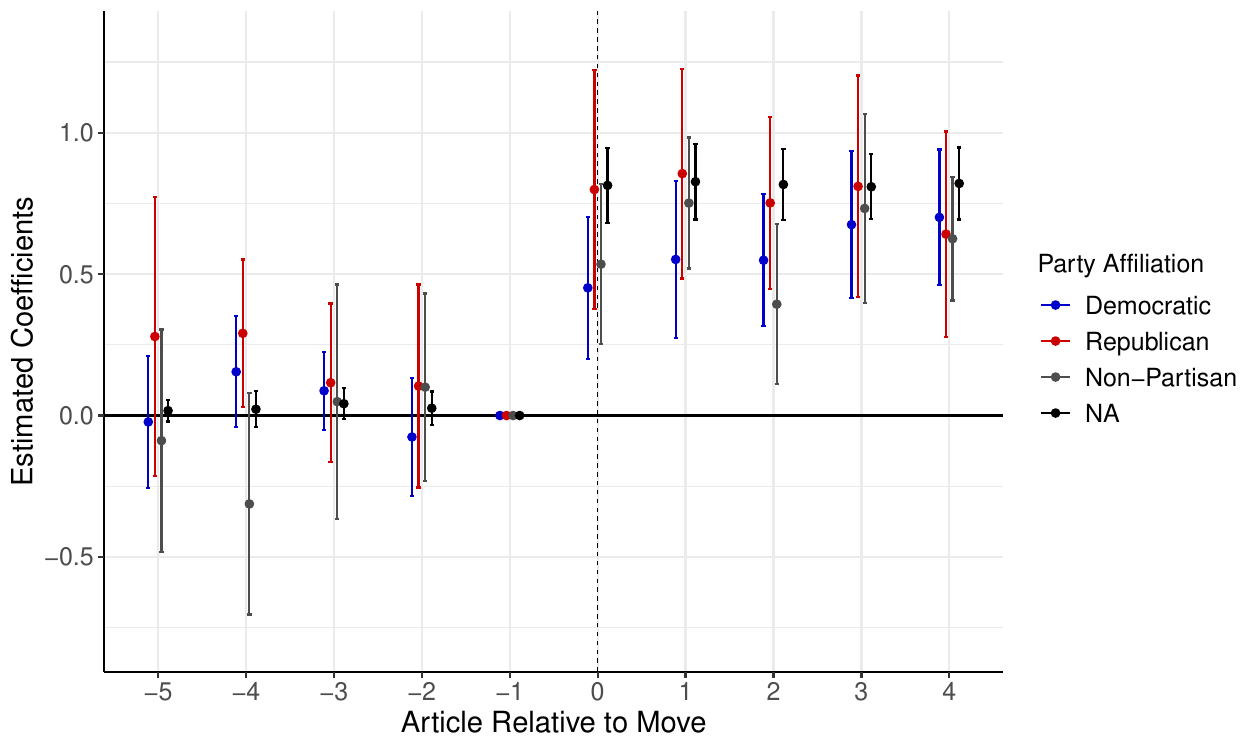}
\par\end{centering}
Notes: Figure follows Figure \ref{fig:eventstudy} in plotting estimated
mover event-study coefficients for the baseline $T=5$ estimation sample,
but estimates the event study separately by journalist party-affiliation
group. Party groups are Democratic, Republican, Non-Partisan, and
NA. Coefficients are normalized and weighted as in the main event-study
specification. Standard errors are two-way clustered at the origin- and
destination-outlet levels.
\end{figure}

\clearpage
\begin{figure}[H]
\caption{Ownership Event Studies\label{fig:ownership_event_study}}

\begin{centering}
\emph{Panel A: Gap-Slope Specification}
\par\end{centering}
\begin{centering}
\includegraphics[width=.86\textwidth,height=.25\textheight,keepaspectratio]{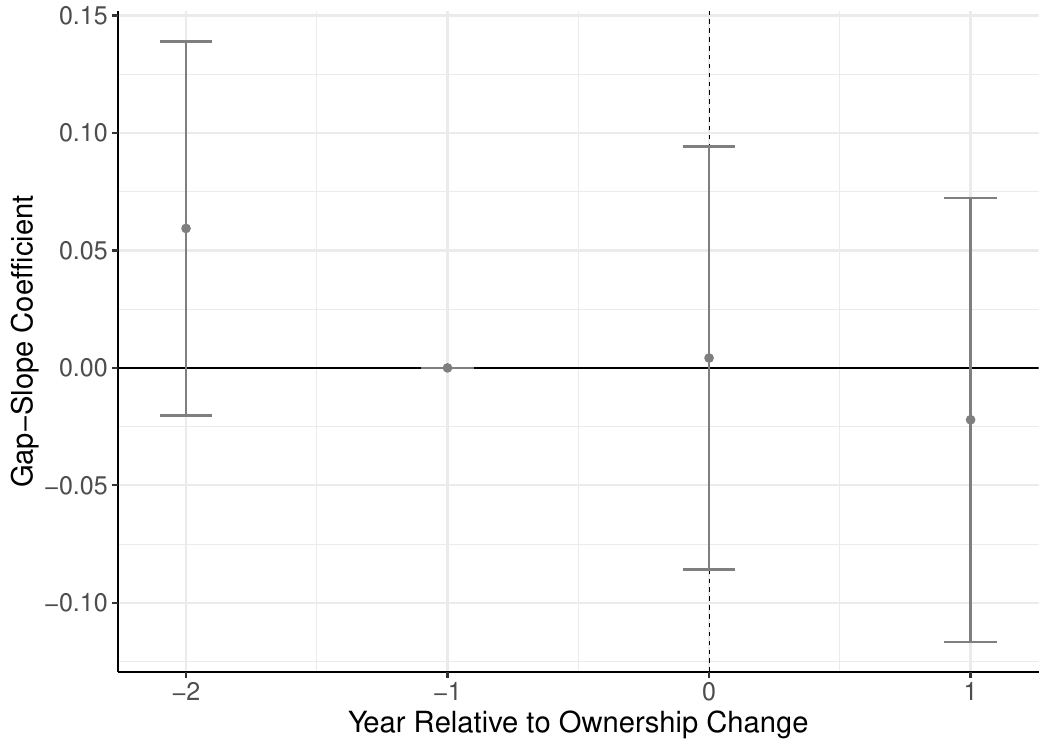}
\par\end{centering}
\begin{centering}
\emph{Panel B: Time-Varying Alignment}
\par\end{centering}
\begin{centering}
\includegraphics[width=.86\textwidth,height=.25\textheight,keepaspectratio]{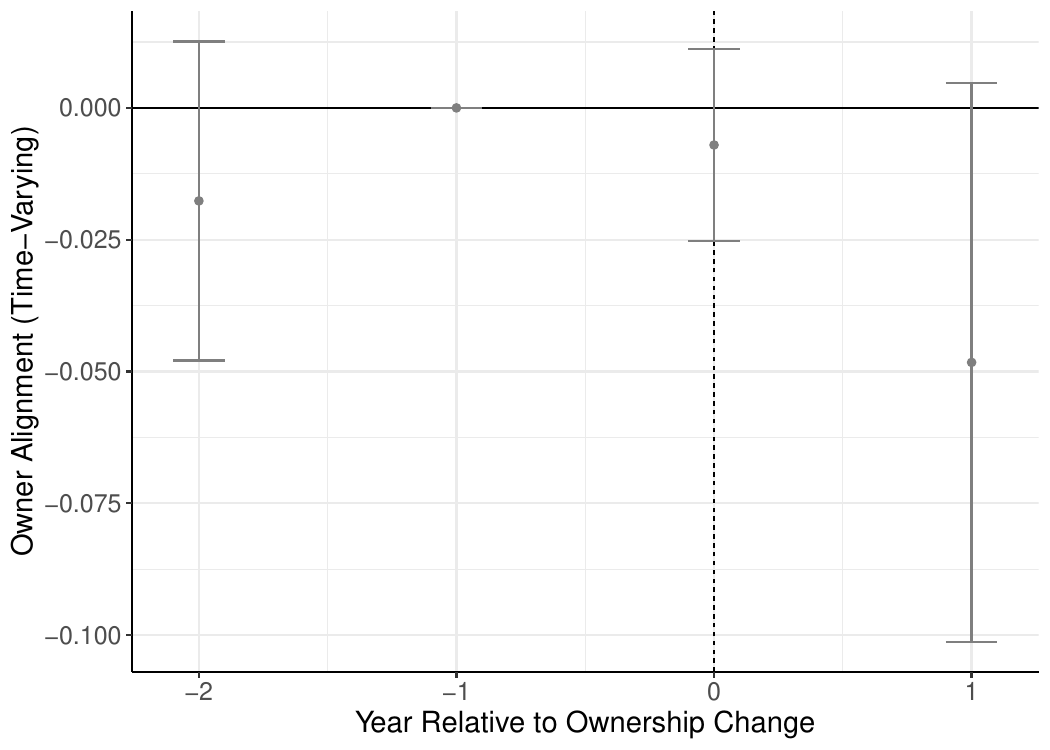}
\par\end{centering}
\begin{centering}
\emph{Panel C: Fixed-Pre Alignment}
\par\end{centering}
\begin{centering}
\includegraphics[width=.86\textwidth,height=.25\textheight,keepaspectratio]{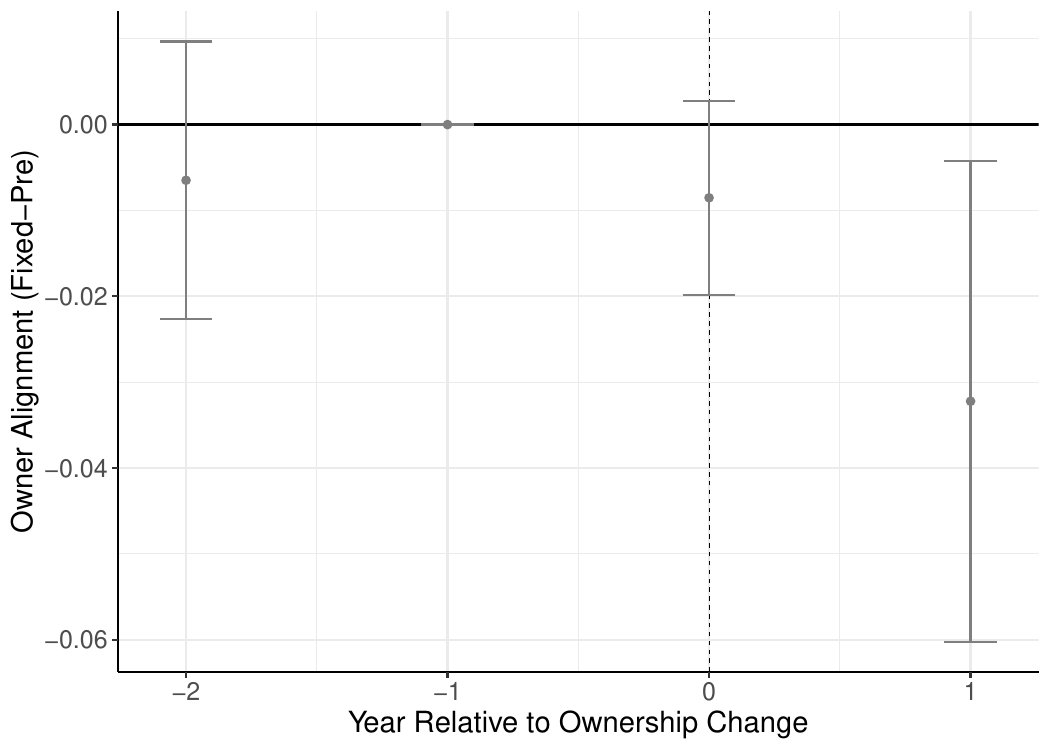}
\par\end{centering}
Notes: Figure reports dynamic event-study estimates around newspaper
ownership changes using the balanced $[-2,+1]$ event window. All
specifications omit the year before acquisition as the base period. Panel A uses the
acquiring-owner pre-period slant gap as the treatment-intensity measure
and weights by the squared gap in a mover-style analysis. Panels B and C report owner-alignment
specifications with equal event-year weights. Standard errors are
clustered at the event level. See Appendix Section \ref{app-sec-ownership-event-study} for detail.
\end{figure}

\clearpage
\begin{figure}[H]
\caption{Ownership Change Movers-Style Variation\label{fig:ownership_event_study_variation}}

\begin{centering}
\includegraphics[width=.86\textwidth,height=.50\textheight,keepaspectratio]{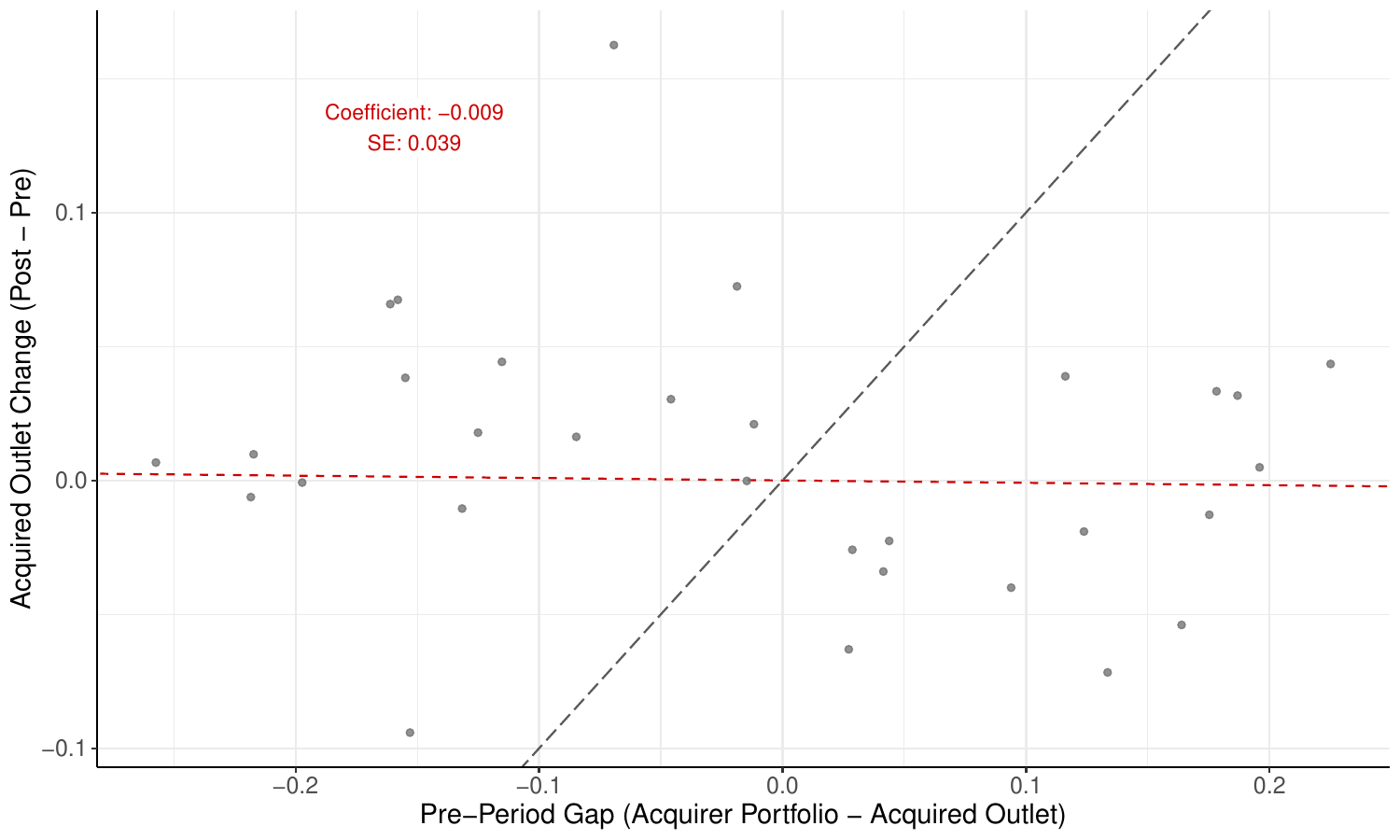}
\par\end{centering}
Notes: Figure reports the event-level variation underlying the pooled estimate for the gap-slope specification in Figure \ref{fig:ownership_event_study} Panel A.
The figure plots the acquiring-owner pre-period slant gap against
the acquired outlet's post-minus-pre change in calendar-year-residualized
slant, where the pre period is event year $-1$ and the post period
averages event years 0 and 1. The red dashed line represents a best-fit line (without an intercept), with slope coefficient and standard error reported in the annotation. Standard errors are
clustered at the event level. The dashed gray line represents a 45-degree line. See Appendix Section \ref{app-sec-ownership-event-study} for detail.
\end{figure}

\clearpage
\begin{figure}[H]
\caption{Ownership Change Alignment Shift Variation\label{fig:ownership_event_study_alignment_variation}}

\begin{centering}
\emph{Panel A: Time-Varying Alignment}
\par\end{centering}
\begin{centering}
\includegraphics[width=.86\textwidth,height=.34\textheight,keepaspectratio]{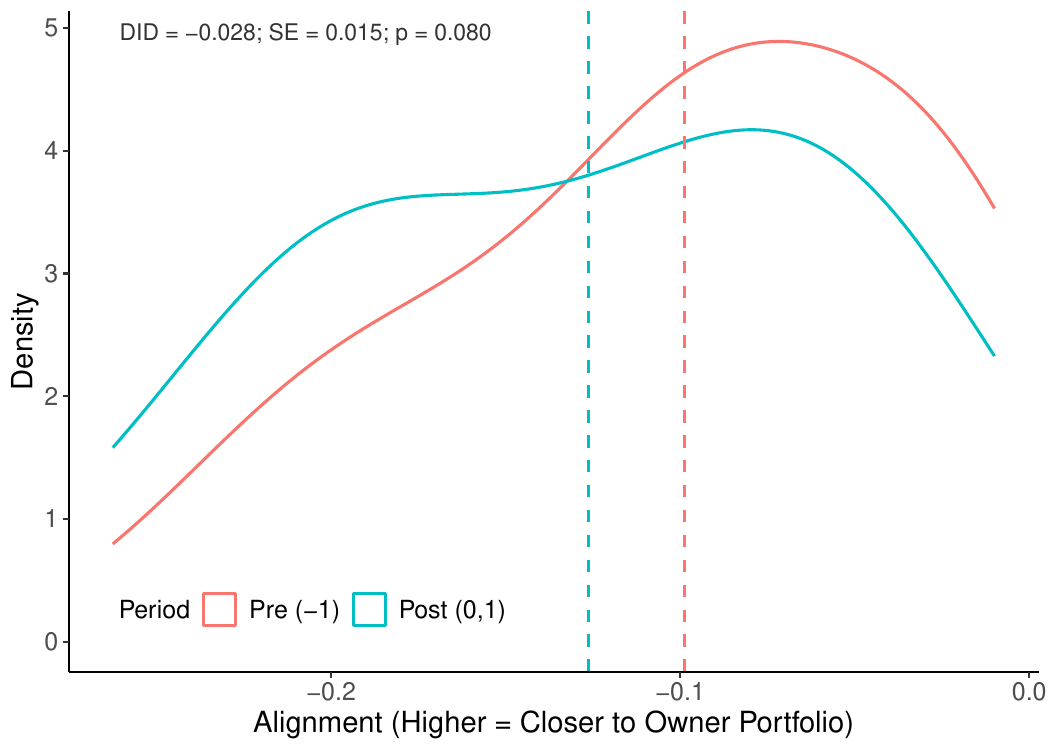}
\par\end{centering}
\begin{centering}
\emph{Panel B: Fixed-Pre Alignment}
\par\end{centering}
\begin{centering}
\includegraphics[width=.86\textwidth,height=.34\textheight,keepaspectratio]{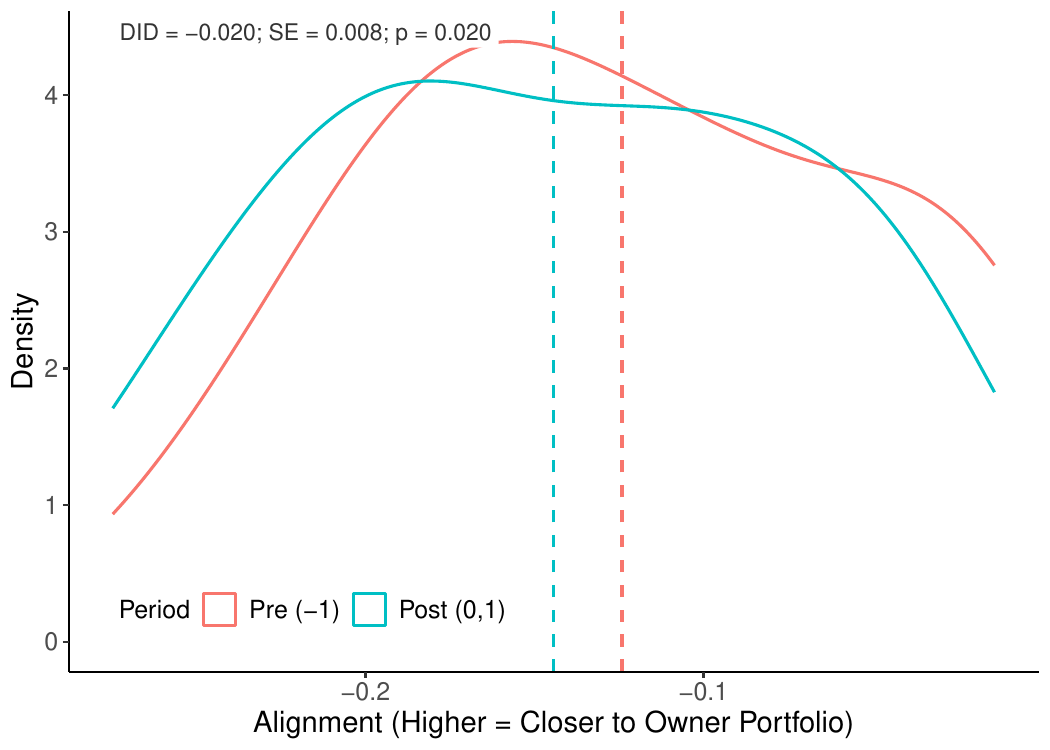}
\par\end{centering}
Notes: Figure reports the event-level alignment variation underlying
the pooled estimates in Figure \ref{fig:ownership_event_study}.
Panels A and B plot pre- and post-period owner-alignment densities
across events, with dashed lines indicating period means. The annotations
report point estimates and standard errors for the difference in means. See Appendix Section \ref{app-sec-ownership-event-study} for detail.
\end{figure}

\clearpage
\begin{figure}[H]
\caption{Counterfactual Mean Slant under Alternative Journalist Party Compositions\label{fig:counterfactual_mean_slant}}

\begin{centering}
\includegraphics[scale=0.62]{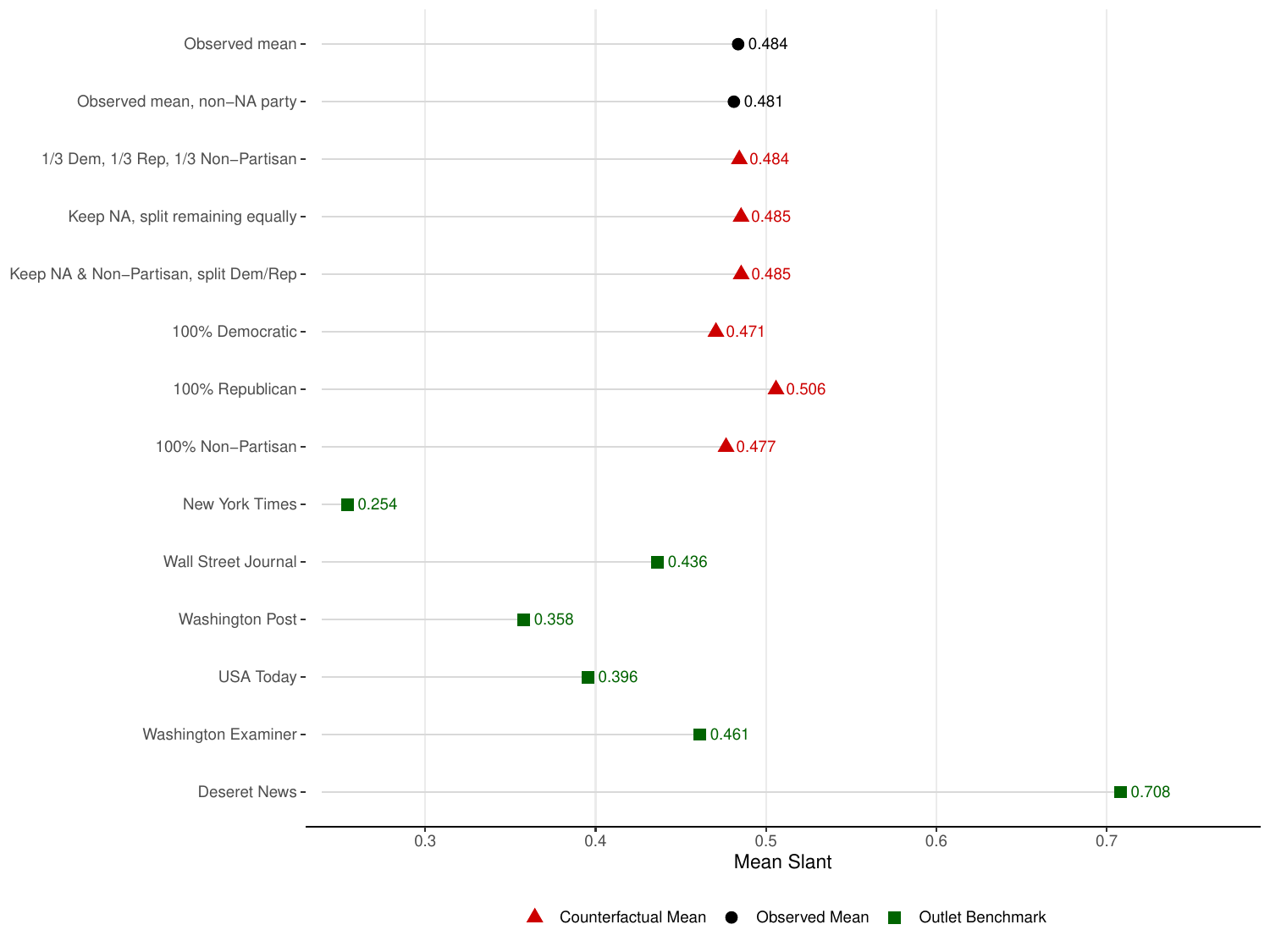}
\par\end{centering}
Notes: Figure plots observed and counterfactual mean slant estimates. Counterfactual
means reweight the distribution of journalist party affiliations (and associated average AKM journalist fixed effects) while holding
the average outlet and time effects fixed. See Section \ref{sec:akm_counterfactuals} for detail.
The figure also reports benchmark observed mean slant for selected outlets, including \emph{The New
York Times}, \emph{The Wall Street Journal}, \emph{The Washington
Post}, \emph{USA Today}, the \emph{Washington Examiner}, and the
\emph{Deseret News}.
\end{figure}

\clearpage
\begin{figure}[H]
\caption{Average Slant by Journalist and Outlet Types\label{appfig-blm-slant}}
\begin{centering}
\includegraphics[width=.8\textwidth,height=.72\textheight,keepaspectratio]{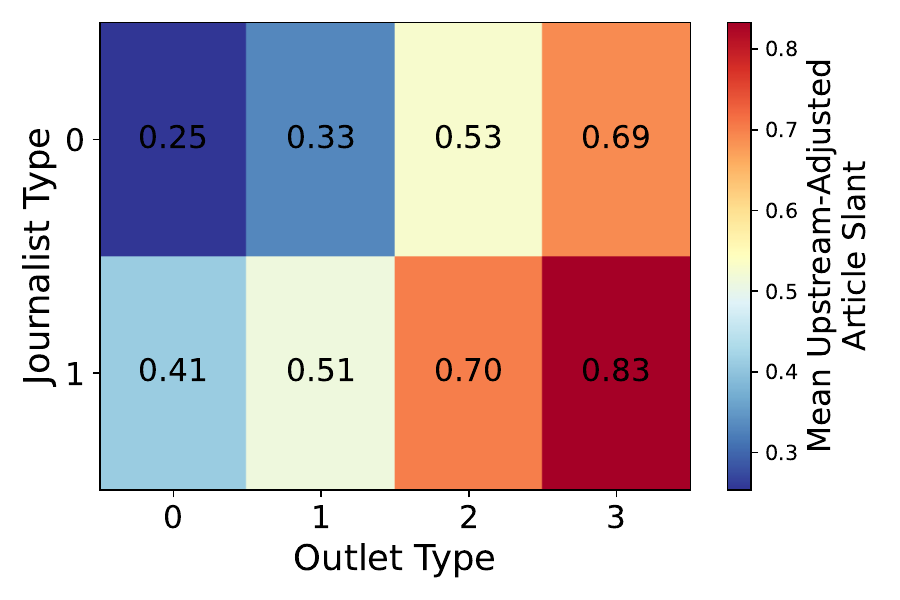}
\par\end{centering}
Notes: Figure shows the estimated mean slant for each combination
of journalist type (rows) and outlet type (columns), from the BLM
model with two journalist types and four outlet types. Outlet types
are ordered by mean slant, from most liberal (type 0) to most conservative
(type 3). The sample contains all credited articles in the two spells around
each journalist's first move. The upstream calendar-adjusted article slant,
with its common level restored, is used to estimate one shared type-by-class
mean matrix across origin and destination roles. These means are displayed directly.
\end{figure}

\clearpage
\begin{figure}[H]
\caption{Distribution of Journalist Type by Outlet Type\label{appfig-blm-sorting}}
\begin{centering}
\includegraphics[width=.8\textwidth,height=.72\textheight,keepaspectratio]{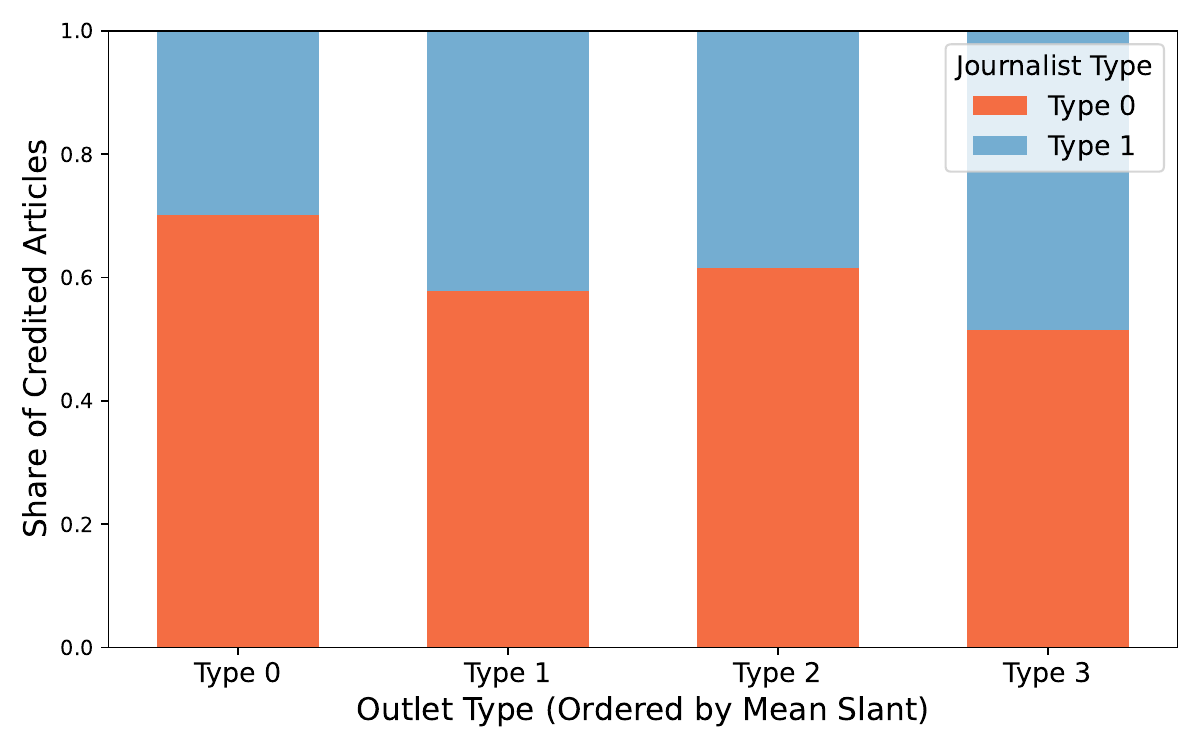}
\par\end{centering}
Notes: Figure shows the share of credited articles attributed to each journalist
type within each outlet type, using soft posterior probabilities. The sample
contains both complete spells around each journalist's first move. Outlet types
are ordered by mean adjusted article slant, from most liberal (type 0) to most conservative (type 3).
\end{figure}

\clearpage
\begin{figure}[H]
\caption{Projection of Journalist Type onto Observable Characteristics\label{appfig-blm-obsevables}}
\begin{centering}
\includegraphics[width=.8\textwidth,height=.72\textheight,keepaspectratio]{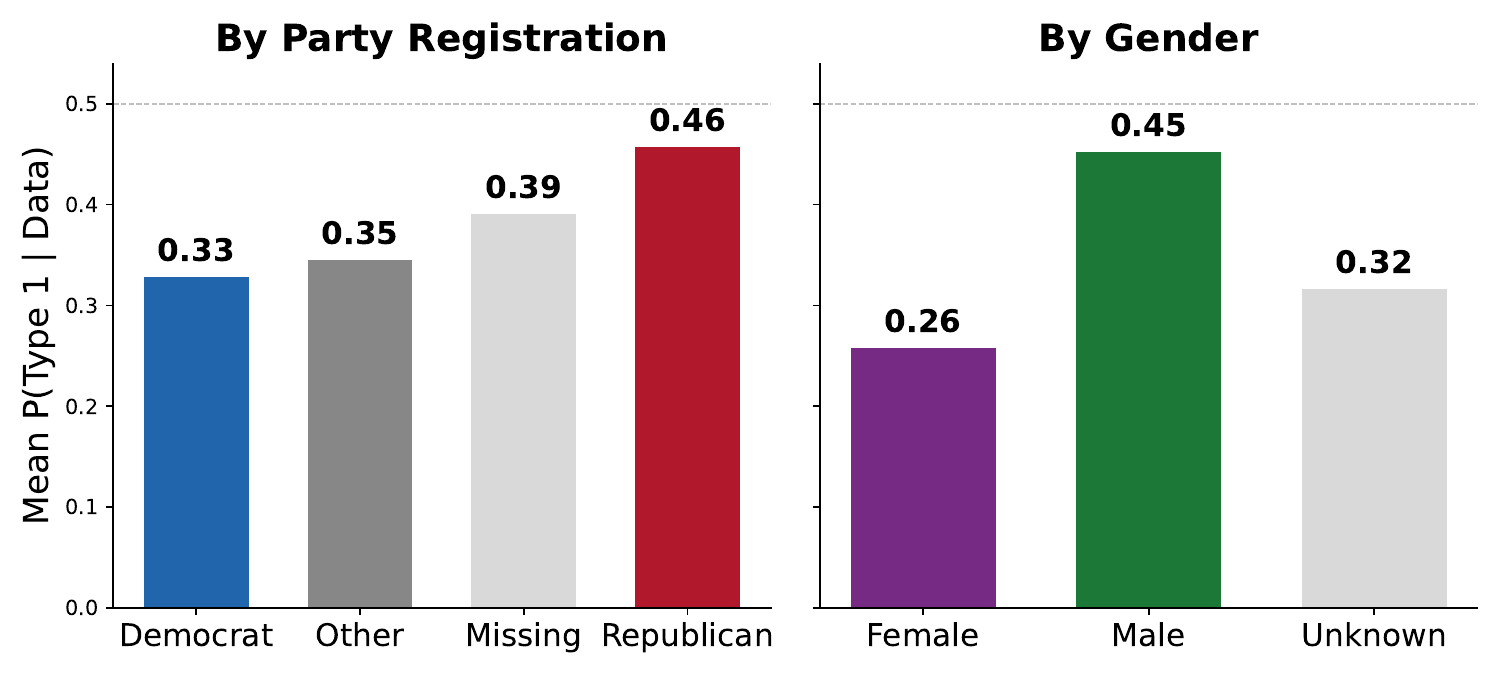}
\par\end{centering}
Notes: Figure shows the mean posterior probability of being classified
as a type-1 (conservative) journalist by party registration (left
panel) and gender (right panel). Posterior probabilities are computed
from the BLM model with two journalist types and four outlet types,
estimated on upstream calendar-adjusted articles in both spells around each
journalist's first move. Each journalist receives equal weight. Party
registration is obtained from voter file records; gender is inferred from first names.
\end{figure}

\clearpage
\begin{figure}[H]
\caption{BLM Variance Decomposition\label{appfig-blm-decomposition}}
\begin{centering}
\includegraphics[width=.8\textwidth,height=.72\textheight,keepaspectratio]{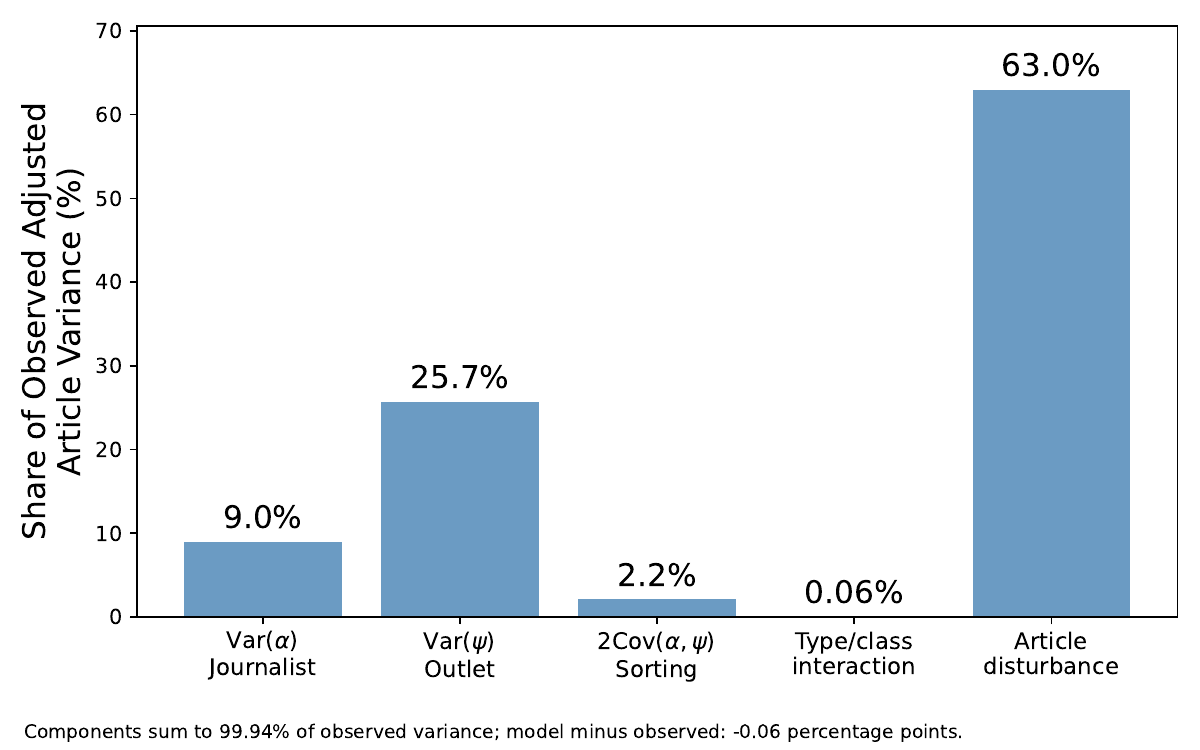}
\par\end{centering}
Notes: Figure reports model variance components as percentages of observed calendar-adjusted
article variance in the two spells around each journalist's first move. The
additive journalist and outlet effects project the shared type-by-class mean
matrix onto additive type indicators using posterior article weights. The five
components are journalist variance, outlet variance, twice their covariance
(sorting), interaction residual variance, and article disturbance variance.
Conditional disturbance variances remain role-specific. Publication-month means
are removed upstream; no additional month effects are estimated. Components are
not rescaled to sum to observed variance; the model-data gap is shown in the plot.
\end{figure}

%% file: linkage_summary_statistics_filled.tex
{\tiny{}

\begin{table}[H]

{\tiny\caption{Journalist Summary Statistics by L2 Linkage Status\label{tab:linkage_summary}}

}{\tiny\par}

\begin{centering}

{\small{}%

\begin{tabular}{lcc}

\hline

 & {\small All} & {\small Linked}\tabularnewline

 & {\small journalists} & {\small to L2}\tabularnewline

\hline

{\small Journalists} & {\small33,571} & {\small10,670}\tabularnewline

{\small Article-author observations} & {\small9,005,189} & {\small2,776,654}\tabularnewline

{\small Outlets represented} & {\small279} & {\small276}\tabularnewline

{\small Mean article slant} & {\small0.475} & {\small0.474}\tabularnewline

{\small SD article slant} & {\small0.285} & {\small0.286}\tabularnewline

{\small Mean outlet slant exposure} & {\small0.473} & {\small0.475}\tabularnewline

{\small Mean articles per journalist} & {\small268.24} & {\small260.23}\tabularnewline

{\small Mean active months per journalist} & {\small25.26} & {\small25.28}\tabularnewline

{\small Mean outlets per journalist} & {\small5.36} & {\small4.07}\tabularnewline

{\small Share eligible movers} & {\small0.093} & {\small0.084}\tabularnewline

{\small Politics and government topic share} & {\small0.100} & {\small0.091}\tabularnewline

\hline

\end{tabular}}{\small\par}

\par\end{centering}

{\small Notes: Table reports summary statistics for cleaned journalist identities in the ProQuest/TDM article sample by linkage status. Linked to L2 indicates that the journalist has a non-empty party affiliation in the author-level L2 output, including affiliations outside the Democratic, Republican, and Non-Partisan recode used in the party-summary table. Article-author observations count all cleaned articles by journalists in each column. The moving share counts journalists with at least one canonical valid move who satisfy the common career cap. Outlet slant exposure is the article-weighted mean outlet slant across each journalist's observed articles. Politics and government topic share is computed among articles with non-missing topic labels.}{\small\par}

\end{table}

}{\tiny\par}

%% file: party_summary_statistics_filled.tex
{\tiny{}

\begin{table}[H]

{\tiny\caption{Journalist Summary Statistics by Party Affiliation\label{tab:party_summary}}

}{\tiny\par}

\begin{centering}

{\small{}%

\begin{tabular}{lcccc}

\hline

 & {\small Democratic} & {\small Republican} & {\small Non-Partisan} & {\small NA}\tabularnewline

\hline

{\small Journalists} & {\small5,438} & {\small2,243} & {\small2,891} & {\small22,999}\tabularnewline

{\small Article-author observations} & {\small1,358,805} & {\small512,070} & {\small879,036} & {\small6,255,278}\tabularnewline

{\small Outlets represented} & {\small270} & {\small261} & {\small264} & {\small279}\tabularnewline

{\small Mean article slant} & {\small0.449} & {\small0.558} & {\small0.466} & {\small0.475}\tabularnewline

{\small SD article slant} & {\small0.277} & {\small0.283} & {\small0.292} & {\small0.285}\tabularnewline

{\small Mean outlet slant exposure} & {\small0.448} & {\small0.538} & {\small0.478} & {\small0.471}\tabularnewline

{\small Mean articles per journalist} & {\small249.872} & {\small228.297} & {\small304.059} & {\small271.980}\tabularnewline

{\small Mean active months per journalist} & {\small25.610} & {\small24.370} & {\small25.416} & {\small25.251}\tabularnewline

{\small Mean outlets per journalist} & {\small4.080} & {\small3.544} & {\small4.480} & {\small5.946}\tabularnewline

{\small Share eligible movers} & {\small0.081} & {\small0.071} & {\small0.101} & {\small0.097}\tabularnewline

{\small Politics and government topic share} & {\small0.072} & {\small0.080} & {\small0.124} & {\small0.105}\tabularnewline

\hline

\end{tabular}}{\small\par}

\par\end{centering}

{\small Notes: Table reports summary statistics for cleaned journalist identities in the ProQuest/TDM article sample by party affiliation. Party groups use the standard analysis recode of journalist party affiliation; NA indicates no observed party affiliation. The moving share counts journalists with at least one canonical valid move who satisfy the common career cap. Outlet slant exposure is the article-weighted mean outlet slant across each journalist's observed articles. Politics and government topic share is computed among articles with non-missing topic labels.}{\small\par}

\end{table}

}{\tiny\par}

%% file: slant_measure_comparison_table_filled.tex
\begin{table}[H]

\centering

\caption{Slant-measure anonymization comparison}

\label{tab:slant_measure_comparison}

\begingroup

\footnotesize

\setlength{\tabcolsep}{3pt}

\renewcommand{\arraystretch}{1.08}

\begin{tabular}{@{}p{0.29\textwidth}>{\centering\arraybackslash}p{0.222\textwidth}>{\centering\arraybackslash}p{0.222\textwidth}>{\centering\arraybackslash}p{0.222\textwidth}@{}}

\toprule

 & \multicolumn{1}{c}{Baseline} & \multicolumn{1}{c}{Geography anonymized} & \multicolumn{1}{c}{Aggressively anonymized} \\

\midrule

\raggedright Validation balanced accuracy & 0.816 & 0.786 & 0.747 \tabularnewline

\raggedright Validation ROC-AUC & 0.901 & 0.876 & 0.833 \tabularnewline

\raggedright Budak et al.\ correlation & 0.307 & 0.279 & 0.217 \tabularnewline

\cmidrule(lr){2-4}

\raggedright $\hat{\rho}_{0}$ (event study, $T=5$) & 0.773 & 0.794 & 0.775 \tabularnewline

\raggedright \textit{95\% CI} & \textit{[0.643, 0.903]} & \textit{[0.668, 0.921]} & \textit{[0.637, 0.914]} \tabularnewline

\raggedright KSS journalist share & 0.085 & 0.102 & 0.145 \tabularnewline

\raggedright Outlet-decomp.\ journalist share & 0.143 & 0.171 & 0.174 \tabularnewline

\raggedright \textit{95\% CI} & \textit{[0.113, 0.189]} & \textit{[0.128, 0.237]} & \textit{[0.126, 0.252]} \tabularnewline

\midrule

\multicolumn{4}{@{}l}{\textit{Top 25 tokens by $|\mathrm{SHAP\ sum}|$ within direction}} \\

\addlinespace[0.25em]

\raggedright Democratic & trump, democrat, democratic, affordable, donald, climate, ibr, rights, gill, care, immigration, civil, violence, tammy, black, ud, repeal, brown, detroit, african, voting, shutdown, cummings, elijah, inequality & trump, democratic, democrat, climate, democrats, care, donald, people, change, women, brown, health, president, immigration, administration, black, affordable, ibr, gill, rights, act, public, violence, schumer, mr & democratic, democrat, democrats, climate, people, change, women, rights, health, president, civil, care, affordable, immigration, trump, environmental, public, group, black, immigrants, violence, racial, income, administration, families \tabularnewline

\cmidrule(lr){2-4}

\raggedright Republican & said, state, republican, news, year, obama, percent, fox, years, house, texas, told, time, just, business, small, city, president, committee, sen, breitbart, gov, tax, government, national & state, republican, said, specific, news, fox, breitbart, obama, business, year, small, reported, committee, rob, jobs, told, abbott, percent, gov, local, paul, border, clinton, businesses, tuesday & republican, said, specific, state, individual, country, governor, conservative, business, year, agency, month, years, senator, small, region, time, news, economy, percent, military, jobs, economic, growth, businesses \tabularnewline

\bottomrule

\addlinespace[0.35em]

\multicolumn{4}{@{}p{0.98\textwidth}@{}}{\scriptsize Notes: The non-anonymized slant measure in the baseline column remains the canonical measure used in the paper. Validation statistics use a common hold-out sample of politician-tweeted articles, and the Budak et al.\ row reports Spearman rank correlations on a common sample of articles rated by blinded coders in \citet{Budak2016FairBalanced}; the samples contain 26,499 and 5,673 articles, respectively. The event-study row reports $\hat{\rho}_{0}$, the event-time-zero gap-closure coefficient from the $T=5$ specification corresponding to Figure \ref{fig:eventstudy}, Panel A. Event-study article outcomes and leave-focal-journalist-out outlet gaps are residualized by publication month separately for each slant measure; squared-gap weights are measure-specific, while structural event rows and estimator rules are held fixed. The KSS row reports the mover-only leave-match-out specification in Table \ref{tab:akm_kss_table}, Panel B, column 2, and has no confidence interval. The outlet-decomposition row reports the calendar-residualized, article-weighted cross-fit specification corresponding to Table \ref{tab:outlet_decomp}, column 1. Comparisons are across slant measures within a row, not across the three distinct event-study, KSS, and outlet-decomposition estimands. Brackets report 95 percent confidence intervals. SHAP uses Shapley values to attribute a model prediction to its input features (\citealp{Lundberg2017UnifiedApproach}). Token lists report the 25 largest absolute signed SHAP sums within each direction and separately trained model after restricting to meaningful tokens observed in at least ten sampled documents.} \tabularnewline

\end{tabular}

\endgroup

\end{table}

%% file: consistent_mover_summary_statistics_filled.tex
{\tiny{}

\begin{table}[H]

{\tiny\caption{Journalist Summary Statistics by Mover Status\label{tab:consistent_mover_summary}}

}{\tiny\par}

\begin{centering}

{\small{}%

\begin{tabular}{lcccc}

\hline

 & {\small All} & {\small Single-outlet} & {\small$T=5$ consistent} & {\small$T=10$ consistent}\tabularnewline

 & {\small journalists} & {\small journalists} & {\small movers} & {\small movers}\tabularnewline

\hline

{\small Journalists} & {\small33,571} & {\small12,248} & {\small3,114} & {\small1,866}\tabularnewline

{\small Article-author observations} & {\small9,005,189} & {\small1,726,099} & {\small1,999,915} & {\small1,302,504}\tabularnewline

{\small Outlets represented} & {\small279} & {\small262} & {\small272} & {\small267}\tabularnewline

{\small Mean article slant} & {\small0.475} & {\small0.488} & {\small0.470} & {\small0.467}\tabularnewline

{\small SD article slant} & {\small0.285} & {\small0.283} & {\small0.286} & {\small0.288}\tabularnewline

{\small Mean outlet slant exposure} & {\small0.473} & {\small0.478} & {\small0.472} & {\small0.472}\tabularnewline

{\small Mean articles per journalist} & {\small268.243} & {\small140.929} & {\small642.233} & {\small698.019}\tabularnewline

{\small Mean active months per journalist} & {\small25.264} & {\small19.706} & {\small40.368} & {\small43.189}\tabularnewline

{\small Mean outlets per journalist} & {\small5.357} & {\small1.000} & {\small9.781} & {\small9.454}\tabularnewline

{\small Share eligible movers} & {\small0.093} & {\small0.000} & {\small1.000} & {\small1.000}\tabularnewline

{\small Politics and government topic share} & {\small0.100} & {\small0.099} & {\small0.107} & {\small0.112}\tabularnewline

\hline

\end{tabular}}{\small\par}

\par\end{centering}

{\small Notes: Table reports summary statistics for cleaned journalist identities in the ProQuest/TDM article sample. Single-outlet journalists publish at one outlet in the broad cleaned sample. The moving share counts journalists with at least one canonical valid move who satisfy the common career cap. The $T=10$ consistent mover column is a subset of the $T=5$ consistent mover column. Article-author observations in mover columns count all cleaned articles by journalists with at least one qualifying consistent move, not only event-window observations. Outlet slant exposure is the article-weighted mean outlet slant across each journalist's observed articles. Politics and government topic share is computed among articles with non-missing topic labels.}{\small\par}

\end{table}

}{\tiny\par}

%% file: move_summary_statistics_filled.tex
{\tiny{}

\begin{table}[H]

{\tiny\caption{Move-Level Summary Statistics\label{tab:move_summary}}

}{\tiny\par}

\begin{centering}

{\small{}%

\begin{tabular}{lcc}

\hline

 & {\small$T=5$} & {\small$T=10$}\tabularnewline

\hline

{\small Move pairs} & {\small5,429} & {\small2,437}\tabularnewline

{\small Unique journalists} & {\small3,114} & {\small1,866}\tabularnewline

{\small Origin outlets} & {\small255} & {\small234}\tabularnewline

{\small Destination outlets} & {\small243} & {\small228}\tabularnewline

{\small Event-window article observations} & {\small54,290} & {\small48,740}\tabularnewline

\hline

{\small Mean origin outlet slant} & {\small0.485} & {\small0.488}\tabularnewline

{\small Mean destination outlet slant} & {\small0.474} & {\small0.466}\tabularnewline

{\small Mean destination-origin outlet slant gap} & {\small-0.011} & {\small-0.022}\tabularnewline

{\small Mean absolute outlet slant gap} & {\small0.117} & {\small0.124}\tabularnewline

{\small Median absolute outlet slant gap} & {\small0.084} & {\small0.091}\tabularnewline

{\small Share moving to more conservative outlet} & {\small0.46} & {\small0.43}\tabularnewline

{\small Share large moves} & {\small0.44} & {\small0.47}\tabularnewline

\hline

{\small Mean pre-move article slant} & {\small0.480} & {\small0.486}\tabularnewline

{\small Mean post-move article slant} & {\small0.465} & {\small0.460}\tabularnewline

{\small Mean post-minus-pre article slant} & {\small-0.015} & {\small-0.026}\tabularnewline

{\small Event-study $t=0$ gap-closure coefficient} & {\small0.773} & {\small0.809}\tabularnewline

\hline

\end{tabular}}{\small\par}

\par\end{centering}

{\small Notes: Table reports move-pair summary statistics for the balanced event-study mover sample. Column $T$ retains move pairs with exactly $T$ articles before the move and $T$ articles at or after the move, using rows with $|t|\leq T$ and excluding $t=T$. Outlet and article slant summary rows use raw slant; outlet rows use leave-focal-author-out means and gaps. More conservative moves and large moves (absolute gap greater than 0.1) are defined using these raw leave-out gaps. The event-study row reports the normalized gap-closure coefficient at $t=0$ from the main weighted specification, which instead residualizes article slant and the leave-out outlet gap by publication month and includes move-pair fixed effects.}{\small\par}

\end{table}

}{\tiny\par}

%% file: appendix_sample_construction_filled.tex
\begin{table}[p]

\centering

\caption{Sample construction counts}

\label{tab:appendix_sample_construction}

\vspace{-0.4em}

\begingroup

\fontsize{10.5}{11.2}\selectfont

\setlength{\tabcolsep}{2.4pt}

\renewcommand{\arraystretch}{1.00}

\begin{tabular}{@{}p{0.345\textwidth}p{0.165\textwidth}rrr@{}}

\toprule

Cleaning step & Unit & Rows & \shortstack{Unique\\articles} & \shortstack{Author--\\articles} \\

\midrule

\multicolumn{5}{l}{\emph{Panel A. Tweet-linked articles}} \\

Tweeted URL occurrences & URL occurrences & 1,196,782 & 1,107,105 & -- \\

Sampled for scraping & scrape rows & 600,000 & 569,777 & 359,588 \\

Join scraped text to tweets & article--tweets & 1,365,407 & 569,777 & 359,588 \\

Complete article/tweet metadata & article--tweets & 844,437 & 328,966 & 355,021 \\

Source text $>$ 1,000 chars & article--tweets & 332,500 & 214,816 & 306,004 \\

Detected article authors & article--tweets & 195,211 & 116,092 & 306,004 \\

Exclude .gov URLs & article--tweets & 192,933 & 114,450 & 303,503 \\

Democratic or Republican shares & article--tweets & 191,616 & 113,719 & 301,668 \\

Final sentiment / cleaning input & article--tweets & 191,584 & 113,704 & 301,631 \\

Drop invalid articles & article--tweets & 189,243 & 112,004 & 297,682 \\

Drop criticizing tweets & article--tweets & 184,384 & 108,986 & 290,317 \\

Post-cleaning text $>$ 1,000 chars & article--tweets & 182,697 & 107,878 & 287,584 \\

Keep English articles & article--tweets & 182,538 & 107,720 & 287,255 \\

Training split & article--tweets & 127,777 & 78,213 & 209,403 \\

Validation candidate & article--tweets & 54,761 & 36,629 & 97,945 \\

Validation after overlap removal & article--tweets & 26,522 & 26,034 & 68,154 \\

Training on/after 2011-06-01 & article--tweets & 127,707 & 78,143 & 209,215 \\

Validation on/after 2011-06-01 & article--tweets & 26,499 & 26,011 & 68,084 \\

\midrule

\multicolumn{5}{l}{\emph{Panel B. Budak articles}} \\

Raw rows with URL and text & article rows & 14,476 & 7,694 & 21,161 \\

Drop invalid / clean non-text & article rows & 14,029 & 7,490 & 20,023 \\

Text $>$ 1,000 chars & article rows & 10,735 & 5,724 & 17,960 \\

Deduplicate URLs / construct slant & article rows & 5,724 & 5,724 & 17,960 \\

\midrule

\multicolumn{5}{l}{\emph{Panel C. ProQuest/TDM analysis data}} \\

Raw merged sample & article rows & 11,006,646 & 11,006,646 & 11,637,445 \\

News articles with authors & article rows & 10,992,582 & 10,992,582 & 11,622,809 \\

Keep clean parsed author names & author--articles & 13,173,727 & 10,992,582 & 11,622,809 \\

Drop duplicate author--articles & author--articles & 11,622,800 & 10,992,582 & 11,622,800 \\

Drop duplicate-mapped articles & author--articles & 9,288,608 & 8,769,535 & 9,288,608 \\

Serious authors with mapped group ID & author--articles & 9,005,189 & 8,521,826 & 9,005,189 \\

\bottomrule

\end{tabular}

\vspace{0.35em}

\begin{minipage}{0.98\textwidth}

\fontsize{9.0}{9.8}\selectfont

\textit{Notes:} Panels A--C show how the tweet-linked, Budak, and ProQuest/TDM samples are built. Counts are measured after each listed step; when a row combines steps, they are measured after the final step. Unit says what the row counts. Unique articles are distinct articles, while author--articles are distinct author--article pairs. Training and validation are separate branches, not successive filters. The common-sample model inputs are the training split and the validation sample left after removing overlap with training. Date-filtered rows include observations on or after June 1, 2011.

\end{minipage}

\endgroup

\end{table}

%% file: budak_article_filled.tex
{\tiny{}

\begin{table}[H]{\tiny\caption{Prediction of Human Perceived Slant from \citet{Budak2016FairBalanced}, Article Level\label{tab:budak_et_al_article}}}

{\tiny\par}

\begin{centering}

{\small{}%

\begin{tabular}{lcccc}

\hline 

{\small Dependent Variable:} & \multicolumn{4}{c}{{\small Human Perceived Slant (SD)}}\tabularnewline

\hline 

 & {\small (1)} & {\small (2)} & {\small (3)} & {\small (4)}\tabularnewline

\hline 

{\small RoBERTa} & {\small0.291} & {\small.} & {\small.} & {\small0.261}\tabularnewline

 & {\small (0.013)} & {\small.} & {\small.} & {\small (0.015)}\tabularnewline

{\small GST (re-estimated)} & {\small.} & {\small0.116} & {\small.} & {\small0.004}\tabularnewline

 & {\small.} & {\small (0.013)} & {\small.} & {\small (0.014)}\tabularnewline

{\small FastText} & {\small.} & {\small.} & {\small0.188} & {\small0.055}\tabularnewline

 & {\small.} & {\small.} & {\small (0.013)} & {\small (0.015)}\tabularnewline

\hline 

{\small Observations} & {\small5,673} & {\small5,673} & {\small5,673} & {\small5,673}\tabularnewline

{\small$R^{2}$} & {\small0.084} & {\small0.013} & {\small0.035} & {\small0.087}\tabularnewline

\hline 

\end{tabular}}{\small\par}

\par\end{centering}

{\small Notes: Table shows the results of regressing blinded-coder human perceived slant from \citet{Budak2016FairBalanced} on various text-based models of slant. Sample is composed of a random sample of articles from \citet{Budak2016FairBalanced}. For some selected articles, we were unable to scrape the underlying text and these articles are removed. In \citet{Budak2016FairBalanced}, each article is scored on a 5-point favorability score indicating the extent to which the article is favorable to Republicans and a separate score for how favorable the article is to Democrats. We take the difference between these two scores as our measure of perceived slant. Each article is rated twice---once by a coder who does not observe the outlet's name, and once by a coder who does observe the outlet's name. Following \citet{Budak2016FairBalanced}, our human perceived slant measure uses ratings from coders who do not observe the outlet's name. The $R^{2}$ from regressing unblinded coder slant on blinded coder slant is \BudakInterRaterRsq. The RoBERTa measure is our main measure of slant. The FastText measure is a measure of slant using Facebook's FastText model and trained on a dataset similar to the RoBERTa model. %

The GST measure is the Republican posterior probability from a phrase-based classifier following \citet{Gentzkow2019MeasuringGroupDifferences}, re-estimated on the same article training sample as RoBERTa, with equal party priors and the training-weighted average month effect. Human perceived slant is standardized to mean zero and standard deviation one in the full validation sample. All text-based model slant estimates are normalized to have a standard deviation of one in that sample. Regressions include an intercept; parentheses report conventional OLS standard errors.}{\small\par}

\end{table}

}{\tiny\par}

%% file: akm_kss_demo_table_filled.tex
{\tiny{}

\begin{table}[H]

{\tiny\caption{AKM Decompositions\label{tab:akm_kss_demo_table}}

}{\tiny\par}

\begin{centering}

{\small{}%

\begin{tabular}{lcccc}

\hline 

 & {\small (1)} & {\small (2)} & {\small (3)} & {\small (4)}\tabularnewline

\hline 

{\small Sample:} & {\small Women} & {\small Men} & {\small White} & {\small Non-White}\tabularnewline

 &  &  &  & \tabularnewline

\hline 

 &  &  &  & \tabularnewline

\multicolumn{5}{c}{{\small\emph{Panel A: Plug-in Estimator}}}\tabularnewline

\hline 

\multicolumn{5}{c}{{\small Share of Variance}{\scriptsize , $x_{ijt}$}}\tabularnewline

\hline 

{\small Journalist, $\text{var}(\gamma_{i})$} & {\small0.180} & {\small0.214} & {\small0.191} & {\small0.272}\tabularnewline

{\small Outlet, $\text{var}(\alpha_{j})$} & {\small0.234} & {\small0.217} & {\small0.232} & {\small0.236}\tabularnewline

{\small Sorting, $2\times\text{cov}(\gamma_{i},\alpha_{j})$} & {\small-0.007} & {\small0.031} & {\small0.029} & {\small-0.072}\tabularnewline

\hline 

\multicolumn{5}{c}{{\small Correlation}}\tabularnewline

\hline 

{\small Sorting, $\text{cor}(\gamma_{i},\alpha_{j})$} & {\small-0.018} & {\small0.072} & {\small0.068} & {\small-0.143}\tabularnewline

\hline 

 &  &  &  & \tabularnewline

\multicolumn{5}{c}{{\small\emph{Panel B: Leave-Out Estimator (\citealt{Kline2020LeaveOutVariance})}}}\tabularnewline

\hline 

\multicolumn{5}{c}{{\small Share of Variance}{\scriptsize , $x_{ijt}$}}\tabularnewline

\hline 

{\small Journalist, $\text{var}(\gamma_{i})$} & {\small0.116} & {\small0.166} & {\small0.152} & {\small0.206}\tabularnewline

{\small Outlet, $\text{var}(\alpha_{j})$} & {\small0.190} & {\small0.182} & {\small0.203} & {\small0.174}\tabularnewline

{\small Sorting, $2\times\text{cov}(\gamma_{i},\alpha_{j})$} & {\small0.074} & {\small0.098} & {\small0.083} & {\small0.041}\tabularnewline

\hline 

\multicolumn{5}{c}{{\small Correlation}}\tabularnewline

\hline 

{\small Sorting, $\text{cor}(\gamma_{i},\alpha_{j})$} & {\small0.251} & {\small0.282} & {\small0.237} & {\small0.107}\tabularnewline

\hline 

\end{tabular}}{\small\par}

\par\end{centering}

Notes: Table reports AKM variance decomposition exercises with publication-month fixed effects across various samples. Both panels use the leave-out connected set: Panel A reports the plug-in estimator and Panel B reports the \citet{Kline2020LeaveOutVariance} estimator. Columns (1)--(4) restrict to various groups of journalists according to name-based predictions of gender and race.

\end{table}

}{\tiny\par}

%% file: akm_table_filled.tex
{\tiny{}

\begin{table}[H]

{\tiny\caption{AKM Decompositions, Full Connected Sample\label{tab:akm_table}}

}{\tiny\par}

\begin{centering}

{\scriptsize{}%

\begin{tabular}{lcccccccccccccc}

\hline 

 & {\scriptsize (1)} & {\scriptsize (2)} & {\scriptsize (3)} & {\scriptsize (4)} & {\scriptsize (5)} & {\scriptsize (6)} & {\scriptsize (7)} & {\scriptsize (8)} & {\scriptsize (9)} & {\scriptsize (10)} & {\scriptsize (11)} & {\scriptsize (12)} & {\scriptsize (13)} & {\scriptsize (14)}\tabularnewline

\cline{2-15}

{\scriptsize Sample:} & {\scriptsize All} & {\scriptsize Dropping} & {\scriptsize Dropping} & {\scriptsize Max. 2} & {\scriptsize Drop 1 Mo.} & {\scriptsize Top 20} & {\scriptsize$T=5$} & {\scriptsize$T=5$} & {\scriptsize$T=10$} & {\scriptsize Parent} & {\scriptsize Women} & {\scriptsize Men} & {\scriptsize White} & {\scriptsize Non-White}\tabularnewline

 &  & {\scriptsize Stayers (Full)} & {\scriptsize Stayers (Large)} & {\scriptsize Outlets} & {\scriptsize Moves} & {\scriptsize Outlets} & {\scriptsize Full} & {\scriptsize Large} & {\scriptsize Full} & {\scriptsize Company} &  &  &  & \tabularnewline

\hline 

 & \multicolumn{14}{c}{{\scriptsize Share of Variance, $x_{ijt}$}}\tabularnewline

\cline{2-15}

{\scriptsize Journalist, $\text{var}(\gamma_{i})$} & {\scriptsize0.194} & {\scriptsize0.167} & {\scriptsize0.185} & {\scriptsize0.201} & {\scriptsize0.193} & {\scriptsize0.266} & {\scriptsize0.235} & {\scriptsize0.272} & {\scriptsize0.215} & {\scriptsize0.352} & {\scriptsize0.179} & {\scriptsize0.218} & {\scriptsize0.191} & {\scriptsize0.305}\tabularnewline

{\scriptsize Outlet, $\text{var}(\alpha_{j})$} & {\scriptsize0.217} & {\scriptsize0.224} & {\scriptsize0.264} & {\scriptsize0.216} & {\scriptsize0.227} & {\scriptsize0.094} & {\scriptsize0.188} & {\scriptsize0.248} & {\scriptsize0.210} & {\scriptsize0.082} & {\scriptsize0.246} & {\scriptsize0.223} & {\scriptsize0.237} & {\scriptsize0.275}\tabularnewline

{\scriptsize Time, $\text{var}(\tau_{t})$} & {\scriptsize0.025} & {\scriptsize0.027} & {\scriptsize0.022} & {\scriptsize0.025} & {\scriptsize0.025} & {\scriptsize0.045} & {\scriptsize0.029} & {\scriptsize0.024} & {\scriptsize0.029} & {\scriptsize0.022} & {\scriptsize0.027} & {\scriptsize0.024} & {\scriptsize0.024} & {\scriptsize0.035}\tabularnewline

{\scriptsize Sorting, $2\times\text{cov}(\gamma_{i},\alpha_{j})$} & {\scriptsize0.038} & {\scriptsize0.040} & {\scriptsize0.001} & {\scriptsize0.030} & {\scriptsize0.029} & {\scriptsize-0.001} & {\scriptsize0.056} & {\scriptsize-0.022} & {\scriptsize0.028} & {\scriptsize0.033} & {\scriptsize-0.012} & {\scriptsize0.022} & {\scriptsize0.025} & {\scriptsize-0.143}\tabularnewline

{\scriptsize Other Components} & {\scriptsize0.526} & {\scriptsize0.542} & {\scriptsize0.529} & {\scriptsize0.527} & {\scriptsize0.526} & {\scriptsize0.596} & {\scriptsize0.492} & {\scriptsize0.477} & {\scriptsize0.518} & {\scriptsize0.510} & {\scriptsize0.561} & {\scriptsize0.513} & {\scriptsize0.524} & {\scriptsize0.528}\tabularnewline

\hline 

 & \multicolumn{14}{c}{{\scriptsize Correlation}}\tabularnewline

\cline{2-15}

{\scriptsize Sorting, $\text{cor}(\gamma_{i},\alpha_{j})$} & {\scriptsize0.091} & {\scriptsize0.103} & {\scriptsize0.001} & {\scriptsize0.073} & {\scriptsize0.069} & {\scriptsize-0.004} & {\scriptsize0.133} & {\scriptsize-0.042} & {\scriptsize0.066} & {\scriptsize0.096} & {\scriptsize-0.029} & {\scriptsize0.050} & {\scriptsize0.058} & {\scriptsize-0.246}\tabularnewline

\hline 

 & \multicolumn{14}{c}{{\scriptsize Sample Size}}\tabularnewline

\cline{2-15}

{\scriptsize Observations} & {\scriptsize7065499} & {\scriptsize1482338} & {\scriptsize489191} & {\scriptsize6790652} & {\scriptsize7021584} & {\scriptsize2437070} & {\scriptsize54290} & {\scriptsize21560} & {\scriptsize48700} & {\scriptsize4127869} & {\scriptsize2355089} & {\scriptsize4512701} & {\scriptsize5684989} & {\scriptsize583770}\tabularnewline

{\scriptsize Journalists (All)} & {\scriptsize30504} & {\scriptsize4091} & {\scriptsize1399} & {\scriptsize29854} & {\scriptsize29991} & {\scriptsize11138} & {\scriptsize3114} & {\scriptsize1399} & {\scriptsize1864} & {\scriptsize20866} & {\scriptsize12135} & {\scriptsize17253} & {\scriptsize23984} & {\scriptsize2635}\tabularnewline

{\scriptsize Journalists (2+ outlets)} & {\scriptsize4091} & {\scriptsize4091} & {\scriptsize1399} & {\scriptsize3460} & {\scriptsize3420} & {\scriptsize594} & {\scriptsize3114} & {\scriptsize1399} & {\scriptsize1864} & {\scriptsize1272} & {\scriptsize1443} & {\scriptsize2556} & {\scriptsize3261} & {\scriptsize411}\tabularnewline

{\scriptsize Outlets} & {\scriptsize266} & {\scriptsize266} & {\scriptsize237} & {\scriptsize265} & {\scriptsize266} & {\scriptsize17} & {\scriptsize260} & {\scriptsize237} & {\scriptsize246} & {\scriptsize33} & {\scriptsize247} & {\scriptsize260} & {\scriptsize262} & {\scriptsize172}\tabularnewline

{\scriptsize Months} & {\scriptsize72} & {\scriptsize72} & {\scriptsize72} & {\scriptsize72} & {\scriptsize72} & {\scriptsize72} & {\scriptsize72} & {\scriptsize72} & {\scriptsize72} & {\scriptsize72} & {\scriptsize72} & {\scriptsize72} & {\scriptsize72} & {\scriptsize72}\tabularnewline

\hline 

\end{tabular}}{\scriptsize\par}

\par\end{centering}

\begin{centering}

\par\end{centering}

{\small Notes: Table reports AKM variance decomposition exercises using the plug-in estimator on the ordinary connected set with various samples. All columns use raw article slant, the raw outcome-variance denominator, and publication-month fixed effects. Column (1) uses the full sample. Column (2) drops journalists observed at only one retained outlet. Column (3) retains only origin and destination spells around moves whose absolute calendar-residualized leave-focal-author-out outlet gap exceeds 0.1, counting each author--article observation once. Column (4) drops journalists that appear in more than two outlets. Column (5) restricts to tenures of longer than one month in duration. Column (6) restricts to the top 20 outlets according to weekend circulation. Column (7) uses the $T=5$ event study sample. Column (8) is the same as Column (7) except that it restricts to moves whose absolute calendar-residualized leave-focal-author-out outlet gap exceeds 0.1. Column (9) uses the $T=10$ event study sample. Column (10) treats all outlets owned by a single ``Parent Company'' as the same outlet and restricts attention to outlets for which we observe the parent company. Columns (11)--(14) restrict to various groups of journalists according to name-based predictions of gender and race.}{\small\par}

\end{table}

}{\tiny\par}

%% file: modeling_selection_full_table.tex
\begin{tabular}{lcc lcc}
\toprule
 & Estimate & SE & & Estimate & SE \\
\midrule
$\beta$ & 0.674 & 0.053 & $\delta$ ideal, Dem & 0.537 & 0.027 \\
$\beta/(\beta+1)$ & 0.403 & 0.019 & $\delta$ ideal, Rep & 0.636 & 0.028 \\
$\tau_\zeta$ & 1.162 & 0.057 & $\delta$ ideal, NP & 0.571 & 0.028 \\
$\tau_{\delta,\text{Dem}}$ & 0.679 & 0.090 & $\delta$ ideal, NA & 0.562 & 0.017 \\
$\tau_{\delta,\text{Rep}}$ & 0.793 & 0.081 & class position $0$ & 0.132 & --- \\
$\tau_{\delta,\text{NP}}$ & 0.937 & 0.184 & class position $1$ & 0.253 & 0.005 \\
$\tau_{\delta,\text{NA}}$ & 0.876 & 0.067 & class position $2$ & 0.391 & 0.015 \\
$\lambda$ & 0.046 & 0.008 & class position $3$ & 0.629 & 0.024 \\
$\gamma$ & 0.019 & 0.004 & class position $4$ & 0.842 & 0.017 \\
$\tilde\kappa_d$ & 0.350 & 0.020 & $\tilde\alpha_1$ & -0.276 & 0.049 \\
$\kappa_n$ & 0.677 & 0.009 & $\tilde\alpha_2$ & -1.033 & 0.064 \\
$s$ & 3.048 & 0.031 & $\tilde\alpha_3$ & -0.881 & 0.071 \\
$s_c$ & 4.526 & 0.037 & $\tilde\alpha_4$ & -1.234 & 0.089 \\
$\sigma_e$ & 0.204 & first step &  &  &  \\
\bottomrule
\end{tabular}

%% file: structural_variance_table.tex
\begin{table}[H]
\caption{Model-Implied Variance of Article Slant\label{tab:structural_article_variance}}
\begin{centering}
\begin{tabular}{lrr}
\toprule
Component & Variance (slant$^2$) & Share (\%) \\
\midrule
Journalist: $\operatorname{Var}(J)$ & \StructuralSlantVarianceJournalist & \StructuralSlantVarianceJournalistTotalPercent \\
Outlet class: $\operatorname{Var}(O)$ & \StructuralSlantVarianceOutletClass & \StructuralSlantVarianceOutletClassTotalPercent \\
Accepted offer deviation: $\operatorname{Var}(X)$ & \StructuralSlantVarianceOfferShock & \StructuralSlantVarianceOfferShockTotalPercent \\
$2\operatorname{Cov}(J,O)$ & \StructuralSlantVarianceTwiceCovJournalistOutletClass & \StructuralSlantVarianceTwiceCovJournalistOutletClassTotalPercent \\
$2\operatorname{Cov}(J,X)$ & \StructuralSlantVarianceTwiceCovJournalistOfferShock & \StructuralSlantVarianceTwiceCovJournalistOfferShockTotalPercent \\
$2\operatorname{Cov}(O,X)$ & \StructuralSlantVarianceTwiceCovOutletClassOfferShock & \StructuralSlantVarianceTwiceCovOutletClassOfferShockTotalPercent \\
Independent article noise: $\sigma_e^2$ & \StructuralSlantVarianceArticleNoise & \StructuralSlantVarianceArticleNoiseTotalPercent \\
\midrule
Systematic variance: $\operatorname{Var}(J+O+X)$ & \StructuralSlantVarianceSystematic & \StructuralSlantVarianceSystematicTotalPercent \\
Total article variance & \StructuralSlantVarianceTotal & 100.0 \\
\bottomrule
\end{tabular}
\par\end{centering}
\smallskip{}

{\small Notes: The table reports fitted-model variance accounting for stationary
published article credits at observed journalist party composition and fixed
outlet-class shares. For an accepted contract from class $k$, define
$J=\frac{\beta}{1+\beta}\delta_i$,
$O=\frac{1}{1+\beta}\delta_k$, and
$X=\frac{1}{1+\beta}\xi_{ikt}$, so $x^*_{ikt}=J+O+X$.
The class baseline $\delta_k=\Lam(b_k)$ is the median preference before selection;
the accepted deviation $X$ need not have mean zero. Weights reflect the stationary
distribution of published credits, including entry selection and spell duration.
All three doubled covariances retain their signs. The first seven rows sum to
total model-implied article variance, including independent article noise
$\sigma_e^2$; this total is the denominator for every percentage. The outlet term
describes class baselines, and the components are accounting terms rather than
causal effects. Entries are point estimates conditional on the fit and numerical
integration grid.}{\small\par}
\end{table}